\documentclass[fleqn,usenatbib]{mnras}

\usepackage{newtxtext,newtxmath}
\usepackage[T1]{fontenc}

\DeclareRobustCommand{\VAN}[3]{#2}
\let\VANthebibliography\thebibliography
\def\thebibliography{\DeclareRobustCommand{\VAN}[3]{##3}\VANthebibliography}

\usepackage{graphicx}	% Including figure files
\usepackage{amsmath}	% Advanced maths commands
\usepackage{hyperref}
\hypersetup{colorlinks=True,citecolor=blue}
\usepackage{xcolor}

\def\ms{\hbox{\,m\,s$^{-1}$}}         %m.s -1
\def\cms{\hbox{\,cm\,s$^{-1}$}}       %cm.s -1
\def\m2s2{\hbox{\,m$^{2}$\,s$^{-2}$}} %m2.s -2
\def\kms{\hbox{\,km\,s$^{-1}$}}       %km.s -1
\def\S/N{$S/N_{cont}$}

\def\logrhk{$\log$(R$^{\prime}_{\text{HK}}$) }
\def\ang{\text{\AA}}

\def\Halpha{H\hspace{0.1ex}$\alpha$ }

\def\CaK{Ca\,II\,\,K }
\def\CaH{Ca\,II\,\,H }
\def\CaHK{Ca\,II\,\,H {\&} K }
\def\K3{K$_3$ } 
\def\K2{K$_2$ }
\def\K2v{K$_{2V}$ }
\def\K2r{K$_{2R}$ }

\def\Sidx{$S_{\text{index}}$}

\title[Stellar surface information from Ca H\&K lines]{Stellar surface information from the Ca II H\&K lines - II. Defining better activity proxies}

\author[M. Cretignier]{M. Cretignier$^{1}$\thanks{E-mail: michael.cretignier@physics.ox.ac.uk},
N.C. Hara$^{2}$,
A.G.M. Pietrow$^{3}$,
Y. Zhao$^{4}$,
H. Yu$^{1}$,
X. Dumusque$^{4}$,
A. Sozzetti$^{5}$,
C. Lovis$^{4}$, \newauthor
and S. Aigrain$^{1}$
\\
$^{1}$Department of Physics, University of Oxford, OX13RH Oxford, UK \\
$^{2}$Aix Marseille Univ, CNRS, CNES, LAM, Marseille, France\\
$^{3}$Leibniz-Institut für Astrophysik Potsdam (AIP), An der Sternwarte 16, 14482 Potsdam, Germany\\
$^{4}$Astronomy Department of the University of Geneva, 51 ch. de Pegasi, 1290 Versoix, Switzerland\\
$^{5}$INAF–Osservatorio, Astrofisico di Torino, Via Osservatorio 20, 10025 Pino Torinese, Italy\\
}

\date{Accepted XXX. Received YYY; in original form ZZZ}

\pubyear{2023}

\begin{document}
\label{firstpage}
\pagerange{\pageref{firstpage}--\pageref{lastpage}}
\maketitle

\begin{abstract}
   In our former paper I, we showed on the Sun that different active regions possess unique intensity profiles on the Ca II H \& K lines. We now extend the analysis by showing how those properties can be used on real stellar observations, delivering more powerful activity proxies for radial velocity correction. More information can be extracted on rotational timescale from the \CaHK lines than the classical indicators: S-index and $\log(R^{'}_{\text{HK}})$. For high-resolution HARPS observations of $\alpha$ Cen B, we apply a principal and independent component analysis on the \CaHK spectra time-series to disentangle the different sources that contribute to the disk-integrated line profiles. While the first component can be understood as a denoised version of the Mount-Wilson S-index, the second component appears as powerful activity proxies to correct the RVs induced by the inhibition of the convective blueshift in stellar active regions. However, we failed to interpret the extracted component into a physical framework. We conclude that a more complex kernel or bandpass than the classical triangular of the Mount Wilson convention should be used to extract activity proxies. To this regard, we provide the first principal component activity profile obtained across the spectral type sequence between M1V to F9V type stars. 
\end{abstract}


%%%%%%%%%%%%%%%%%%%%%%%%%%%%%%%%%%%%%%%%%%%%%%%%%%

%%%%%%%%%%%%%%%%% BODY OF PAPER %%%%%%%%%%%%%%%%%%

\section{Introduction}

Stellar activity is today the main limitation for the upcoming ultra-stable spectroscopic era that aims to an radial velocity (RV) precision below 50 \cms . This era has begun with a new generation of extremely precise spectrograph such as ESPRESSO \citep{Pepe(2014b)}, NEID \citep{Schwab(2018)}, EXPRES \citep{Jurgenson(2016)}, HARPS-N \citep{Cosentino(2012)} or KPF \citep{Gibson(2016)}, that are a legacy from previous instruments such as HARPS \citep{Mayor(2003)}. At such level, the stellar activity dominates the RV measurements which makes more difficult the detection of Earth-like planets around Sun-like stars \citep{Meunier(2020),Crass(2021)b}, and in the worst case induces a false detection (e.g \citet{Robertson(2015),Faria(2020)}). Among the stellar activity components, faculae/plages and spots are known to dominate in the RV budget of Sun-like stars \citep{Meunier(2022)}. 

Stellar activity does not only affect the Doppler shift, but the shape of the spectrum. A general review about stellar activity and its link with the RV problem was done recently in \citet{Hara(2023b)}. This shape change can be diagnosed by extracting so-called indicators from the spectra, such as the variation of strength of Ca II H \& K emission line. However, it is not clear whether the information in this line is best exploited by current data analysis techniques.

In our first paper \citep{Cretignier(2024)} and hereafter paper I, we investigated in detail the behaviour of the \CaHK lines on the Sun. A pressing question to solve was to investigate if such deformation of the lines was also detectable for other stars and not only for our bright Sun. We investigated if this larger amount of information hidden into the \CaHK lines could help to better correct the radial velocity (RV) signal of stellar activity.

Interestingly enough, in the RV community, the analysis of the \CaHK lines has not changed from almost half a century and is always performed by an equivalent-width (EW) measurement via a triangular bandpass integration on the core of the lines (see \citet{Oranje(1983b),Dineva(2022)}). This bandpass being an old heritage from the Mt Wilson survey \citep{Wilson(1978)} that has monitored the chromospheric emission for thousands of stars and which is now used as a reference in several contemporaneous studies for long-term monitoring of stellar activity cycles \citep{Vaughan(1980),Baliunas(1995),Duncan(1991),Wright(2004),Hall(2007),Isaacson(2010),Lovis(2011),Hempelmann(2016),Costes(2021)}. Similar methods have been used to calculate activity based on other strong lines like \Halpha, the \ion{Mg}{I} triplet and the \ion{Na}{I} D doublet \citep{West2008,Silva2011}.

Recently, \citet{Gomes(2022)} remarked that simply changing the bandpass' width on the $H_\alpha$ line was able to solve for part of the discrepancy observed between this proxy and the S-index \citep{Meunier(2009)}. Similarly, \citet{Pietrow(2024)} observed how changing the bandpass can improve the signal extraction for a solar flare. This example shows that investigations must be conducted on how those proxies are extracted and the optimal way to compute them in order to better understand the star or correct the RVs. 

\section{The RV signature of active regions}

On rotational timescale, the RV is dominated by active regions (ARs) inhomogeneities that both 1) break the flux balance of the receding and approaching stellar hemisphere\footnote{Sometimes refereed as the "flux effect"} \citep{Saar(1997),Boisse(2012),Hatzes(2002),Desort(2007),Hebrard(2014)}, and 2) inhibit the convective blueshift (CB) by their magnetic fields \citep{Kaisig(1983),Cavallini(1985),Keil(1989),Title(1989),Meunier(2010),Meunier(2013),Hathaway(2015),Cretignier(2020a)}. The final RV signal due to magnetic active regions being the superposition of both effects: 

\begin{equation}    
\label{eq:tot}
\Delta\text{RV}_{\text{AR}}(t) = \Delta\text{RV}_{\text{flux}}(t) + \Delta\text{RV}_{\text{ICB}}(t),
\end{equation}

Each of them tend to contribute at different level depending if the stars are plage or spot-dominated \citep{Shapiro(2014)}, inclined \citep{Shapiro(2016)} or fast-rotating \citep{Dumusque(2014)}. But both contributions are above 10 \cms\  in the final RV budget \citep{Moulla(2022)}, which is the semi-amplitude of an Earth-like planet on a Sun-like star. 

For slow-rotating Sun-like stars, the RV budget is dominated by the inhibition of the convective blueshift (ICB) $\Delta\text{RV}_{\text{ICB}}(t)$ as shown in \citet{Meunier(2010),Dumusque(2014)}. In standard stellar atmosphere, a net total convective blueshift is induced by the up-welling blueshifted bright granules that dominate in the intensity budget compared to the down-flowing redshifted intergranules \citep{Dravins(1982),Beeck(2013a)}. Usually, the flows in the quiet Sun are approximated to be purely radial with velocity $v_c$, such that the radial velocity component across the surface for an observer is given by:
\begin{equation}
\label{eq:radial}
v_r(\mu)=v_c\cdot \mu,    
\end{equation}

with $\mu=\cos(\theta)$ and $\theta$ the angle between the observer and the normal to the surface vector ($\mu=1$ at center and $\mu=0$ at the limb). The magnetic field in ARs is known to strongly reduce the convection (see e.g \citet{Cavallini(1985), Brandt(1990),Lohner(2018b)}) towards a slower velocity $v_{\text{mag}}$, introducing a velocity shift of $\Delta v = (v_{\text{mag}}-v_c)$, with $v_c \simeq -370\ms{}$ and $v_{\text{mag}}\simeq-125$\ms{} according to \citet{Title(1989)}.  Therefore, assuming the active region is small enough to be considered point-wise, the RV signal of the ICB for a single AR as a function of time can be written as: 

\begin{equation}  
\label{eq:rvcb}
\Delta\text{RV}_{\text{ICB}}(t) \propto \Delta v \cdot \mu(t) \cdot f_{\text{AR}}(t) \cdot \frac{I(\mu)}{I(\mu=1)}\,
\end{equation}

with $f_{\text{AR}}(t)$ the filling factor of a given AR defined as the sky-projected area ratio between the active region and the stellar hemisphere, where the same center-to-limb variation (CLV) was implicitly assumed for both the quiet atmosphere and the AR. This assumption naturally simplifies the problem by removing higher-order dependency between RVs and $\mu$, but it should be noted that the contrast of active regions is poorly known for other stars that the Sun today due to fundamental degeneracies when the stellar surface is unresolved (see comment below). The CLV of the light intensity can be approximated by the linear relation\footnote{Note that the relation is slightly chromatic (see \citet{Neckel(1984)})} $I(\mu)/I(\mu=1)=0.8+0.2\mu$ with less than 2\% of error up for $\mu>0.3$ \citep{Kervella(2017)}. Finally, since $\mu$ ranges between 0 and 1, the constant term will dominates over the linear term and we remain at first order with:

\begin{equation}  
\label{eq:rvcb2}
\Delta\text{RV}_{\text{ICB}}(t) \propto \Delta v \cdot \mu(t) \cdot f_{\text{AR}}(t),
\end{equation}

Note that we used a proportional sign instead of an equality to highlight that extra properties could play a role as well, such as the magnetic field intensity and density itself \citep{Hathaway(2015b)} , and that the convective flows geometry themselves are spectral type dependent \citep{Bauer(2018)}. 

The time variation of the filling factor $f_{\text{AR}}(t)$ is the intrinsic change of the AR size $f_{\tau}(t)$\footnote{Where the $\tau$ notation is used to highlight the intrinsic evolution of the active regions on their "lifetime".} multiplied by the geometric projection on the stellar surface: 

\begin{equation}    
\label{eq:time}
f_{\text{AR}}(t) = f_{\tau}(t) \cdot \mu(t)
\end{equation}

Often in literature, it is assumed that the lifetime of the ARs is longer than the rotational timescale, such that the temporal variation is mainly induced by the change of the projection of the AR along the rotational phase and not due to their intrinsic evolution. Therefore, their unprojected size is assumed to be constant $f_{\tau}(t)\simeq f_0$, which leads to the usual $f_{\text{AR}}(t)^2$ term in the FF' framework (see Eq.11 in \citet{Aigrain(2012)}) by combining Eq.\ref{eq:rvcb2} and Eq.\ref{eq:time}: 

\begin{equation}  
\label{eq:ff2}
\Delta\text{RV}_{\text{ICB}}(t) \propto \Delta v \cdot \frac{f_{\text{AR}}(t)^2}{f_0}
\end{equation}

We note that this assumption is not necessarily perfect for slow-rotating Sun-like stars for which the AR lifetime is similar or shorter than the rotational period \citep[e.g.][]{Solanki(2003),ForgcsDajka2021,Tlatov(2024)}. For this reason, Eq.\ref{eq:rvcb} is more general than Eq.\ref{eq:ff2} and the total contribution from several ARs is just the sum of the individual components:

\begin{equation}  
\label{eq:mono}
\Delta\text{RV}_{\text{ICB}}(t) \simeq  \Delta v \cdot \sum_i \mu_i(t) \cdot f_{i}(t) = \Delta v \cdot \bar{\mu}(t) \cdot f_{\text{tot}}(t)\,,
\end{equation}

where the last equality is obtained by using the global distribution at the stellar surface $\mu_i$\footnote{the term $f_i$ can be understood as a weighting coefficient} by its $0^{\rm th}$ and $1^{\rm st}$ statistical moment, namely its integral $f_{\text{tot}}$ and its mean $\bar{\mu}$. Such approximation is equivalent to assume that a complex distribution of ARs can be approximated by a "point-like" description\footnote{A mistake would be to interpret the "point" in the approximation as a single rotating AR, which is not what the approximation says. The "point" is free to move temporally at the stellar surface and can even move in a retrograde way compared to the stellar rotation. The approximation only stipulates that a point-like description produces the same observable for each individual epoch.}. 

It is important to remark that our model does not say anything about the time-covariance of the signals (such an effort has been made recently in \citet{Hara(2023)}) and mainly describes the problem in terms of instantaneous quantities. Our model is extremely simplistic, but has the advantage to highlight the dominant ingredients needed to correct the $\Delta\text{RV}_{\text{ICB}}$, namely $f_{\text{tot}}(t)$ and $\bar{\mu}(t)$. We investigated in Appendix.\ref{app:nonlinear} the approximation viability of our model which is very good as long as the convective flows are purely radial and the ICB law in Eq.\ref{eq:radial} is linear. 

For slow-rotating Sun-like stars, $\Delta \text{RV}_{\text{AR}}$ depends at first order of approximation linearly on the filling factor $f_{\text{AR}}$ and not quadratically as sometimes thought in the community due to the application of Eq.\ref{eq:ff2} for a single non-evolving rotating AR. A linear model was also chosen by \citet{Milbourne(2021)} to fit the solar RVs (see their Eq.13) with the filling factors of active regions and such a model seems to closely match the solar RVs as observed by HARPS-N \citep{Collier(2019),Dumusque(2021)}. However, it should be noted that the authors used data coming from the data reduction software (DRS) in development and the solar result between the old and new version of the pipeline have changed from this time \citep{Meunier(2024)}.

The non-linear dependency between RVs and the product of $\bar{\mu}(t)$ and $f_{\text{tot}}(t)$ is introducing some complex relations that depends on the Butterfly diagram (in particular the maximum latitudes) and the stellar inclination. Those non-linear dependencies (sometimes referred as hysteresis or elliptical correlations) require extra adjustment in the model as mentioned in \citet{Meunier(2019e)}.

In paper I, we already mentioned the consequent amount of work existing to derive different activity proxies. Interestingly enough, if a huge effort has been made to extract the $f_{\text{AR}}(t)$ component in Eq.\ref{eq:rvcb}, almost nothing can be found regarding $\mu(t)$ that requires the use of CLV signatures to be obtained. While, such CLV variations have been measured on the \CaHK lines \citep{Labonte(1986),Pietrow(2023),Cretignier(2024)}. Another way to obtain $\mu(t)$ would be to reconstruct the stellar surface using for instance Doppler tomography methods, but such attempt on slow-rotating stars have failed so far. 

As shown in Appendix.\ref{app:nonlinear}, an interesting property of Eq.\ref{eq:mono} is that the temporal variation of $\bar{\mu}(t)$ saturates very rapidly toward a constant value as soon as several ARs are present simultaneously (which is the case on the Sun excepted at the solar minima of activity where single and isolated AR are found). We note that such saturation effect also happens for the flux effect RV$_{\text{flux}}(t)$, since ARs situated simultaneously on both the receding and approaching stellar hemisphere will cancel out their RV signal as much as bright and dark ARs will, while their ICB contribution always summed up. Then, even if $\bar{\mu}(t)$ likely plays a role, the dominant contribution of the signal modulation is driven by $f_{\text{tot}}(t)$. Therefore, a huge problem today consists in being able to extract the filling factors of the different components for different spectral types. 

It is important to highlight that the filling factor extracted here is a \textit{universal} filling factor. By universal, we mean that it is obtained by assuming that all active regions share some universal and unique properties $\Delta X$ ($X$ being the temperature $T$, the magnetic field $B$ or the velocity flows $v$) compared to the quiet atmosphere and there is no difference in the properties of two different spots or two different faculae\footnote{In other words, the disk-average of the properties is larger than their spatial standard deviation.}. It is well known that for disk-integrated observations, there is a fundamental degeneracy such that a small and compact magnetic region with large deviations $\Delta X$, produce the same observable as a larger active region with a smaller deviation $\Delta X$. The brake of such a degeneracy invoke either the use of non-linear proxies \citep{Labonte(1986)}, the detailed analysis of line profiles \citep{Brandt(1990)} or the occultation of active regions by transiting planet \citep{Morris(2017)}. 

\section{Methods}
\label{sec:method}

We propose here below to extract the different activity components on the spectra time-series of chromospheric lines such as the \CaHK lines by using algorithms designed to solve the blind source separation (BSS) problem. To understand why different activity signal are presenting different emission profile, we briefly remind the theory behind the line formation. 

A stellar spectrum is always the superposition of all the individual stellar surface spectrum present simultaneously at the stellar surface and such superposition can a priori be inverted from the disk-integrated spectrum \citep{Davis(2017),Sharaf(2023)}. Theoretically, each individual spectrum is shifted by the local RV at the stellar surface where it was emitted, but for slow-rotating stars, the shift is smaller than the intrinsic width of the stellar lines and we can consider that the "disk spatial information" is strongly diluted. However, if the sky-projected disk locations are lost during the integration, the same is not true for the atmospheric depth coordinates.  

Indeed, each wavelength element $\lambda_i$ of a line profile is formed at a different depth given by its contribution function (see for instance Fig.13.2 in \citet{Gray(2005)}). A single wavelength never solely probes an infinitesimally thin specific layer, but rather probes the stellar atmosphere given on broad range of formation depth. This effect is even further enhanced by the non-infinite spectral resolution of the spectrograph that convolves and shares the information of consecutive wavelengths sampling elements. As a consequence, we can consider the flux emitted at a wavelength $I_{\lambda_i}(t)$ as the "thermometer" signal $S_i(t)$ which probe the "temperature" in a stellar atmospheric layer. Such stellar "depth signature" of ARs was for instance also shown by using different photospheric stellar lines \citep{Cretignier(2020a)} or subpart of the photospheric line profiles \citep{Moulla(2022),Moulla(2024)}.

\begin{figure}
	
	\centering
	\includegraphics[width=8.5cm]{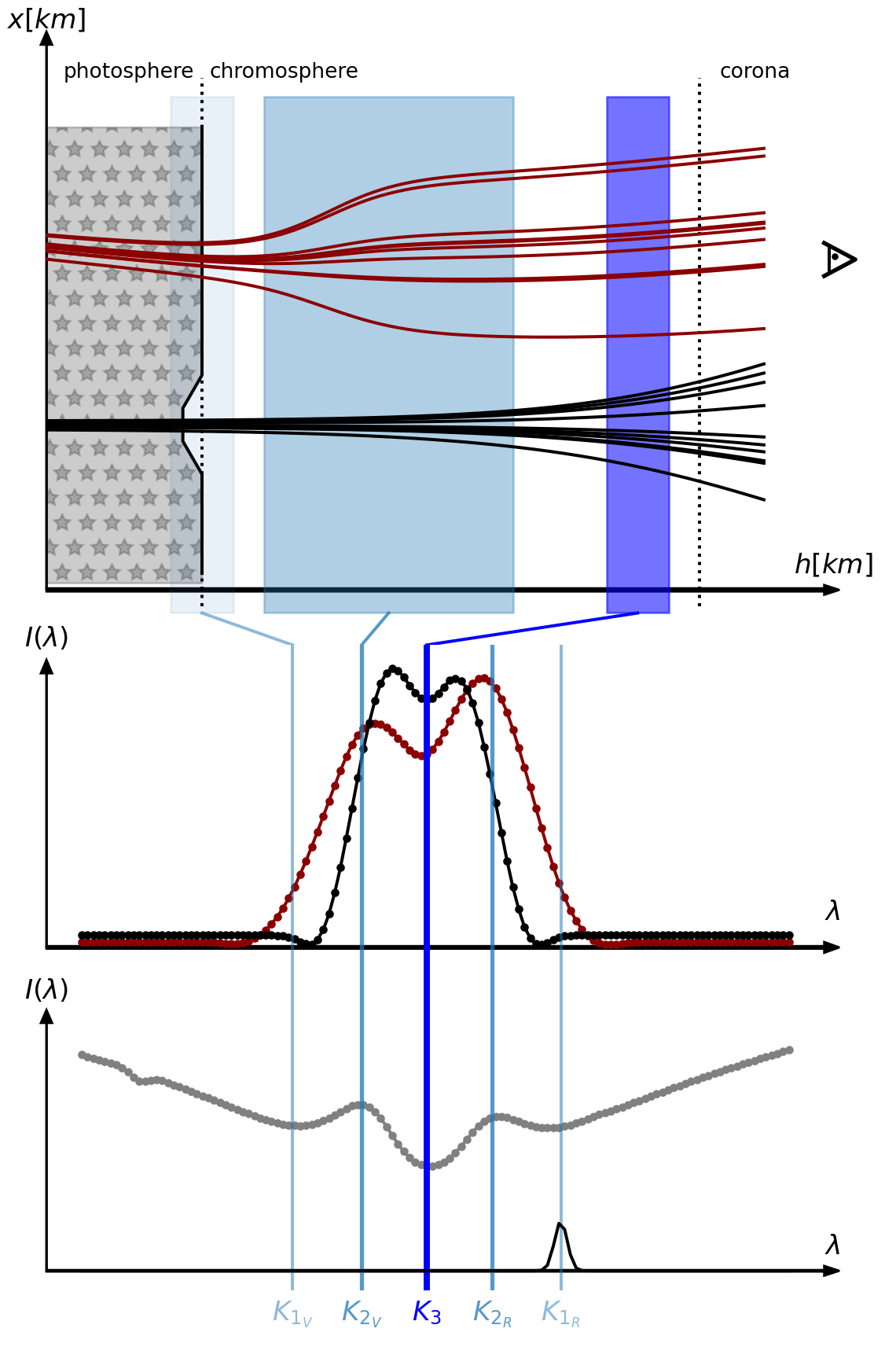}
	\caption{Schematic illustration to explain how the stellar depth information is preserved in a line profile. \textbf{Bottom panel:} Illustration of the \CaK line profile used to highlight the peculiar K1 minima, K2 maxima and K3 minimum location. The point spread function for a $R\sim$ 100'000 resolution spectrograph is shown centered on $K_{1_R}$. \textbf{Middle panel:} Intensity emission profile  induced by different kind of active regions (plage in red and spot in black) from \citet{Cretignier(2024)}. \textbf{Top panel:} Equivalent surface representation of two different active regions (plage at top and spot below). The Wilson depression of the spot is highlighted by the deep in the photosphere/chromosphere ($h=500$ \kms) delimitation. The equivalent atmospheric layers produced at the K1, K2 and K3 wavelength according to \citet{Vernazza(1981)} are also drawn as shaded area. The K3 layer (shaded dark blue) being formed at $h=2000$ km above the photosphere. The magnetic field loops are purely for illustrative purpose, but follow qualitative prescriptions of \citet{Przybylski(2015)} and \citet{Kuridze(2024)}. The observer eye is drawn on the right.}
	\label{FigSchema}
 
\end{figure}

A peculiar property of the resonant scattering \CaHK lines is that, among all the stellar lines in the visible, their core emission probes the largest span of stellar chromospheric layers \citep{Vernazza(1981)}, with atmospheric heights between 500 km and 2000 km above the photosphere. The schematic illustration of the line and its formation depth is shown in Fig.\ref{FigSchema}. The $M$ wavelength $\lambda_i$ elements\footnote{We drop from here the $\lambda_i$ designation that is now implicit until the end of the paper} of the line profile can be understood as the $M$ thermometers recording the stellar intensity spectrum at different depth in the chromosphere.

In paper I, we modeled the residual intensity spectra time-series $\delta S_{\lambda}(t) = I(\lambda,t) - I_{\text{ref}}(\lambda)$ of the \CaHK lines as the multi-linear superposition of three activity components:

\begin{equation}
\label{EqFund2}
\delta S(\lambda,t) = f_{\text{spot}}(t) \cdot I_{\text{spot}}(\lambda) + f_{\text{plage}}(t) \cdot I_{\text{plage}}(\lambda)+f_{\text{ntwk}}(t) \cdot I_{\text{ntwk}}(\lambda)
\end{equation}

and solved for the $I_{\text{AR}}(\lambda_i)$ components by using SDO segmentation of direct imaging of the Sun. Unfortunately, the same profiles have a priori no reason to be identical for all the spectral types since the stellar atmosphere itself is different. This difference will be demonstrated in Sect.~\ref{sec:cenb}. 

We now propose here below a solution to solve for the BSS when both the emission profiles and the filling factors time-series are unknown. 

\subsection{Solving the BSS with a PCA/ICA}
\label{sec:BSS}

We began from the only information in our possession, namely that the problem is multi-linear and can be expressed as: 

\begin{align}
\label{EqICA}
\delta S_{\text{ref}}(\lambda,t) = I(\lambda,t) - I_{\text{ref}}(\lambda) &= (\cdots, F_{j}(t), \cdots) 
\begin{pmatrix}
       \vdots \\
       E_{j}(\lambda) \\
       \vdots \\
\end{pmatrix} &= \boldsymbol{F} \boldsymbol{E}
\end{align}

To solve Eq.\ref{EqICA}, two popular methods are to use principal component analysis (PCA) or independent component analysis (ICA). The main difference between the two being that for the ICA, the number of components used for the decomposition is known and components are rather statistically independent than orthogonal. For this work, we used the FastICA algorithm \citet{Hyvarinen(2000)} implemented in the Python package \textit{Scikit-learn} \citep{Pedregosa(2011),Grisel(2021)}.

There would be several arguments to justify why ICA should be preferred over PCA for the present decomposition of the \CaHK lines time-series. A first unsuitable feature of PCA is the orthogonal basis that is not physically motivated. Indeed, in paper I, we showed that the emission profiles of active regions were showing some co-linearity. A second unsuitable behaviour is that PCA is rather an algorithm that tries to "compress" the information stored in data, whereas in the present situation the dimension of the latent space is rather well known ($N\sim3$). This precious information would be a strong argument in favor of the ICA that precisely requires such prior knowledge to constrain the model.  

%However, there are a few inconveniences that made the ICA unusable in practice. First, we notice that the algorithm is not deterministic, such that two consecutive calls of the function on the exact same data produce different results. Some of these issues are not serious, such as components length and total weights unfixed or randomness in the order of the components (since there is no order or ranking as for a PCA), but in the most dramatic cases the ICA converges to a totally different solution that fully change the interpretation of the results. Such behaviour is not very clear but may be related to complex numerical instabilities of the underlying algorithm. Those difficulties made the ICA unusable in practice and we therefore stuck to the PCA decomposition hereafter. 

At this stage, it is important to point out a few properties related to the BSS decomposition and its implication. First of all, any linear combination of the PCA/ICA vectors is trivially also a solution of the Eq.\ref{EqICA} or, in other words, the real profiles certainly belong to the subspace limited by the components, but are not necessarily the components themselves. 

For this reason, the profiles obtained have not to be strictly interpreted as their real physical equivalence. In order to avoid any confusion, we will use hereafter the convention $E_j(\lambda)$ and $F_j(t)$ to highlight the difference with their physical counterparts $I_j(\lambda)$ and $f_j(t)$ in Eq.\ref{EqFund2}. 

We note that if the real profiles belongs to the subspace defined by the PCA components, it means that there exists a matrix $\boldsymbol{W}$ of change of basis such that the real profiles can be obtained: 

\begin{align}
\delta S_{\text{ref}}(\lambda,t) &= (\cdots, f_{j}(t), \cdots) 
\begin{pmatrix}
       \vdots \\
       I_{j}(\lambda) \\
       \vdots \\
\end{pmatrix} & = \boldsymbol{f} \boldsymbol{I} &= \boldsymbol{F} \boldsymbol{W}^{-1} \boldsymbol{W} \boldsymbol{E} 
\label{EqTransform}
\end{align}

%Secondly, the transpose of Eq.\ref{EqICA} is also a multilinear superposition problem that can also be solved by PCA. As a consequence we have to define which elements are the components (or vector basis) and which elements are the weights (or coefficients) in our problem. We chose $E_j(\lambda)$ to be our vector basis and $F_j(t)$ the coefficients because we assumed that ARs are described by universal profiles and because we are also aiming ultimately to fit an intensity profile basis in order to recover the filling factors time-series. 

%Also doing so, the number of element in our sample (given by the number of observations $T$) will be larger than the dimensional space given by the number of wavelength sampling element $M$\footnote{We typically expect $T\sim300$ and $M\sim50$}. %The space used hereafter and the main hypothesis of this work are schematised in Fig.\ref{FigSchema2}.

\begin{figure*}
	
	\centering
	\includegraphics[width=18cm]{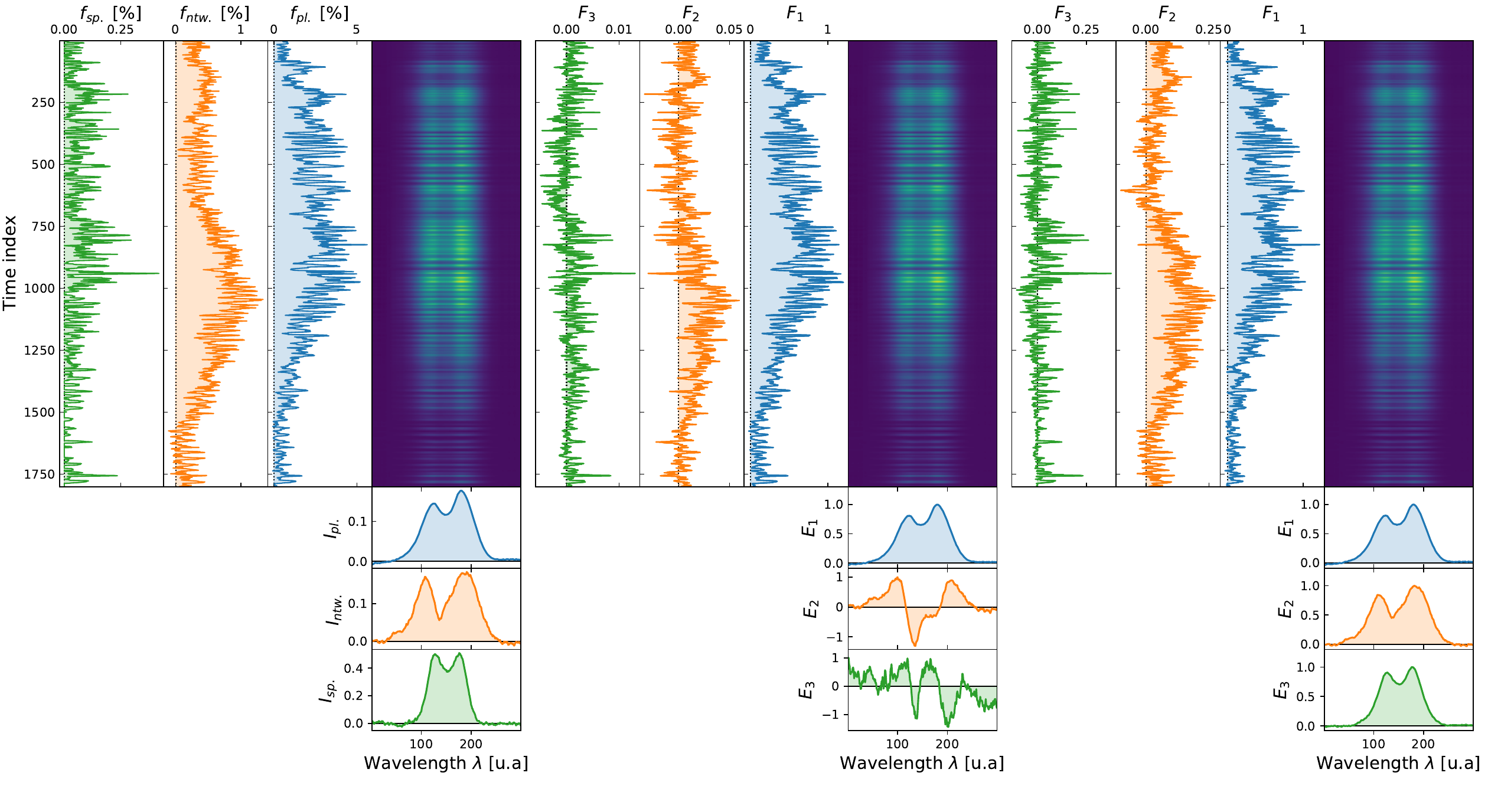}
	\caption{Comparison between the real physical spectra time-series decomposition of the \CaK line (left) with a 3-component PCA decomposition (middle) and 3-component ICA (right). \textbf{Left:} Same analysis as paper I. The spectra time-series of the ISS spectra is decomposed by the three filling factors time-series of SDO. Both spectra and filling factors have been recentered by some quiet reference epochs to mimic real stellar observations. \textbf{Middle:} PCA decomposition of the previous model. If the first component appears very close to the plage components because mainly dominated by their contribution, the second and third component become a mixture of the profiles even if each of them seems to keep a dominance from a single kind of AR. \textbf{Right:} \textbf{ICA decomposition of the previous model. The profiles are closer to the intrinsic physical components.}}
	\label{FigCaIIPCA}
 
\end{figure*}

Secondly, components and coefficients are unsigned which could terribly affect the interpretation of the components afterwards (see Sect.~\ref{sec:physical-interpretation}). Since $F_j(t)$ coefficients as expected to represent filling factors that follows the magnetic cycle long trend, we used as sign convention than a positive correlation is expected between the $F_j(t)$ and the S-index, where the correlation was computed with the low-pass filter of the \Sidx{} to enhance the correlation on magnetic cycle signal. 

Furthermore, by construction, components and weights are unitless, since we can arbitrarily increase the norms of the vectors and decrease the coefficients due to the degeneracy of the multi-linear model in Eq.\ref{EqICA}. Consequently, it is not possible to extract the absolute filling factor of the components, but only their unitless time modulation. We chose to normalise all the $E_j(\lambda)$ components by the maximum absolute value of their profiles.

Last but not least, because the reference spectrum always contains to some extend active regions, except if this later is given by simulations or models, the $f_j(t)$ filling factors measured here above are only the relative ones compared to the reference epoch and a filling factor bias $\boldsymbol{f_{\text{ref}}}=(..., f_j(t_{\text{ref}}), ...)$ is needed to provide a perfectly accurate time-series: 

\begin{align}
\label{EqFull}
\delta S_{\text{0}}(\lambda,t) = I(\lambda,t) - I_{0}(\lambda) &= (\boldsymbol{f_{\text{ref}}}+\boldsymbol{F} \boldsymbol{W}^{-1}) \boldsymbol{W} \boldsymbol{E}
\end{align}

In practice, since such a stellar spectrum free of any active regions $I_0(\lambda)$ is not yet provided by any model in NLTE conditions and for all the spectral types and can neither be obtained via observations, we stick to the Eq.\ref{EqTransform} as the reference equation for this paper.

\subsection{BSS of the solar \CaK line}

\subsubsection{Comparison between the PCA and ICA}
\label{sec:sunpca}

We analysed the ISS disk-integrated spectra of paper I with a PCA and ICA to understand how the data-driven information were related to their true physical quantities.

We reproduced the results obtained from the paper I. As a recall, the $I_j(\lambda)$ profiles were obtained by solving the Eq.\ref{EqFund2} using SDO filling factor of three solar activity components, namely plage, network and spot. 

In order to see if a PCA was able to recover the same profiles and time-series, we extracted the spectra time-series obtained by the previous fit in the left panel of the Fig.\ref{FigCaIIPCA} and applied a PCA/ICA decomposition on it. Rather than fitting the PCA/ICA on the observed ISS spectra, we chose to perform it on the three-component model in order to ensure that the dimensional latent space is exactly 3 and the question of extra missing components from the model will not affect the interpretation. We also centered spectra and filling factor time-series by the average observations at the 10 percentile most quiet epochs of the time-series in order to mimic the lack of absolute accurate quantities. 

\begin{figure}
	\centering
	\includegraphics[width=9cm]{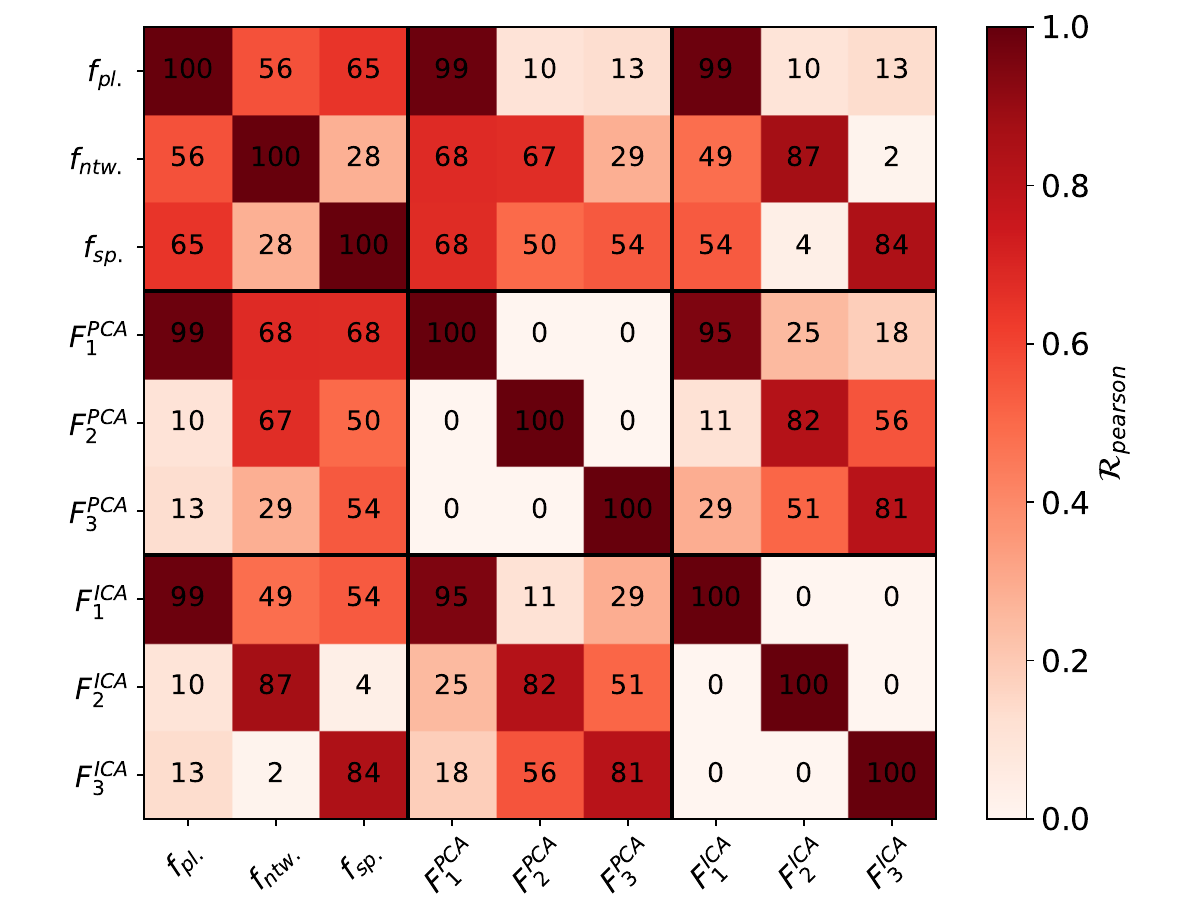}
	\caption{Symmetric correlation matrix between the filling factors time-series used to generate the \CaHK time-series and the data-driven components recovered by the PCA and ICA. While both ICA and PCA provide orthogonals components, visible by the null coefficients in the diagonal block matrices, ICA components are closer to the real filling factor (top right block matrix) compared to the PCA ones (top center block matrix).}
	\label{FigCaIIPCA2}
\end{figure}

We can notice a few properties of the PCA/ICA decomposition when compared with the physical decomposition. Trivially, the components do not deliver the physical quantities, but rather a linear mixture of them as expected. This is a consequence of the wrong statistical assumptions (such as orthogonality) used by the models to solve the BSS problem. This element is highlighted in the correlation matrix of the components $F_j$ with the physical filling factors in Fig.\ref{FigCaIIPCA2}.

To understand the difference of behaviour between the PCA and the ICA, it is useful to derive the explained variance ratio of the components in the model. The respective variance in the data introduced by plage, network and spots are of 93.7, 5.6 and 0.7\% respectively. For the PCA, the three components explain 99.8, 0.2 and less than 0.01\% of the variance. This is very different from the physical model and makes perfectly sense, since the three activity components are not orthogonal and share some moderate co-linearity (top left block matrix in Fig.\ref{FigCaIIPCA2}). Therefore, the first PCA component creates a mixture of the three activity components that absorbs the variance more rapidly. For the ICA, the explained variance is about 93.3, 5.2 and 1.5 \% that is closer to the physical model values, since the algorithm is not optimizing over the explained variance as for the PCA.

No matter if a PCA or an ICA is used, we observe that generally the first component $E_1(\lambda)$ is extremely close to the plage profile $I_{\text{plage}}(\lambda)$. This element is a consequence of the result found in paper I, where we showed that the S-index (and thus the integrated flux of the residual spectra time-series) is dominated by plage, while the network plays a minor role and sunspot a negligible one. As a consequence, since PCA is fitting for the variance in the data as fast as possible, the first component naturally gathers mainly the plage information since they dominate the information contained in the spectra. 

Main differences arise for the subsequent components between the algorithms. While the network profile can be guessed in the $E_2(\lambda)$ component for the ICA, this later looks for the PCA rather like a broadening kernel (or second statistical moment of a Gaussian). This indicates that for the PCA, the component is rather the difference of the network and plage profile than the network profile itself. Third component becomes even more a mixture of the three components for the PCA even if the time-series $F_3(t)$ likely exhibit sunspot information as visible by the sharp peak around the time index $\sim$ 900. 

From this exercise, we clearly see that ICA has recovered a solution closer to the truth with component correlated at 0.99, 0.87 and 0.84 with the physical time-series (top right block matrix in Fig.\ref{FigCaIIPCA2}). After this partial success, we investigated the sensitivity of each algorithms to the white noise level to evaluated the performance of the algorithm with the S/N of the observations. 

\subsubsection{Analysis of the photonic S/N dependency on the PCA and ICA components}
\label{sec:SNcomp}

We performed photon noise injection on the spectra in order to quantify the minimum value required for data-driven method to work properly. Here below, we will mainly use the continuum S/N in the extreme blue (4000 $\ang{}$) as the reference wavelength, recalling that this S/N value is often twice lower than the one in the visible (e.g around 5500 $\ang{}$) due to coatings, fibre efficiency and filters in the spectrograph (see e.g red curve in Fig.C.1 from \citet{Cretignier(2022)}). Also, because the flux level in the core is around 5\%, the S/N in the core of the lines ($S/N_{\text{core}}$ hereafter) is $\sqrt{0.05} \simeq 22\%$ the value of the continuum at 4000 \ang{} ($S/N_{\text{cont}}$ hereafter).

\begin{figure}
	\centering
	\includegraphics[width=8.5cm]{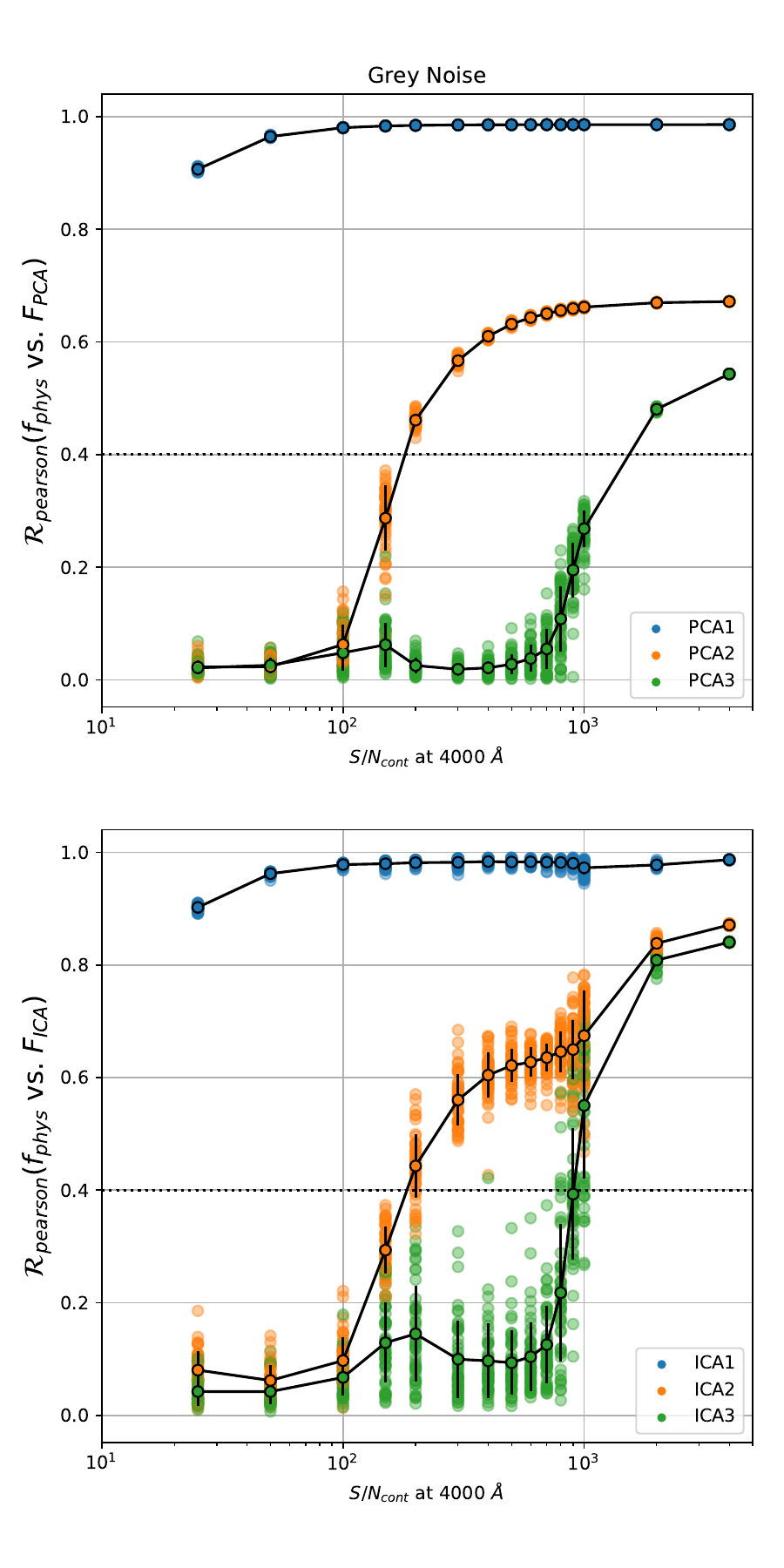}
	\caption{Correlation matrix coefficients of the PCA (top) and ICA (bottom) as a function of the continuum $S/N_{\text{cont}}$ at $\lambda=4000$ $\ang{}$ for the solar BSS decomposition. A dramatic loss is observed for both algorithms on the component $F_3(t)$ for $S/N_{\text{cont}}<1000$ and similarly for the component $F_2(t)$ at $S/N_{\text{cont}}<200$. An horizontal dashed line depicted the $\mathcal{R}<0.40$ unsignificance arbitrary level. As an example, flat-field limited observations on HARPS or HARPS-N have a typical $S/N_{\text{cont}}\sim400-500$, while an observation at $S/N_{\text{cont}}=250$ at $5500$ $\ang{}$ will be at half this value, hence around $S/N_{\text{cont}}\simeq125$. }
	\label{FigSNR1}
\end{figure}

We injected white noise value corresponding to different S/N values. In order to simulate a more realistic noise, we included $\pm15\%$ of S/N variations randomly in the time-domain to account for various weather conditions, but the results were quantitatively very similar in the case of a pure time-independent white noise simulations. We measured the $\mathcal{R}$ Pearson coefficient of the correlation matrix (respectively the diagonal of the top center and top right block matrix in Fig.\ref{FigCaIIPCA2}). For each S/N, 100 independent noise realisations were done. The Pearson correlation coefficients between the components and the filling factors time-series are displayed as a function of the $S/N_{\text{cont}}$ in Fig.\ref{FigSNR1}.

For both the PCA and ICA, the third component $F_3(t)$ loses all information for $S/N_{\text{cont}}<1000$. The same occurs for the second component $F_2(t)$ for $S/N_{\text{cont}}<200$. While initially, the ICA was thought to provide more reliable results, we noted that ICA and PCA provide nearly identical results as soon as $S/N_{\text{cont}}<800$. In practice, ICA results are even worse, as visible by the larger scatter among the different noise realisations, due to the lower numerical stability of the algorithm. 
We displayed in Appendix.\ref{app:SN} how the emission profiles $E_j(\lambda)$ are changing with $S/N_{\text{cont}}$ for both algorithms.

Interestingly, we can compare this critical $S/N_{\text{cont}}=200$ value with another PCA methodology developed in \citet{Cretignier(2022)} where we found that a dramatic loss of information was occurring around $S/N_{\text{cont}}=250$, where this value was the chromatic average continuum S/N and caution should be taken in the comparison since the bandpass and datasets used were different.

\begin{figure*}
	\centering
	\includegraphics[width=18cm]{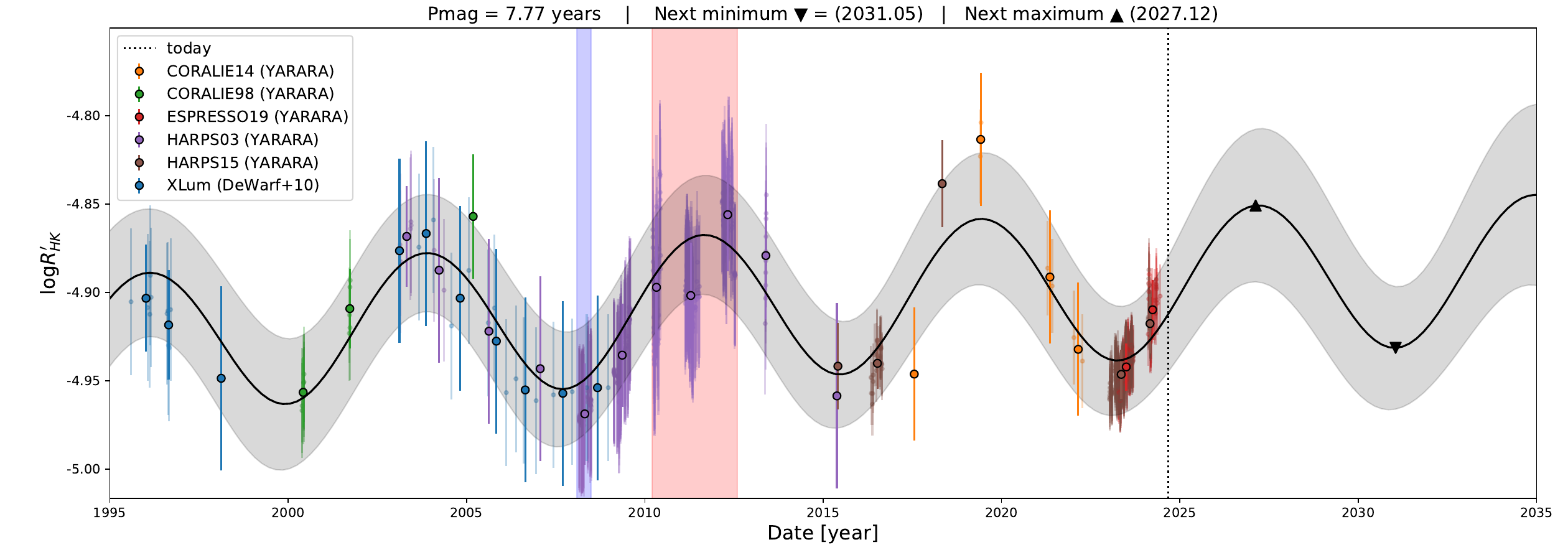}
	\caption{$\alpha$ Cen B archival analysis of the CaII H\&K lines over 30 years of CORALIE, HARPS and ESPRESSO revealing the magnetic cycle periodicity of $P_{\text{mag}}\simeq7.8$ years. Marker with a black edge color are the seasonal mean value for different instruments. Uncertainties are accounting for the expected amplitude of the stellar rotation modulation. The X-ray luminosity from \citet{Dewarf(2010)} is also displayed and scaled in units by using time-overlapping regions. The best fit of a Gaussian Process with a quasi-periodic kernel is displayed in gray shaded. The quiet period of time in 2008 corresponding to HARPS observations is represented with the blue shaded region, while the high activity period from 2010 to 2012 is shown in red.}
	\label{FigMag}
\end{figure*}

We also observed the robustness of the first component $F_1(t)$ for both PCA and ICA even down to $S/N_{\text{cont}}=50$. This is explained by the predominance of a single component (plage component) in the data. In practice, we know that our simulations are slightly optimist since observations tend to contain some heterogeneous noise on top of the white noise in particular at very low S/N, but simulating a realistic instrumental heterogeneous noise is not easy to implement in practice. As a consequence, our results have to be understood has the most optimistic S/N prescription.

As a final comment, note that we here only used one single line while in practice the \CaHK lines is a doublet. But the spectral resolution is also twice better, so in definitive the same S/N is expected for HARPS observations using the two lines simultaneously.

This result on the Sun is critical given that, without any prior information about the solar surface, we were able to recover both with an ICA and PCA the approximated plage emission profile and a correspondent adimensional plage filling factor time-series as well as a mixture of other extra deformation moments. Based on this partial success, we investigated if such a multi-component behaviour of the lines could be detected for real stellar observations as well.

\section{Results}
\label{sec:results}

\subsection{Analysis of the \CaHK lines on $\alpha$ Cen B}
\label{sec:cenb}

We now demonstrate, on stellar observations of a K-dwarf moderate active star, that similar extra components can be extracted from the \CaHK profiles. Furthermore, those new proxies provide better activity proxies than the usual S-index to correct the RVs.

We tested the PCA/ICA decomposition methodology on HARPS observations of $\alpha$ Cen B (HD128621), a moderate active K-dwarf star \logrhk $\simeq$ $-4.90$ \citep{Dumusque(2012b)}. This dataset has been extensively used for stellar activity studies in RVs due to the high signal-to-noise ratio of the observations and the clear activity signature visible in 2010 \citep{Thompson(2017),Dumusque(2018), Wise(2018),Simola(2019),Ning(2019),Cretignier(2020a),Cretignier(2021),Cretignier(2022),Wise(2022),Moulla(2022)}.

$\alpha$ Cen B is a bright K dwarf with a radius $R_*=0.863\pm0.003 \, R_{\odot}$ \citep{Kervella(2017)} and a known rotational period $P_{\text{rot}}$ of $\sim$$36-40$ days \citep{Jay(1997),Dewarf(2010)}. The star is a slow rotating-star with a $v \sin i <1.2$ \kms , where this low projected equatorial velocity is also induced by the low stellar inclination $i \simeq 45$ deg \citep{Dumusque(2014b)}. 

HARPS has intensively observed the star from 2008 to 2012 until the star got too close in sky of its binary companion $\alpha$ Cen A. Such a temporal baseline has efficiently probed the rising side of the stellar magnetic cycle from minimum to maximum.

We demonstrate it by performing a homogeneous analysis of the $\log R^{\prime}_{\text{HK}}$ over 30 years from 4 different instruments. We processed all the HARPS observations publicly available on the ESO archive, as well as 16 nights from the old CORALIE spectrograph (CORALIE98) and 7 nights obtained more recently with the new upgraded instrument (CORALIE14). We also used unpublic data obtained recently with HARPS and ESPRESSO under programs 110.24BB.001 and 111.252R.001 respectively. Finally, we used the X-ray luminosity reported in \citet{Dewarf(2010)}, removed one outlier around 2001 and scaled it in units using the overlapping time period with HARPS observations. Data were ultimately binned by instrument and seasonal observations. The uncertainties were defined to absorb the rotational modulated signal based on the dispersion measured from the 2010 season of HARPS. The table is provided in Appendix.\ref{app:Table}.

The $\log R^{\prime}_{\text{HK}}$ time-series is displayed in Fig.\ref{FigMag} and reveals the $7.8\pm0.2$ years periodicity of the magnetic cycle. Among all the historical datasets existing for stellar activity, the HARPS dataset of the year 2010, 2011 and 2012 is among the highest S/N and most well sampled ones that probes a maximum activity phase of a non-solar type star (K1V). We point out that the three seasons should be studied and not only 2010 as often done in the literature, even if this latter contains the best rotational modulated signal.

\subsubsection{Data description and preprocessing}
\label{sec:preprocess}

%became too close of its binary companion $\alpha$ Cen A. 

Due to the exquisite stellar brightness ($m_v=1.33$) and the large telescope size of 3.6\,m, the nightly stacked observations reach a $S/N_{\text{cont}} \sim 580$ in the continuum at 5500 \ang{}, which is the maximum S/N achievable on HARPS since observations are flat field limited (see \citet{Cretignier(2020b),Cretignier(2021)}). The $S/N_{\text{cont}}$ at 4000 \ang{} is about\footnote{Since the core of the lines reaches a normalised flux around $\sim$ 5\%, the S/N in the core is reduced to one fifth ($\sqrt{0.05}\sim22\%$) of this value ($S/N_{\text{core}}\sim100$).} $S/N_{\text{cont}} \sim 450$ (see blue curve in Fig.C.1 from \citet{Cretignier(2022)}). Based on our noise-injection recovery tests in Sect.~\ref{sec:SNcomp} (see Fig.\ref{FigSNR1}), this S/N value would be sufficient to detect a second deformation component on the CaII H\&K lines for a solar-like activity signal, but not a third component.

We used the 1D spectra order-merged produced by the official DRS (v3.5). Spectra were first continuum normalised using RASSINE \citep{Cretignier(2020b)} and corrected of systematics by YARARA \citep{Cretignier(2021)}. Spectra with anomalous residuals after the YARARA processing were rejected of the analysis. Five observations were thus rejected (2009-04-30, 2011-03-02, 2011-03-16, 2011-06-01, 2011-07-09) which led ultimately to 298 nightly observations over the 5 seasons. Those rejected observations were all presenting either a very low and anomalous S/N or anomalous flux residuals likely related to instrumental issues during the flux acquisition or bad weather conditions and clouds. 

Note that because YARARA is correcting for activity, we reinjected back the activity flux correction on the spectra and extracted the $\text{RV}(t)$ with a classical Gaussian fit on a CCF obtained by a tailored line selections as in \citet{Cretignier(2020a)}. Such tailored line selections were shown to produce a more optimal RV extraction than generic mask, as much in terms of photonic RV precision than for the short-term jitter observed in CCF moments \citep{Lafarga(2020),Bourrier(2021)}. We shifted the spectra to cancel the quadratic trend of the companion $\alpha$ Cen A by fitting a Keplerian solution using the astrometric solution of \citet{Pourbaix(2016)}. The Keplerian orbital parameters for the binary trend were 
$P=79.91$ years, $K=5553.92$ m/s, $e=0.524 $,  $\omega=232.3^\circ$ and $\lambda_0=120.9^\circ$ using as reference $t_0$ epoch BJD = 2'455'500. 

\begin{figure*}
	
	\centering
	\includegraphics[width=18cm]{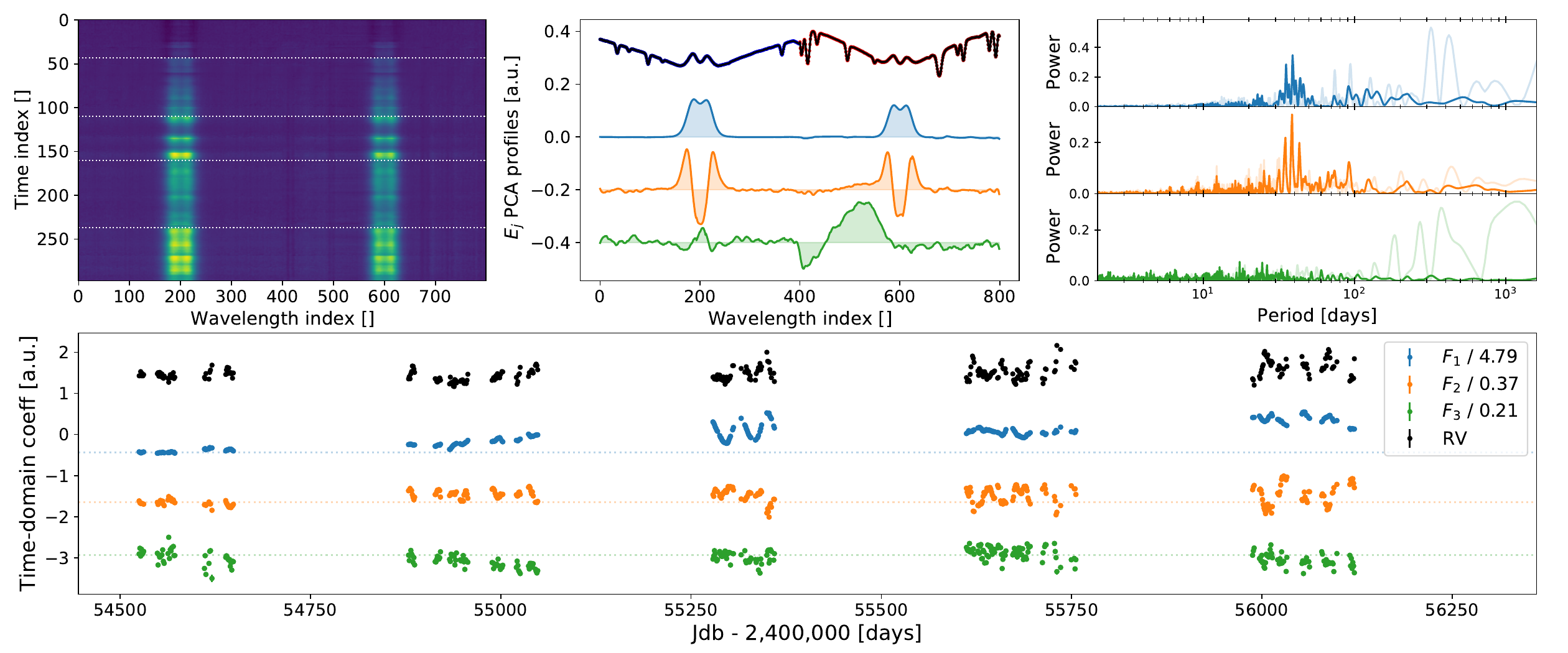}
	\caption{Extraction of two chromospheric components on $\alpha$ Cen B spectra performing a PCA. \textbf{Top left}: Spectra time-series of the \CaHK lines. The different observational seasons are split by the white dotted horizontal lines. \textbf{Top middle}: The PCA components are plotted and compared with the intensity profile of the \CaK line (blue dots) and \CaH line (red dots) with arbitrary offset. The wavelength sampling of the HARPS spectra is 0.01 \ang{}. The components exhibit similar responses from the H \& K lines except for the third component related to an instrumental systematic. \textbf{Bottom}: Time-series of the PCA components in arbitrary units (scaling indicated in the legend). The RV signal (black dots) of $\alpha$ Cen B is plotted as a comparison. The stellar activity is mainly visible in the third, fourth and fifth season, respectively in 2010, 2011 and 2012. \textbf{Top right}: GLS periodogram of the PCA coefficients. Both the periodogram of the time-series (light curve) and the high-pass filter of them are displayed (plain line). The first and second components present a long trend likely related to the stellar magnetic cycle and a rotational modulation, while the third component is likely an instrumental systematic with 1-year power.}
	\label{FigPCA2}
 
\end{figure*}

The PCA and ICA were performed on the spectra time-series $\delta S_{\text{ref}} (\lambda,t) = I(\lambda,t) - I_{\text{ref}}(\lambda)$ of the joint \CaHK lines for the 2008, 2009, 2010, 2011 and 2012 season, with $I_{\text{ref}}(\lambda)$ taken as the median of the spectra in 2008 season, which is lowest part of the stellar magnetic cycle (see blue shaded region in Fig.\ref{FigMag}) where no rotational modulation is visible. The bandpass used was defined as $\pm 2\ang{}$ around the core of the lines.

The number of components for the PCA can be obtained with the Elbow method of the explained variance of the components \citep{Satopaa(2011)}. The PCA delivered a number of components around $N\sim 3$. This number makes sense since we would not expect more than three components to fit plages, spots and network based on our paper I \citep{Cretignier(2024)} and the S/N of the observations is limiting the number of components that can be detected as demonstrated in Sect.~\ref{sec:SNcomp}. %By fitting three components with the ICA, we observed that two families of solutions seemed to exist. This behaviour can be understood if not three, but four components are needed in the model and the ICA oscillates between the components selected. By fixing the number of components to $N=4$, we observed that only one stable solution was existing confirming that such a number was the good value. The reason for this fourth component on top of the three activity ones is likely instrumental as showed in Sect.\ref{sec:alphacenb}.

As already raised, the signs of the $E_j(\lambda)$ weighting profiles are not determined due to the mentioned degeneracy in Eq.\ref{EqICA}. To fix the sign, we imposed a positive mean value for the $F_j(t)$ components, since physically, filling factor are strictly positive. We also arbitrary scaled the $E_j(\lambda)$ weighting profiles to an amplitude of 1\% in order to interpret the $F_j(t)$ vectors directly in term of normalised flux variation. Note that the sign or amplitudes are without consequence for the RV correction that will be performed, but can affect the physical interpretation of the components (see Sect.~\ref{sec:physical-interpretation}).

\subsubsection{Extraction of new activity proxies for RVs correction}
\label{sec:alphacenb}

We first present the results obtain with the PCA before to describe the ICA results.

The PCA weighting profiles $E_j(\lambda)$ and time-components $F_{j}(t)$ are represented in Fig.\ref{FigPCA}. The spectra time-series of the CaII H \& K lines is represented in the top left panel. By looking at the components that represent the expected emission profiles (top middle panel), we notice a similar behaviour of the \CaK and \CaH lines, the latter having a response 10\% smaller, except for the third component. An almost identical response of both lines is expected \citep{Labonte(1986)} since both lines probe almost the same atmospheric depth \citep{Bjorgen(2018)}. Hence any significant difference between both lines is the signature of instrumental systematics. 

The third PCA component shows precisely such a differential behaviour of the lines and is related to the residual of the ghost located on the left side of the core of the CaII H line. As a recall, ghosts are spurious reflection of the main spectral orders on the detector producing a secondary image of the spectrum (see Fig.5 in \citet{Cretignier(2021)}). Even if a ghost correction is implemented in YARARA, some residuals were known (see Fig.6 bottom panel in the same paper) due to the complexity of the contamination signal. As proof that our signal is related to the ghosts residuals, we displayed in Appendix.\ref{app:bss} the same PCA decomposition on the spectra after that the YARARA ghost correction has been injected back (see Fig.\ref{FigPCA}). The weighting profile $E_3(\lambda)$ is taking most of its power at the exact same wavelength location, namely on the left side of the core of the H line. We use this analysis as an opportunity to assess the relevance of the ghost contamination on the \CaHK lines.

%\begin{figure*}
%	\centering
%	\includegraphics[width=18cm]{Images/CenB_seasons_pca_M3S5.pdf}
%	\caption{Correlations plots between the RV time-series and the different activity proxies time-series delivered by the PCA (from left to right: $F_1(t)$, $F_2(t)$ and the season-by-season multilinear model ($\alpha F_1(t)+\beta F_2(t)+\gamma$) fit on the RVs. In each subplot, the Pearson coefficient $\mathcal{R}$ are indicated. The uncertainties on the $\mathcal{R}$ is obtained by 1000 random realisations of white noise on the RVs. \textbf{First row:} Analysis on the full time-series. \textbf{Second row:} Season-by-season analysis where 2008 (blue), 2009 (orange), 2010 (green), 2012 (red) and 2013 (purple), are investigated independently. If for the 2010 season, $\text{RV}(t)$ is clearly dominated by the $F_1(t)$ component ($\mathcal{R}_{2010}=0.64$), the 2011 and 2012 seasons are rather dominated by the $F_2(t)$ component ($\mathcal{R}_{2011}=-0.55$ and $\mathcal{R}_{2012}=-0.73$). }
%	\label{FigSeasons}
 
%\end{figure*}

From our analysis, the amplitude of the ghost is $\sim$15 times smaller (4.77/0.31) than the stellar magnetic cycle mainly captured by the $F_1(t)$ component, but is only 1.26 times smaller (0.39/0.31) in amplitude than the second PCA component. In our case, since the ghost was not covering the core of the line and was centered on the left wing of the CaII H line, the orthogonality between the activity and instrumental components was enhanced, reducing their mixing. But this is a fortunate configuration that depends on the systemic $RV_{\text{sys}}$ of the stars observed (see Fig.B.1 in \citet{Cretignier(2021)}). With the YARARA correction (Fig.\ref{FigPCA2}), the situation is slightly better with a relative amplitude 1.76 times smaller (0.37/0.21 from bottom panel) and the residual signal in the time-domain does not present anymore the characteristic 1-year signal. 

Such an example is a good illustration of the ability of the PCA to highlight or disentangle phenomena related to stellar activity from instrumental systematics. As a consequence, we advise to fit jointly both lines\footnote{Or reversely to extract an independent S-index for the H and K lines in the DRS.} in order to identify more easily such instrumental effects. 

\begin{figure*}
	
	\centering
	\includegraphics[width=18cm]{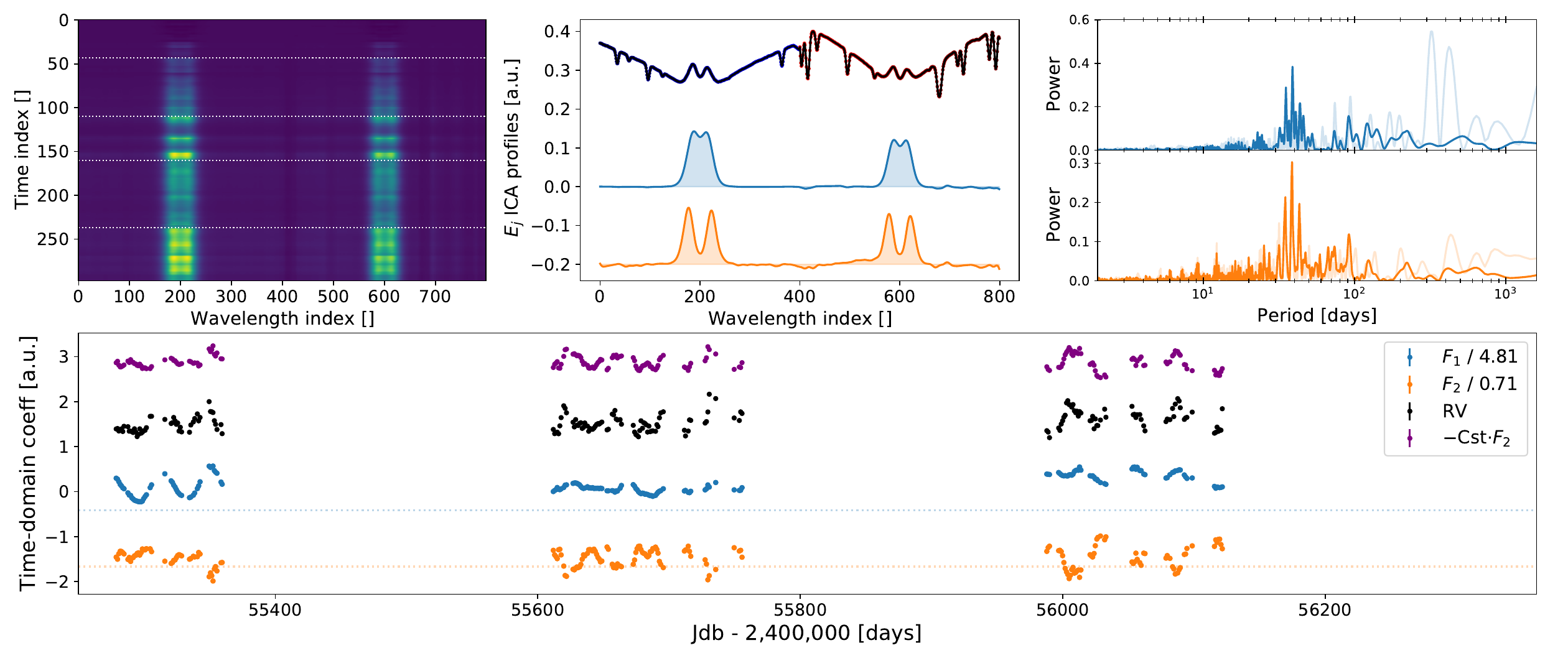}
	\caption{Same as Fig.\ref{FigPCA2} but performing a 2-component ICA decomposition on the sub-dimensional space of the two firsts PCA components. The previous PCA weighting profile $E_2(\lambda)$, associated to a broadening of the lines, is now visible as emission around the K1V minima. We also decreased the timespan of the $x$-axis to focus on the highest activity-level seasons (2010, 2011 and 2012) and improve the visual comparison between the RVs and $-F_2(t)$ for the year 2011 and 2012 (purple dots).}
	\label{FigICA}
 
\end{figure*}

We then investigated the $F_j(t)$ coefficients time-series (bottom panel in Fig.\ref{FigPCA2}). We displayed the Generalised Lomb-Scargle (GLS) periodogram \citep{Zechmeister(2009)} for each time-series in the top right panel of the Fig.\ref{FigPCA2}. Both $F_1(t)$ and $F_2(t)$ coefficients clearly exhibit the rotational power of the star, however the magnetic cycle is mainly visible in $F_1(t)$ and not in $F_2(t)$. 

Looking at the coefficients themselves, we notice that $F_1(t)$ is close to a perfect linear transform of the \Sidx{} time-series (Pearson correlation coefficient $\mathcal{R}=0.996$). This makes sense given that $E_1(\lambda)$ is the component with the largest EW. However, instead of extracting the core intensity with a triangular bandpass similarly to the Mount-Wilson S-index, the PCA produces the double-peak kernel which is the optimal extraction of the signal. We can understand this $F_1(t)$ component as a denoised version of the Mount-Wilson S-index performed by the PCA extraction, even if for $\alpha$ Cen B the proxy is not photon-noise limited.

We also performed an ICA decomposition with a 3-component model (see Appendix.\ref{app:bss} Fig.\ref{FigICA2}), but found that the algorithm was providing less reliable components. After investigations, we found that the ICA was not able to detect the ghost residual contamination and therefore, to reproduce the PCA results, only two components were required and adding a third component was providing meaningless and noisier decomposition. However, if the ICA is not able to see the ghost, it implies that the ghost contaminate in an unpredictable way the ICA components. This lower performance of the ICA could be related to its S/N sensitivity and its higher intrinsic instability as revealed in Sect.~\ref{sec:SNcomp}. 

Interestingly, since PCA captures the main direction of variance in the data and compress the information in a lower dimensional space, it could be used as a denoising tool \citep{Martinez2008,Routray2019}. We therefore, projected the data into the sub-dimensional space described by the truncated basis of the two first PCA components (in order to remove the residual ghost contamination) and fit a 2-component ICA on this data. The result of the decomposition is displayed in Fig.\ref{FigICA}. Very similar $F_j(t)$ coefficient time-series were found between the PCA and the ICA ($\mathcal{R}>0.99$). The main difference observed is related to the $E_2(\lambda)$ weighting profile that became a purely positive emission profile opposite to the PCA second component, which is similar to the decomposition observed for the Sun in Sect.~\ref{sec:sunpca}.

Because the $E_j(\lambda)$ weighting profiles are purely positive for the ICA, we can more easily interpret the flux variations amplitudes induced by the different components. As a recall, the solar magnetic cycle was found in paper I to induce a peak-to-peak variation around 1.5\% in normalised flux units. The magnetic cycle of $\alpha$ Cen B, mainly captured by the first component $F_1(t)$, is three times larger with a peak-to-peak of 4.81\% in normalised flux units. On the other side, the secondary order deformation, captured by $F_2(t)$ is 6.8 times weaker with a signature about 0.71\% in flux units. The larger amplitude of the activity signal on $\alpha$ Cen B is partially due to the higher activity level of the star and a more favorable chromospheric contrast for K-dwarf compared to G-dwarf, since the photospheric continuum is lower.

Because the signal is larger than for the Sun, we naturally expect that the S/N needed to detect the components is lower. We performed similar noise-injection for $\alpha$ Cen B as the ones described in Sect.~\ref{sec:SNcomp} and found that for the K-dwarf, the critical S/N below which the second component vanished is around $S/N_{\text{cont}}=75$ at 4000 $\ang{}$ ($S/N_{\text{core}} = 17$).

\subsubsection{RVs correction at stellar rotational timescale}
\label{sec:rot}

Detecting a multi-component variation of the line profiles is already an interesting result, never used or mentioned so far in the RV literature to our knowledge, but the most interesting part is certainly to observe how the PCA/ICA coefficients compared with the $\text{RV}(t)$ time-series. Since both algorithms provides very similar time-series, we only analyse the PCA components hereafter. By eye, it can be observed that RVs are better correlated with $F_1(t)$ or $F_2(t)$ depending on the season, this very clear in the bottom panel of Fig.\ref{FigICA} by comparing the black dots with blue or purple ones. %This is further confirmed in the season-by-season correlations plots in Fig.\ref{FigSeasons}. In this figure, the RV time-serie is compared with $F_1(t)$ and $F_2(t)$ time-series obtained by the PCA. 

We can easily guess that a multi-linear regressions of $F_j(t)$ is more powerful to fit the $\text{RV}(t)$ time-series than the classical S-index\footnote{which is a simple linear transform of our $F_1(t)$ vector as a recall}, which by construction is a weighted mixture of different filling factors. For this reason, we fit a multi-linear model of $F_1(t)$ and $F_2(t)$ on the RVs season-by-season including a parabolic trend: 

\begin{equation}
\label{eq:rvmodel}
    RV(t) = \alpha_s \cdot F_1(t_s)+\beta_s \cdot F_2(t_s) + \gamma_1 \cdot t + \gamma_2 \cdot t^2 ,
\end{equation}

with $\alpha_s$, $\beta_s$ the values obtained at epochs $t_s$ for each of the five seasons $s$ independently. The reason for the season-by-season analysis is that the first component is dominated by the stellar magnetic cycle and therefore contains a significant long-term power signal, while such power is missing in the second component. Since our main goal is to study stellar rotational timescale signal, any fit performed on the full baseline would be affected by this difference of long-trend power in both proxies. Furthermore, the present dataset contains a binary trend that may be incorrectly fit at the precision level we are aiming here, introducing a long-trend parasitic signal. As an example, fitting the two proxies on the RVs and adding an extract term for a linear drift indicates that a long-term trend of 50 cm/s/year may be necessary which would implies an accuracy better than $\sim$ 0.3\% on the binary $K$ semi-amplitude.  

Last but not least, studying the time variation of the parameters in the model may indicate that the used model is not the correct one or that some fundamental changes are observed for the active regions physical properties at the stellar surface.

\begin{figure*}
	
	\centering
	\includegraphics[width=18cm]{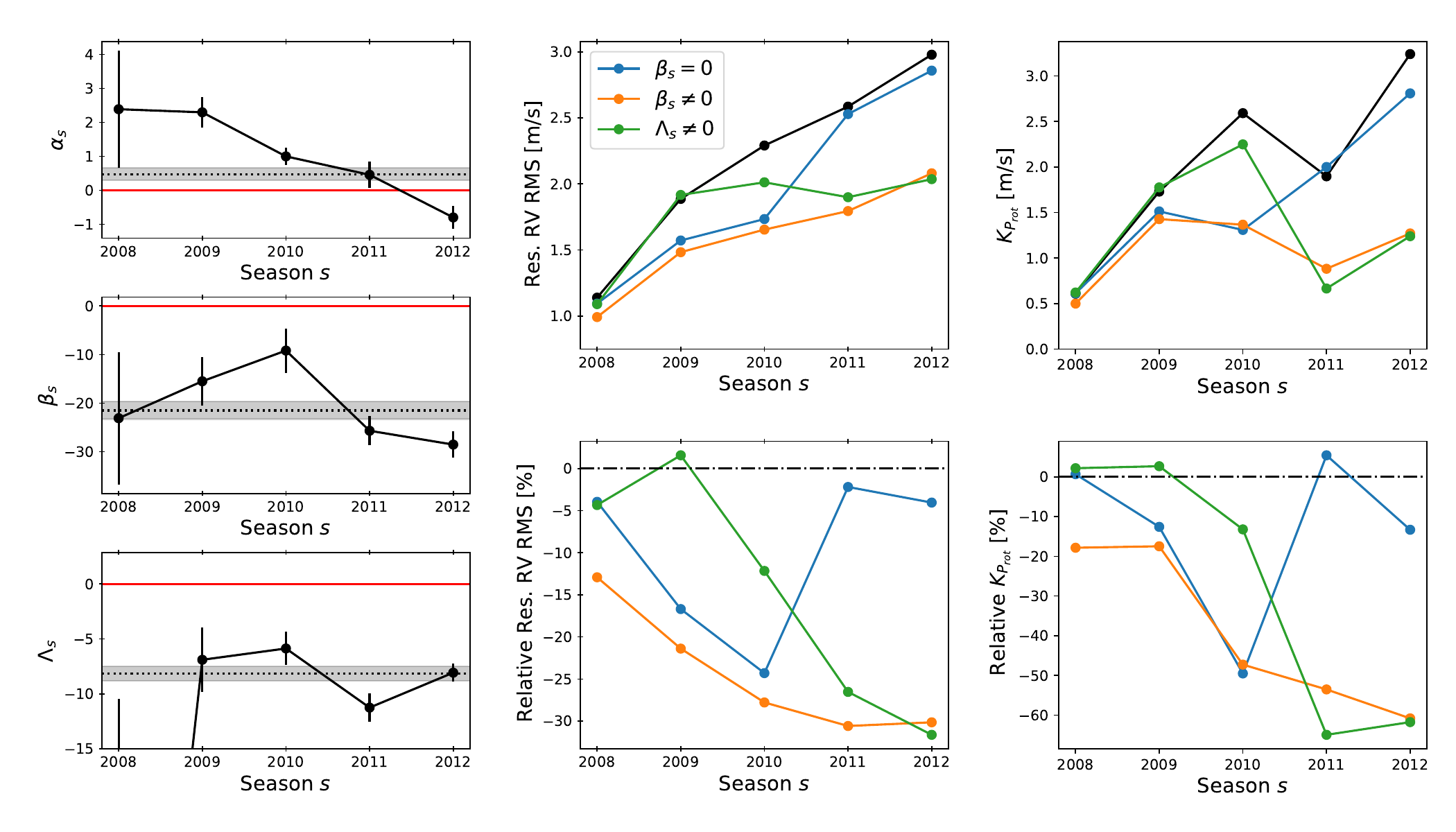}
	\caption{Season-by-season analysis of the RV model fit with Eq.\ref{eq:rvmodel} (related to $\alpha_s$ and $\beta_s$) and Eq.\ref{eq:rvmodel2} (related to $\Lambda_s$). \textbf{Top left:} $\alpha_s$ parameter (related in-distinctively to $F_1(t)$ or $S_{\text{MW}}(t)$) as a function of the season $s$. The parameter $\alpha$ obtained from the full baseline fit and its uncertainty is indicated by the horizontal dotted line and shaded area. 
     \textbf{Middle left:} Same as top left panel for the $\beta_s$ parameter (related to $F_2(t)$). 
        \textbf{Bottom left:} Same as top left panel for the $\Lambda_s$ parameter (related to $F_2(t)\cdot F_1(t)$). 
     \textbf{Top center:} RV RMS of each season fitting the RVs with (orange dots) or without (blue dots) the new proxy $F_2(t)$. Nominal seasonal RV RMS are displayed as reference (black) and bottom $y$-axis limit matches the instrumental precision expected of HARPS around 75 cm/s. \textbf{Bottom center:}  Same as top in relative difference with the RMS values of the uncorrected RVs. \textbf{Top right:} K semi-amplitude related to $P_{\text{rot}}$ periodicities between 32 and 42 days (see main text).  \textbf{Bottom right:} Same as top in relative difference with the uncorrected RVs.}
	\label{FigNewProxy}
 
\end{figure*}

The $\alpha_s$ and $\beta_s$ coefficient as a function of the season $s$ are displayed in the first and second panel of the Fig.\ref{FigNewProxy}. We also displayed in the panels the values obtained by requiring the same parameter over the five seasons (that we called $\alpha$ and $\beta$). A rapid comparison between the full baseline and season-by-season coefficients indicate that their value are not consistent over time and are evolving over the cycle. We will discuss in Sect.~\ref{sec:physical-interpretation} this aspect.

In order to compare the improvement brought by the new proxy $F_2(t)$, we computed in the centered panels of the Fig.\ref{FigNewProxy} the RV root-mean-square (RMS) for a model that includes ($\beta_s\neq 0$) or exclude ($\beta_s= 0$) the usage of this proxy. We notice that $F_1(t)$ mainly improve the 2010 season (by 25\%) but has a poor effect on 2011 or 2012, while those two seasons are as much as active as 2010 (see Fig.\ref{FigMag} as a reminder) and have a larger RV dispersion. When using both proxies, the improvement for 2010, 2011 and 2012 is about 30\%. 

While the RV RMS is often mentioned as a metric of merit, it has a limited meaning since it does not say anything about the periodicities affected by the corrections. In order to develop a metric more "stellar rotationally" driven, we extracted the highest peak in amplitude between the periodicities 32 and 44 days of a GLS for each season. The $K$ semi-amplitude  at $P_{\text{rot}}$ is displayed on the rights panels and demonstrated an improvement by 55\% at the stellar rotation period for the three seasons using both proxies. 

\begin{figure}
	
	\centering
	\includegraphics[width=8.5cm]{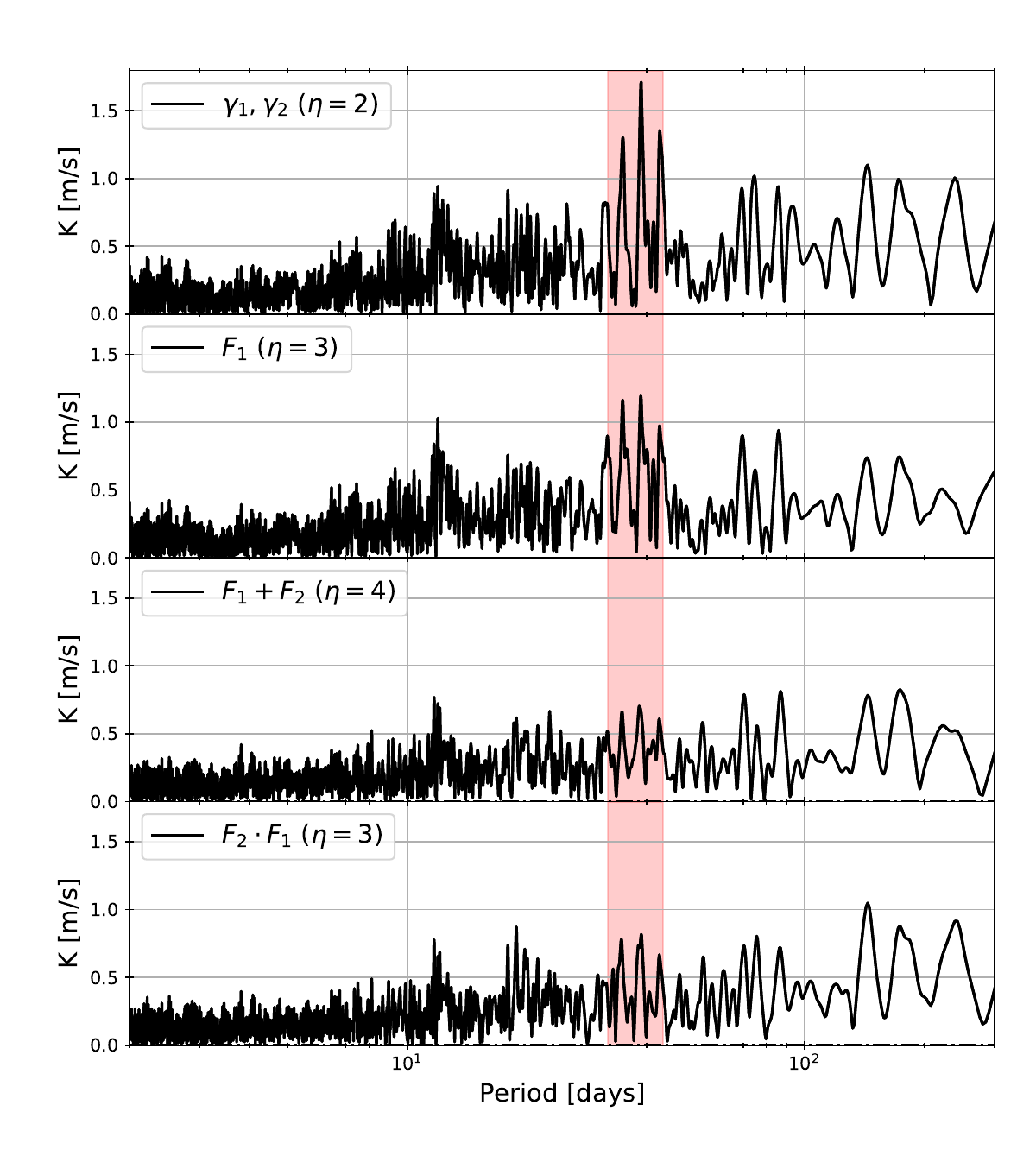}
	\caption{GLS of the different corrected RV residual time-series to highlight the correction performance at the rotational period (32-42 days) indicated by the red shaded region. The number of $\eta$ free parameters in the model is indicated in the legend of each panel. \textbf{First raw:}  RV corrected with the parabolic fit only ($\alpha=0$, $\beta=0$) and used as the reference uncorrected RVs exhibiting a 1.7 m/s signal at the rotational period. \textbf{Second raw:} RV residual obtained by using $F_1(t)$ only ($\alpha\neq0$, $\beta=0$) which represents the current fit ability when using only the $S_{\text{MW}}(t)$ proxy. The activity signal decreased down to 1.2 m/s. \textbf{Third raw:} RV residuals using  $F_1(t)$ and $F_2(t)$ in a multi-linear framework ($\alpha\neq0$, $\beta\neq0$). Rotational signal is decreased down to 0.7 m/s and similarly for its second harmonics around 12 days. \textbf{Fourth raw:} RV residuals using  $F_1(t)$ and $F_2(t)$ in a non-linear framework ($\Lambda\neq0$). Results are similar to the multi-linear framework except for the first harmonic that is boosted to 0.8 m/s.}
	\label{FigResiduals}
\end{figure}

We finally displayed in the Fig.\ref{FigResiduals} the GLS periodogram over the five years baseline. Because the periodogram were very similar if a season-by-season model was used compared to the full baseline fit, we only displayed here the full model fit that contains fewer $\eta$ free parameters. As reference, we compared the polynomial detrended RVs (top panel) with: 1) the model excluding (second panel) and 2) including the new $F_2(t)$ proxy (third panel). the analysis shows that the 2.0 m/s rotational signal is decreased down to 1.0 m/s with $F_1(t)$, but down to 0.50 m/s with the combination of $F_1(t)$ and $F_2(t)$.

%In the top panels of Fig.\ref{FigSeasons} are represented the full baseline correlations for which $F_1(t)$ seems to be the best activity proxy ($\mathcal{R}=0.48\pm0.01$). The multi-linear model season-by-season lead to a correlation of $\mathcal{R}=0.72\pm0.01$. The full potential of $F_2(t)$ becomes notably visible in the season-by-season analysis, displayed in the bottom panels, in particular for the most active period in 2011 and 2012. The correlation is better with $F_1(t)$ in 2010 ($\mathcal{R}=0.64\pm0.04$), but with $F_2(t)$ in 2011 ($\mathcal{R}=-0.53\pm0.03$) and 2012 ($\mathcal{R}= -0.73\pm0.02$). 

We note that opposite to the Sun for which $F_1(t)$ is the better proxy to fit the RVs, since dominated by the plage information, the relevance of $F_2(t)$ likely indicates that for the K-dwarf star another component has a major contribution in the emission in the core of the line and that this component is strongly affecting the RVs. We try to give a physical interpretation in the following section.

\subsubsection{Physical interpretation of the $E_j(\lambda)$ weighting profiles and $F_j(t)$ time coefficients}
\label{sec:physical-interpretation}

In general, any dimensional reduction algorithm is extremely difficult to interpret. This is due to the large number of degeneracies existing in the model and the incorrect assumptions about the data used by these algorithms to obtain a unique solution. As a simple example of this issue, we take the interpretation of the $E_j(\lambda)$ weighting profile and $F_j(t)$ time-coefficients. This section has for purpose to be purely qualitative. 

The results on $\alpha$ Cen B reveals that at least two profiles with two different characteristic widths are detected, but such result does not provide any information regarding the nature of the active regions producing those profiles. 

To interpret the components, it could be interesting to first compare the PCA components $E_j(\lambda)$ with the ones obtained on the Sun in Sect.~\ref{sec:sunpca}. We can identify that $E_1(\lambda)$ is close to the plage profile, $E_2(\lambda)$ is similar to second solar profile that was a mixture of the plage, network and sunspot. 

It has been shown that the network was affecting the RVs in a negligible way \citep{Milbourne(2021)} such that, while this later can contribute in the \Sidx{}, it does not contribute in the RV budget. Then the negative correlation between $F_2(t)$ and the RVs would simply be explained if $F_2(t)$ represents a difference of filling factor between network and plages. %Such an interpretation would also explain why the correlations are stronger in season-by-season analysis compared to global time-series fit and why the correlations change over time (see Fig.\ref{FigNewProxy}). 
The combination of $F_1(t)$ and $F_2(t)$ allowing to provide a better estimate of $f_{\text{plage}}(t)$ than $F_1(t)$ alone than contains some $f_\text{ntwk}(t)$ contamination. 

However, we have to point out that if some components represent difference of profiles, the sign convention used in Sect.~\ref{sec:BSS} to fix the components does not hold. As a consequence, if the sign of $E_1(\lambda)$ and $F_1(t)$ can be fixed without ambiguity by assuming that active regions represent a brightening of the chromosphere, determining the sign of $E_2(\lambda)$ and $F_2(t)$ is not so easy with the PCA. This degeneracy is better constrained with the ICA that only exhibit emission profiles. 

Using the ICA decomposition, we conclude that the second component $E_2(\lambda)$ is related to a profile component that is broader than the plage profile and takes most of its power around the K1 minima in the core of the lines (see Fig.\ref{FigSchema} as a recall). The most natural candidate would be the network as obtained from paper I. A broader profile contribution from the network was already point out in several studied on the Sun (see e.g bottom left panel in Fig.6 of \citet{Rezaei(2007)} or \citet{Sindhuja(2015)}). 

However, if so, our result therefore shows that the RV signal of $\alpha$ Cen B is dominated at high activity level by the network and not by plage. Moreover, in order to explain the anti-correlation with the RV signal, rather than the inhibition of the convective blueshift (ICB) as for the Sun, the signal would be an inhibition of the convective redshift (ICR). This is in contradiction with the result of \citet{Liebing(2021)} that did not find any evidence of convective redshift for such spectral type. The only hints of convective redshift found in the literature occurs towards the stellar limb (see e.g \citet{Reiners(2016)}). Moreover, it is unclear why no magnetic cycle signal is visible on $F_2(t)$, while such missing signature is only visible for the spot and not for the network on the Sun (see Fig.\ref{FigCaIIPCA} or paper~I).

Furthermore, if the components really represent the filling factor of different kind of active regions, it is unclear why the $\alpha_s$ and $\beta_s$ coefficients are changing along the stellar cycle and why $\alpha_s$ is even reverting its sign on the last season. One explanation could be that the weighting mixture of the convective blueshift and flux effect (that we neglected all along this work) is changing over time which would be the case if the faculae/spots ratio is changing from season-to-season. As an example, such long-term evolution in the structure of the RV activity signal has been highlighted recently for the Sun and interpreted as the spot/faculae ratio variation (see Fig.14 in \citet{Klein(2024)}). However, our residuals RVs demonstrate that no significant activity signal remains at the rotational period and we do not expect to be able to correct for both the inhibition of the convective blueshift and the flux effect given that their signature are very orthogonal to each other in the time-domain (see Fig.3 in \citet{Dumusque(2014)}). 

A last explanation could be related to another peculiar behaviour of the line profiles mentioned in paper I, namely that the \CaHK plage profiles are broader close to the limb compared to star center. If so, $F_2(\lambda)$ could be proportional to $F_2(t) \propto -\bar{\mu}(t)$ which would explain the anti-correlation due to Eq.\ref{eq:mono}. Indeed, assuming also that $F_1(t) \propto f_{\text{tot}}(t)$, and recalling that the PCA vectors are dimensionless, the application of Eq.\ref{eq:mono} to our dataset can be rewritten as:

\begin{equation}
\label{eq:rvmodel2}
    RV(t) = \Lambda_s \cdot F_2(t_s) \cdot F_1(t_s) + \gamma_1 \cdot t + \gamma_2 \cdot t^2 ,
\end{equation}

Such an explanation would also be coherent with the fact that no hint of magnetic cycle is visible in $F_2(t)$ which is similar to the behaviour of $\bar{\mu}(t)$ described in Appendix.\ref{app:nonlinear}. To check this eventuality, we fit the model described by the Eq.\ref{eq:rvmodel2} on the RV time-series similarly to the exercice done with the multi-linear framework in Sect.~\ref{sec:rot} with the Eq.\ref{eq:rvmodel}. This time, we observed that all the seasonal $\Lambda_s$ coefficients were consistent\footnote{Note that the very first season is not displayed due the large uncertainty attached to the value ($\Lambda_{2008}=-67\pm54$)} over time. Consequently, fitting a season-by-season $\Lambda_s$ model or a full baseline $\Lambda$ model provide very similar results. While at first glance this result may appear very promising, the performance comparison of this model (green curve in Fig.\ref{FigNewProxy}) are more  ambiguous, with a similar performance for 2011 and 2012 but a clear loss for the rotational period correction in 2010.

Furthermore, we showed in Appendix.\ref{app:nonlinear} that $\bar{\mu}(t)$ should play a minor role compared to $f(t)$ to correct the RVs and its relevance for the season 2011 and 2012 is therefore unexplained. We could imagine a situation at high-activity level where the star is covered by active regions more homogeneously such that the total filling factor $f(t)$ remains unchanged during several rotation, while the average $\bar{\mu}(t)$ location change, but we have to admit that such a case never occurred geometrically in any of our simulations.

We point out that all those conclusions are obtained from the Rosetta stone given by the solar results, but their extrapolation for cooler spectral types may be wrong. In definitive, for other spectral types than G2 stars and without direct stellar surface counterpart, it is hard to interpret data-driven components with physical quantities. In particular, if the first component $F_1(t)$ is likely representing the signal of the dominant "specie" of ARs, the second PCA component $F_2(t)$, that is a promising activity proxy to decorrelate the RVs, is more difficult to interpret physically. 

Nevertheless, this exercise showed that higher level line profile deformations are detectable in the \CaHK lines and those new distortions are powerful indicators to decorrelate the RVs. Therefore, stopping the extraction at the "Mount-Wilson" level could miss a significant part of the information revelant for the RVs.

\subsubsection{Extracting the main chromospheric component across spectral types}

A huge difference between the Mount-Wilson S-index ($S_{\text{MW}}$) and our index $F_1(t)$ is that our index is fit on the data by the $E_{1}(\lambda)$ profile, which implies that it is not a simple flux integration of the spectra. This extraction acts as a denoising process during the proxy extraction stage. $S_{\text{MW}}$ and any other similar pseudo equivalent-width measurements, are flux-integrated metrics very sensitive to local outliers and anomalous spikes. Such anomalous flux measurements are also not accounted in the uncertainties of the indexes that mainly assumes white noise. On the opposite fitting a profile on data is less affected by local outliers and could even be completely mitigated by an iterative fitting method when combined with an usual sigma-clipping rejection method on the residuals of the fit. 

We therefore decided to extract the first principal component $E_1(\lambda)$ across the spectral type sequence, such that this empirical profile could be used to extract a more precise activity proxy in the future (Cretignier et. al in prep).

\begin{figure}
	
	\centering
	\includegraphics[width=8cm]{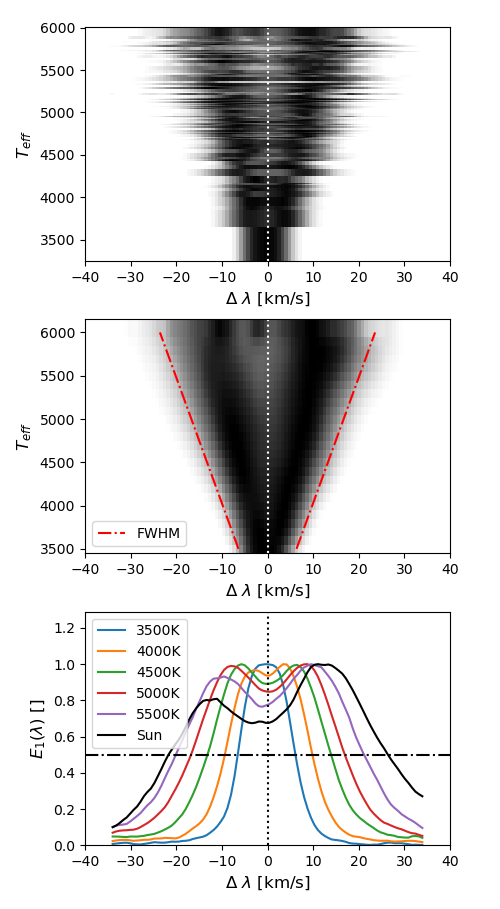}
	\caption{Extraction of the emission profile library $E_{1}(\lambda,T_{\text{eff}})$ for the CaII lines. \textbf{Top:} $E_1(\lambda)$ component extracted from PCA on diverse stellar datasets. \textbf{Middle:} Smoothed model by polynomial fit on each velocity bin. A linear regression was fit (red dotted-dashed line) for the FWHM as a function of $T_{\text{eff}}$ (see main text). \textbf{Bottom:} Comparison of five emission profiles from the library $E_{1}(\lambda,T_{\text{eff}})$ with the solar $E_1(\lambda)$ component. The effective temperature are given in the labels.}
	\label{FigE1}
 
\end{figure}

We gathered some main sequence stars intensively observed with HARPS and HARPS-N for which a clear magnetic cycle was visible in the Mount-Wilson S-index. The data are made of 39 HARPS and 14 HARPS-N stellar datasets and were selected to probe effective temperature between 4000K up to 6200K in order to provide a calibration from M1V up to F9V spectral type. The stellar atmospheric temperature are obtained directly by YARARA as described in the Appendix.\ref{app:atmos}. All the spectra were post-processed in an identical way to $\alpha$ Cen B as described earlier (see Sect.~\ref{sec:preprocess}) and only the first principal component $E_\text{1}(\lambda)$ was studied since this component was the only one detectable (see Sect.~\ref{sec:SNcomp}) even for moderated S/N values ($S/N_{\text{cont}}<150$). The wavelength of the lines were replaced by the velocity variable and both lines H and K averaged together.
Finally, a smooth model over the effective temperature $E_\text{1}(\lambda,T_{\text{eff}})$ was obtained by fitting a seventh degree polynomial function for each velocity bin. The results are displayed in Fig.\ref{FigE1}. 

In the top panel, star-to-star variations can be observed which may be explained either by different activity level of the stars, different metallicity or inaccurate effective temperature value. However, even if some star-to-star variation is observed, it is clear that the main dependency is globally due to the effective temperature. This confirms that stellar activity studies done on the Sun \textbf{have limitations in their application} for other spectral types. 

We identify that the FWHM of the emission profile $E_\text{1}(\lambda,T_{\text{eff}})$ is decreasing from $\sim$45 \kms{} to $\sim$10 \kms{} toward cooler spectral type. The linear relation with $T_{\text{eff}}$ goes as: 
\begin{equation}
\text{FWHM}(T_{\text{eff}}) = 12.6+31.7 \cdot \left(\frac{T_{\text{eff}}-3500K}{5778K - 3500K}\right) \text{km/s}.
\end{equation}

This effect reduces the number of independent wavelength sampling elements we have to split the different active regions contributions for cool M-dwarf. Also, the peak asymmetry between the two K2 maxima almost vanishes and no separation is observed below 4000K. Very often, it is argued in literature than because of the lower S/N in the extreme blue for M-dwarfs, $H_\alpha$ tend to be a less noisy activity proxy. Our result also reveal that this is because the width of the emission profile itself becomes thinner than the signal in the \CaHK lines becomes more difficult to characterized on late-type stars. 

\section{Conclusions}

\label{sec:conclusion}

We showed in this paper that a strong revisit of the current paradigm for activity proxies extraction is needed. The "Mount Wilson" S-index convention, widely used in the RV community as an old legacy, is an imperfect activity proxy since the triangular bandpass used for its computation is likely mixing the intensity signals of plage, network and spots plus their eventual CLV signatures (paper I) and a multi-component analysis of the profiles should be conducted when signal-to-noise ratio permits.

For other spectral types than the Sun, data-driven activity profiles can be extracted using a PCA or ICA on the spectra time-series of the CaII H\&K lines. As a proof of concept, we applied the decomposition on $\alpha$ Cen B, a moderate active K-dwarf star intensively observed by HARPS that contains strong activity signals, and extracted three PCA components. One component $F_3(t)$ was related to a 1-year systematics, which demonstrates the usefulness of PCA to disentangle independent effect such as stellar activity and instrumental systematics by fitting the two lines simultaneously. The two other components $F_1(t)$ and $F_2(t)$ were related to stellar activity components. Those two time-series have been shown to be  powerful activity proxies for RVs, when combined, in particular to correct the rotational modulation in 2010, 2011 and 2012. The second ICA component taking most of its power around the K1 minima, while the first ICA component takes most of its power at the core center around the K3 minimum.

We then pointed out the difficulty to interpret the time-series extracted due to the dimensionless nature of PCA algorithms and the lack of a simultaneous photometric proxy such as those provided by light curves. In particular, all our interpretations for the second activity proxy $F_2(t)$ failed to fit in the current mainstream ICB paradigm of faculae and this component, while promising for RV correction remains a mystery. 

Our work could be further improved by thinking how the real intrinsic activity profiles could be extracted (by determining the matrix $\boldsymbol{W}$ in Eq.\ref{EqTransform}) which would avoid their expected mixing in our proxy. This could also be achieved by applying some filtering in the time-domain or by using more chromospheric or photospheric lines to disentangle their signatures. 

%A main disadvantage of data-driven methods is that the amplitude of the signals is unconstrained. As a consequence, while they can be used to denoise the data in order to increase the precision, they are always poorly accurate. Such an element is extremely concerning given that Eq.\ref{eq:mono} is non-linear and a bias or offset on the the filling factor time-series will introduce a wrong $\Delta \text{RV}_{\text{ICB}}(t)$ model ultimately.

Naturally, our work does not represent at all the state-of-art in RV correction \citep{Zhao(2021)} and far better results can be achieved. Actually, we already processed this dataset in \citet{Cretignier(2023)} and obtained better results than here. However, opposite to our previous work and similar methods that require a spectrum on a wide bandpass, the work presented here only requires the careful analysis of two lines in a restricted bandpass which is a massive difference for the design of an instrument. Because of it, our method is closer to a physically-driven model that is easier to interpret when using a single line than when mixing thousand of photospheric lines in a CCF for instance.

In definitive, since the \CaHK lines are likely the best "single line activity proxy", we strongly advise that high-resolution spectra extraction should carefully protect them of any contamination by instrumental systematic. As an example, both HARPS and HARPS-N suffer from contaminations by ghosts that cross the cores of the lines or that are close to them \citep{Cretignier(2021),Dumusque(2021)}, which makes the extraction of the signal very challenging.  Extraction methods at the raw frames level should be considered.

There exist no other stellar lines in the visible range that contains a richness of information as large as the one provided by the \CaHK lines to monitor the active regions at the stellar surfaces. The absence of those lines from the spectral range of a spectrograph is a dramatic loss since no other line characterize the upper photosphere/chromosphere at such good S/N. In that context, we highlight that their inclusions has been considered as \textit{not crucial} for the ANDES instrument at ELT \citep{Palle(2023)}, a statement that we do not share.     

We demonstrated with photon noise injection on the Sun that for $S/N_{\text{cont}}<200$ in the continuum at 4000 \ang{} ($S/N_{\text{core}}<45$), the multi-components variation vanished in the noise. This critical S/N was lower for $\alpha$ Cen B ($S/N_{\text{cont}}<75$ or $S/N_{\text{core}}<17$). The reason for it was the larger amplitude of the activity signal compared to the Sun (three times larger), while the precise reason for this higher amplitude signal is likely related to a higher stellar activity level, but also perhaps to a more favorable spectral type of the K1 star in comparison to our G2 host star.

For low S/N observations, it is very likely that the PCA will not be able to return any meaningful component. For such cases, fitting directly the empirical $E_1(\lambda,T_{\text{eff}})$ profile provided in this work could be the very last option to avoid the usual EW core integration that would deliver an S-index metric. 

Also, since the S/N in the extreme blue ($\sim$4000 $\ang{}$) is often low and degraded due to the presence of filters, coating or throughout the fibers, it may be more interesting for spectrographs to split the violet part of the spectrum sooner in the optical path in order to optimize the flux in the core of the lines. This is even more justified since that wavelength range contains a very small amount of RV information (see for instance Fig.2.6 right panel in \citet{Cretignier(thesis)}) and precise RV precision is not required to extract the \CaHK profiles. The most relevant factor being accurate, or at least precise, flux measurements at high signal-to-noise ratio. 

In that context, a spectrophotometric space mission at high spectral resolution ($R\sim 100'000$), but in a restricted bandpass ($\sim 10$ \ang{}) taking daily observations of the CaII K line\footnote{Or CaII H line to monitor simultaneously the Balmer $H_{\varepsilon}$ line} for a large sample of stars could be a game changer for the RV method. 

\section*{Acknowledgements}

We thank the anonymous referee for their precious and constructive comments regarding the paper and Pr. Christopher Watson for the fruitful discussion we had with him. 
M.C. acknowledges the SNSF support under grant P500PT\_211024.

This research has made use of NASA's Astrophysics Data System (ADS) bibliographic services. 
We acknowledge the community efforts devoted to the development of the following open-source packages that were used in this work: numpy (\href{http:\\numpy.org}{numpy.org}), matplotlib (\href{http:\\matplotlib.org}{matplotlib.org}), and astropy (\href{http:\\astropy.org}{astropy.org}).

%%%%%%%%%%%%%%%%%%%%%%%%%%%%%%%%%%%%%%%%%%%%%%%%%%
\section*{Data Availability}

The data underlying this article are available on the ESO archive for HARPS (\url{http://archive.eso.org/cms.html}) and HARPS-N (\url{http://archives.ia2.inaf.it/tng/}) spectra. Useful datasets such as the emission profiles $E_{\text{1}}(\lambda,T_\text{eff})$ will be provided as online supplementary material on CDS (TBD: provide link later).



\bibliographystyle{mnras}
\bibliography{mnras_template} % if your bibtex file is called example.bib

\begin{thebibliography}{}
\makeatletter
\relax
\def\mn@urlcharsother{\let\do\@makeother \do\$\do\&\do\#\do\^\do\_\do\%\do\~}
\def\mn@doi{\begingroup\mn@urlcharsother \@ifnextchar [ {\mn@doi@}
  {\mn@doi@[]}}
\def\mn@doi@[#1]#2{\def\@tempa{#1}\ifx\@tempa\@empty \href
  {http://dx.doi.org/#2} {doi:#2}\else \href {http://dx.doi.org/#2} {#1}\fi
  \endgroup}
\def\mn@eprint#1#2{\mn@eprint@#1:#2::\@nil}
\def\mn@eprint@arXiv#1{\href {http://arxiv.org/abs/#1} {{\tt arXiv:#1}}}
\def\mn@eprint@dblp#1{\href {http://dblp.uni-trier.de/rec/bibtex/#1.xml}
  {dblp:#1}}
\def\mn@eprint@#1:#2:#3:#4\@nil{\def\@tempa {#1}\def\@tempb {#2}\def\@tempc
  {#3}\ifx \@tempc \@empty \let \@tempc \@tempb \let \@tempb \@tempa \fi \ifx
  \@tempb \@empty \def\@tempb {arXiv}\fi \@ifundefined
  {mn@eprint@\@tempb}{\@tempb:\@tempc}{\expandafter \expandafter \csname
  mn@eprint@\@tempb\endcsname \expandafter{\@tempc}}}

\bibitem[\protect\citeauthoryear{{Adibekyan}, {Sousa}, {Santos}, {Delgado
  Mena}, {Gonz{\'a}lez Hern{\'a}ndez}, {Israelian}, {Mayor}  \&
  {Khachatryan}}{{Adibekyan} et~al.}{2012}]{Adibekyan(2012)}
{Adibekyan} V.~Z.,  {Sousa} S.~G.,  {Santos} N.~C.,  {Delgado Mena} E.,
  {Gonz{\'a}lez Hern{\'a}ndez} J.~I.,  {Israelian} G.,  {Mayor} M.,
  {Khachatryan} G.,  2012, \mn@doi [\aap] {10.1051/0004-6361/201219401}, \href
  {https://ui.adsabs.harvard.edu/abs/2012A&A...545A..32A} {545, A32}

\bibitem[\protect\citeauthoryear{{Aigrain}, {Pont}  \& {Zucker}}{{Aigrain}
  et~al.}{2012}]{Aigrain(2012)}
{Aigrain} S.,  {Pont} F.,   {Zucker} S.,  2012, \mn@doi [mnras]
  {10.1111/j.1365-2966.2011.19960.x}, \href
  {https://ui.adsabs.harvard.edu/abs/2012MNRAS.419.3147A} {419, 3147}

\bibitem[\protect\citeauthoryear{{Al Moulla}, {Dumusque}, {Cretignier}, {Zhao}
  \& {Valenti}}{{Al Moulla} et~al.}{2022}]{Moulla(2022)}
{Al Moulla} K.,  {Dumusque} X.,  {Cretignier} M.,  {Zhao} Y.,   {Valenti}
  J.~A.,  2022, arXiv e-prints, \href
  {https://ui.adsabs.harvard.edu/abs/2022arXiv220507047A} {p. arXiv:2205.07047}

\bibitem[\protect\citeauthoryear{{Al Moulla}, {Dumusque}  \& {Cretignier}}{{Al
  Moulla} et~al.}{2024}]{Moulla(2024)}
{Al Moulla} K.,  {Dumusque} X.,   {Cretignier} M.,  2024, \mn@doi [\aap]
  {10.1051/0004-6361/202348150}, \href
  {https://ui.adsabs.harvard.edu/abs/2024A&A...683A.106A} {683, A106}

\bibitem[\protect\citeauthoryear{{Baliunas} et~al.,}{{Baliunas}
  et~al.}{1995}]{Baliunas(1995)}
{Baliunas} S.~L.,  et~al., 1995, \mn@doi [\apj] {10.1086/175072}, \href
  {https://ui.adsabs.harvard.edu/abs/1995ApJ...438..269B} {438, 269}

\bibitem[\protect\citeauthoryear{{Bauer}, {Reiners}, {Beeck}  \&
  {Jeffers}}{{Bauer} et~al.}{2018}]{Bauer(2018)}
{Bauer} F.~F.,  {Reiners} A.,  {Beeck} B.,   {Jeffers} S.~V.,  2018, \mn@doi
  [aap] {10.1051/0004-6361/201731227}, \href
  {https://ui.adsabs.harvard.edu/abs/2018A&A...610A..52B} {610, A52}

\bibitem[\protect\citeauthoryear{{Beeck}, {Cameron}, {Reiners}  \&
  {Sch{\"u}ssler}}{{Beeck} et~al.}{2013}]{Beeck(2013a)}
{Beeck} B.,  {Cameron} R.~H.,  {Reiners} A.,   {Sch{\"u}ssler} M.,  2013,
  \mn@doi [aap] {10.1051/0004-6361/201321343}, \href
  {http://adsabs.harvard.edu/abs/2013A%26A...558A..48B} {558, A48}

\bibitem[\protect\citeauthoryear{{Bj{\o}rgen}, {Sukhorukov}, {Leenaarts},
  {Carlsson}, {de la Cruz Rodr{\'\i}guez}, {Scharmer}  \&
  {Hansteen}}{{Bj{\o}rgen} et~al.}{2018}]{Bjorgen(2018)}
{Bj{\o}rgen} J.~P.,  {Sukhorukov} A.~V.,  {Leenaarts} J.,  {Carlsson} M.,  {de
  la Cruz Rodr{\'\i}guez} J.,  {Scharmer} G.~B.,   {Hansteen} V.~H.,  2018,
  \mn@doi [aap] {10.1051/0004-6361/201731926}, \href
  {https://ui.adsabs.harvard.edu/abs/2018A&A...611A..62B} {611, A62}

\bibitem[\protect\citeauthoryear{{Boisse}, {Bonfils}  \& {Santos}}{{Boisse}
  et~al.}{2012}]{Boisse(2012)}
{Boisse} I.,  {Bonfils} X.,   {Santos} N.~C.,  2012, \mn@doi [aap]
  {10.1051/0004-6361/201219115}, \href
  {https://ui.adsabs.harvard.edu/abs/2012A&A...545A.109B} {545, A109}

\bibitem[\protect\citeauthoryear{{Borgniet}, {Meunier}  \&
  {Lagrange}}{{Borgniet} et~al.}{2015}]{Borgniet(2015)}
{Borgniet} S.,  {Meunier} N.,   {Lagrange} A.-M.,  2015, \mn@doi [aap]
  {10.1051/0004-6361/201425007}, \href
  {http://adsabs.harvard.edu/abs/2015A%26A...581A.133B} {581, A133}

\bibitem[\protect\citeauthoryear{{Bourrier} et~al.,}{{Bourrier}
  et~al.}{2021}]{Bourrier(2021)}
{Bourrier} V.,  et~al., 2021, \mn@doi [aap] {10.1051/0004-6361/202141527},
  \href {https://ui.adsabs.harvard.edu/abs/2021A&A...654A.152B} {654, A152}

\bibitem[\protect\citeauthoryear{{Brandt} \& {Solanki}}{{Brandt} \&
  {Solanki}}{1990}]{Brandt(1990)}
{Brandt} P.~N.,  {Solanki} S.~K.,  1990, aap, \href
  {https://ui.adsabs.harvard.edu/abs/1990A&A...231..221B} {231, 221}

\bibitem[\protect\citeauthoryear{{Brewer}, {Fischer}, {Valenti}  \&
  {Piskunov}}{{Brewer} et~al.}{2016}]{Brewer(2016)}
{Brewer} J.~M.,  {Fischer} D.~A.,  {Valenti} J.~A.,   {Piskunov} N.,  2016,
  \mn@doi [\apjs] {10.3847/0067-0049/225/2/32}, \href
  {https://ui.adsabs.harvard.edu/abs/2016ApJS..225...32B} {225, 32}

\bibitem[\protect\citeauthoryear{{Cavallini}, {Ceppatelli}  \&
  {Righini}}{{Cavallini} et~al.}{1985}]{Cavallini(1985)}
{Cavallini} F.,  {Ceppatelli} G.,   {Righini} A.,  1985, aap, \href
  {http://adsabs.harvard.edu/abs/1985A%26A...143..116C} {143, 116}

\bibitem[\protect\citeauthoryear{{Chen} \& {Guestrin}}{{Chen} \&
  {Guestrin}}{2016}]{Chan(2016)}
{Chen} T.,  {Guestrin} C.,  2016, \mn@doi [arXiv e-prints]
  {10.48550/arXiv.1603.02754}, \href
  {https://ui.adsabs.harvard.edu/abs/2016arXiv160302754C} {p. arXiv:1603.02754}

\bibitem[\protect\citeauthoryear{{Collier Cameron} et~al.,}{{Collier Cameron}
  et~al.}{2019}]{Collier(2019)}
{Collier Cameron} A.,  et~al., 2019, \mn@doi [mnras] {10.1093/mnras/stz1215},
  \href {https://ui.adsabs.harvard.edu/abs/2019MNRAS.487.1082C} {487, 1082}

\bibitem[\protect\citeauthoryear{{Cosentino} et~al.,}{{Cosentino}
  et~al.}{2012}]{Cosentino(2012)}
{Cosentino} R.,  et~al., 2012, in Ground-based and Airborne Instrumentation for
  Astronomy IV. p. 84461V, \mn@doi{10.1117/12.925738}

\bibitem[\protect\citeauthoryear{{Costes} et~al.,}{{Costes}
  et~al.}{2021}]{Costes(2021)}
{Costes} J.~C.,  et~al., 2021, \mn@doi [mnras] {10.1093/mnras/stab1183}, \href
  {https://ui.adsabs.harvard.edu/abs/2021MNRAS.505..830C} {505, 830}

\bibitem[\protect\citeauthoryear{{Crass} et~al.,}{{Crass}
  et~al.}{2021}]{Crass(2021)b}
{Crass} J.,  et~al., 2021, arXiv e-prints, \href
  {https://ui.adsabs.harvard.edu/abs/2021arXiv210714291C} {p. arXiv:2107.14291}

\bibitem[\protect\citeauthoryear{{Cretignier}}{{Cretignier}}{2022}]{Cretignier(thesis)}
{Cretignier} M.,  2022, PhD thesis, University of Geneva, Switzerland

\bibitem[\protect\citeauthoryear{{Cretignier}, {Dumusque}, {Allart}, {Pepe}  \&
  {Lovis}}{{Cretignier} et~al.}{2020a}]{Cretignier(2020a)}
{Cretignier} M.,  {Dumusque} X.,  {Allart} R.,  {Pepe} F.,   {Lovis} C.,
  2020a, \mn@doi [aap] {10.1051/0004-6361/201936548}, \href
  {https://ui.adsabs.harvard.edu/abs/2020A&A...633A..76C} {633, A76}

\bibitem[\protect\citeauthoryear{{Cretignier}, {Francfort}, {Dumusque},
  {Allart}  \& {Pepe}}{{Cretignier} et~al.}{2020b}]{Cretignier(2020b)}
{Cretignier} M.,  {Francfort} J.,  {Dumusque} X.,  {Allart} R.,   {Pepe} F.,
  2020b, \mn@doi [aap] {10.1051/0004-6361/202037722}, \href
  {https://ui.adsabs.harvard.edu/abs/2020A&A...640A..42C} {640, A42}

\bibitem[\protect\citeauthoryear{{Cretignier}, {Dumusque}, {Hara}  \&
  {Pepe}}{{Cretignier} et~al.}{2021}]{Cretignier(2021)}
{Cretignier} M.,  {Dumusque} X.,  {Hara} N.~C.,   {Pepe} F.,  2021, \mn@doi
  [aap] {10.1051/0004-6361/202140986}, \href
  {https://ui.adsabs.harvard.edu/abs/2021A&A...653A..43C} {653, A43}

\bibitem[\protect\citeauthoryear{{Cretignier}, {Dumusque}  \&
  {Pepe}}{{Cretignier} et~al.}{2022}]{Cretignier(2022)}
{Cretignier} M.,  {Dumusque} X.,   {Pepe} F.,  2022, \mn@doi [aap]
  {10.1051/0004-6361/202142435}, \href
  {https://ui.adsabs.harvard.edu/abs/2022A&A...659A..68C} {659, A68}

\bibitem[\protect\citeauthoryear{{Cretignier}, {Dumusque}, {Aigrain}  \&
  {Pepe}}{{Cretignier} et~al.}{2023}]{Cretignier(2023)}
{Cretignier} M.,  {Dumusque} X.,  {Aigrain} S.,   {Pepe} F.,  2023, \mn@doi
  [\aap] {10.1051/0004-6361/202347232}, \href
  {https://ui.adsabs.harvard.edu/abs/2023A&A...678A...2C} {678, A2}

\bibitem[\protect\citeauthoryear{{Cretignier}, {Pietrow}  \&
  {Aigrain}}{{Cretignier} et~al.}{2024}]{Cretignier(2024)}
{Cretignier} M.,  {Pietrow} A.~G.~M.,   {Aigrain} S.,  2024, \mn@doi [\mnras]
  {10.1093/mnras/stad3292}, \href
  {https://ui.adsabs.harvard.edu/abs/2024MNRAS.527.2940C} {527, 2940}

\bibitem[\protect\citeauthoryear{{Davis}, {Cisewski}, {Dumusque}, {Fischer}  \&
  {Ford}}{{Davis} et~al.}{2017}]{Davis(2017)}
{Davis} A.~B.,  {Cisewski} J.,  {Dumusque} X.,  {Fischer} D.~A.,   {Ford}
  E.~B.,  2017, \mn@doi [apj] {10.3847/1538-4357/aa8303}, \href
  {http://adsabs.harvard.edu/abs/2017ApJ...846...59D} {846, 59}

\bibitem[\protect\citeauthoryear{{DeWarf}, {Datin}  \& {Guinan}}{{DeWarf}
  et~al.}{2010}]{Dewarf(2010)}
{DeWarf} L.~E.,  {Datin} K.~M.,   {Guinan} E.~F.,  2010, \mn@doi [apj]
  {10.1088/0004-637X/722/1/343}, \href
  {http://adsabs.harvard.edu/abs/2010ApJ...722..343D} {722, 343}

\bibitem[\protect\citeauthoryear{{Delgado Mena}, {Tsantaki}, {Adibekyan},
  {Sousa}, {Santos}, {Gonz{\'a}lez Hern{\'a}ndez}  \& {Israelian}}{{Delgado
  Mena} et~al.}{2017}]{Delgado(2017)}
{Delgado Mena} E.,  {Tsantaki} M.,  {Adibekyan} V.~Z.,  {Sousa} S.~G.,
  {Santos} N.~C.,  {Gonz{\'a}lez Hern{\'a}ndez} J.~I.,   {Israelian} G.,  2017,
  \mn@doi [\aap] {10.1051/0004-6361/201730535}, \href
  {https://ui.adsabs.harvard.edu/abs/2017A&A...606A..94D} {606, A94}

\bibitem[\protect\citeauthoryear{{Desort}, {Lagrange}, {Galland}, {Udry}  \&
  {Mayor}}{{Desort} et~al.}{2007}]{Desort(2007)}
{Desort} M.,  {Lagrange} A.~M.,  {Galland} F.,  {Udry} S.,   {Mayor} M.,  2007,
  \mn@doi [aap] {10.1051/0004-6361:20078144}, \href
  {https://ui.adsabs.harvard.edu/abs/2007A&A...473..983D} {473, 983}

\bibitem[\protect\citeauthoryear{{Dineva}, {Pearson}, {Ilyin}, {Verma},
  {Diercke}, {Strassmeier}  \& {Denker}}{{Dineva} et~al.}{2022}]{Dineva(2022)}
{Dineva} E.,  {Pearson} J.,  {Ilyin} I.,  {Verma} M.,  {Diercke} A.,
  {Strassmeier} K.~G.,   {Denker} C.,  2022, \mn@doi [Astronomische
  Nachrichten] {10.1002/asna.20223996}, \href
  {https://ui.adsabs.harvard.edu/abs/2022AN....34323996D} {343, e23996}

\bibitem[\protect\citeauthoryear{{Dravins}}{{Dravins}}{1982}]{Dravins(1982)}
{Dravins} D.,  1982, \mn@doi [araa] {10.1146/annurev.aa.20.090182.000425},
  \href {https://ui.adsabs.harvard.edu/abs/1982ARA&A..20...61D} {20, 61}

\bibitem[\protect\citeauthoryear{{Dumusque}}{{Dumusque}}{2014}]{Dumusque(2014b)}
{Dumusque} X.,  2014, \mn@doi [apj] {10.1088/0004-637X/796/2/133}, \href
  {http://adsabs.harvard.edu/abs/2014ApJ...796..133D} {796, 133}

\bibitem[\protect\citeauthoryear{{Dumusque}}{{Dumusque}}{2018}]{Dumusque(2018)}
{Dumusque} X.,  2018, \mn@doi [aap] {10.1051/0004-6361/201833795}, \href
  {https://ui.adsabs.harvard.edu/abs/2018A&A...620A..47D} {620, A47}

\bibitem[\protect\citeauthoryear{{Dumusque} et~al.,}{{Dumusque}
  et~al.}{2012}]{Dumusque(2012b)}
{Dumusque} X.,  et~al., 2012, \mn@doi [nat] {10.1038/nature11572}, \href
  {http://adsabs.harvard.edu/abs/2012Natur.491..207D} {491, 207}

\bibitem[\protect\citeauthoryear{{Dumusque}, {Boisse}  \& {Santos}}{{Dumusque}
  et~al.}{2014}]{Dumusque(2014)}
{Dumusque} X.,  {Boisse} I.,   {Santos} N.~C.,  2014, \mn@doi [apj]
  {10.1088/0004-637X/796/2/132}, \href
  {http://adsabs.harvard.edu/abs/2014ApJ...796..132D} {796, 132}

\bibitem[\protect\citeauthoryear{{Dumusque} et~al.,}{{Dumusque}
  et~al.}{2021}]{Dumusque(2021)}
{Dumusque} X.,  et~al., 2021, \mn@doi [aap] {10.1051/0004-6361/202039350},
  \href {https://ui.adsabs.harvard.edu/abs/2021A&A...648A.103D} {648, A103}

\bibitem[\protect\citeauthoryear{{Duncan} et~al.,}{{Duncan}
  et~al.}{1991}]{Duncan(1991)}
{Duncan} D.~K.,  et~al., 1991, \mn@doi [\apjs] {10.1086/191572}, \href
  {https://ui.adsabs.harvard.edu/abs/1991ApJS...76..383D} {76, 383}

\bibitem[\protect\citeauthoryear{{Faria} et~al.,}{{Faria}
  et~al.}{2020}]{Faria(2020)}
{Faria} J.~P.,  et~al., 2020, \mn@doi [aap] {10.1051/0004-6361/201936389},
  \href {https://ui.adsabs.harvard.edu/abs/2020A&A...635A..13F} {635, A13}

\bibitem[\protect\citeauthoryear{{Flores R.}, {Corral}  \&
  {Fierro-Santill{\'a}n}}{{Flores R.} et~al.}{2021}]{Flores(2021)}
{Flores R.} M.,  {Corral} L.~J.,   {Fierro-Santill{\'a}n} C.~R.,  2021, \mn@doi
  [arXiv e-prints] {10.48550/arXiv.2105.07110}, \href
  {https://ui.adsabs.harvard.edu/abs/2021arXiv210507110F} {p. arXiv:2105.07110}

\bibitem[\protect\citeauthoryear{{Forg{\'a}cs-Dajka}, {Dobos}  \&
  {Ballai}}{{Forg{\'a}cs-Dajka} et~al.}{2021}]{ForgcsDajka2021}
{Forg{\'a}cs-Dajka} E.,  {Dobos} L.,   {Ballai} I.,  2021, \mn@doi [\aap]
  {10.1051/0004-6361/202140731}, \href
  {https://ui.adsabs.harvard.edu/abs/2021A&A...653A..50F} {653, A50}

\bibitem[\protect\citeauthoryear{{Gibson}, {Howard}, {Marcy}, {Edelstein},
  {Wishnow}  \& {Poppett}}{{Gibson} et~al.}{2016}]{Gibson(2016)}
{Gibson} S.~R.,  {Howard} A.~W.,  {Marcy} G.~W.,  {Edelstein} J.,  {Wishnow}
  E.~H.,   {Poppett} C.~L.,  2016, in {Evans} C.~J.,  {Simard} L.,   {Takami}
  H.,  eds,  Society of Photo-Optical Instrumentation Engineers (SPIE)
  Conference Series Vol. 9908, Ground-based and Airborne Instrumentation for
  Astronomy VI. p. 990870, \mn@doi{10.1117/12.2233334}

\bibitem[\protect\citeauthoryear{{Gomes da Silva}, {Santos}, {Bonfils},
  {Delfosse}, {Forveille}  \& {Udry}}{{Gomes da Silva}
  et~al.}{2011}]{Silva2011}
{Gomes da Silva} J.,  {Santos} N.~C.,  {Bonfils} X.,  {Delfosse} X.,
  {Forveille} T.,   {Udry} S.,  2011, \mn@doi [\aap]
  {10.1051/0004-6361/201116971}, \href
  {https://ui.adsabs.harvard.edu/abs/2011A&A...534A..30G} {534, A30}

\bibitem[\protect\citeauthoryear{{Gomes da Silva} et~al.,}{{Gomes da Silva}
  et~al.}{2021}]{Gomes(2021)}
{Gomes da Silva} J.,  et~al., 2021, \mn@doi [aap]
  {10.1051/0004-6361/202039765}, \href
  {https://ui.adsabs.harvard.edu/abs/2021A&A...646A..77G} {646, A77}

\bibitem[\protect\citeauthoryear{{Gomes da Silva}, {Bensabat}, {Monteiro}  \&
  {Santos}}{{Gomes da Silva} et~al.}{2022}]{Gomes(2022)}
{Gomes da Silva} J.,  {Bensabat} A.,  {Monteiro} T.,   {Santos} N.~C.,  2022,
  \mn@doi [\aap] {10.1051/0004-6361/202244595}, \href
  {https://ui.adsabs.harvard.edu/abs/2022A&A...668A.174G} {668, A174}

\bibitem[\protect\citeauthoryear{{Gray}}{{Gray}}{2005}]{Gray(2005)}
{Gray} D.~F.,  2005, {The Observation and Analysis of Stellar Photospheres}

\bibitem[\protect\citeauthoryear{{Grisel} et~al.,}{{Grisel}
  et~al.}{2021}]{Grisel(2021)}
{Grisel} O.,  et~al., 2021, {scikit-learn/scikit-learn: scikit-learn 0.24.2},
  Zenodo, \mn@doi{10.5281/zenodo.4725836}

\bibitem[\protect\citeauthoryear{{Hall}, {Lockwood}  \& {Skiff}}{{Hall}
  et~al.}{2007}]{Hall(2007)}
{Hall} J.~C.,  {Lockwood} G.~W.,   {Skiff} B.~A.,  2007, \mn@doi [\aj]
  {10.1086/510356}, \href
  {https://ui.adsabs.harvard.edu/abs/2007AJ....133..862H} {133, 862}

\bibitem[\protect\citeauthoryear{{Hara} \& {Delisle}}{{Hara} \&
  {Delisle}}{2023}]{Hara(2023)}
{Hara} N.~C.,  {Delisle} J.-B.,  2023, \mn@doi [arXiv e-prints]
  {10.48550/arXiv.2304.08489}, \href
  {https://ui.adsabs.harvard.edu/abs/2023arXiv230408489H} {p. arXiv:2304.08489}

\bibitem[\protect\citeauthoryear{{Hara} \& {Ford}}{{Hara} \&
  {Ford}}{2023}]{Hara(2023b)}
{Hara} N.~C.,  {Ford} E.~B.,  2023, \mn@doi [Annual Review of Statistics and
  Its Application] {10.1146/annurev-statistics-033021-012225}, \href
  {https://ui.adsabs.harvard.edu/abs/2023AnRSA..10..623H} {10, 623}

\bibitem[\protect\citeauthoryear{{Hathaway}}{{Hathaway}}{2015}]{Hathaway(2015)}
{Hathaway} D.~H.,  2015, \mn@doi [Living Reviews in Solar Physics]
  {10.1007/lrsp-2015-4}, \href
  {http://adsabs.harvard.edu/abs/2015LRSP...12....4H} {12, 4}

\bibitem[\protect\citeauthoryear{{Hathaway}, {Teil}, {Norton}  \&
  {Kitiashvili}}{{Hathaway} et~al.}{2015a}]{Hathaway(2015b)}
{Hathaway} D.~H.,  {Teil} T.,  {Norton} A.~A.,   {Kitiashvili} I.,  2015a,
  \mn@doi [\apj] {10.1088/0004-637X/811/2/105}, \href
  {https://ui.adsabs.harvard.edu/abs/2015ApJ...811..105H} {811, 105}

\bibitem[\protect\citeauthoryear{{Hathaway}, {Teil}, {Norton}  \&
  {Kitiashvili}}{{Hathaway} et~al.}{2015b}]{Hathaway(2015)b}
{Hathaway} D.~H.,  {Teil} T.,  {Norton} A.~A.,   {Kitiashvili} I.,  2015b,
  \mn@doi [apj] {10.1088/0004-637X/811/2/105}, \href
  {https://ui.adsabs.harvard.edu/abs/2015ApJ...811..105H} {811, 105}

\bibitem[\protect\citeauthoryear{{Hatzes}}{{Hatzes}}{2002}]{Hatzes(2002)}
{Hatzes} A.~P.,  2002, \mn@doi [Astronomische Nachrichten]
  {10.1002/1521-3994(200208)323:3/4\textless{}392::AID-ASNA392\textgreater{}3.0.CO;2-M},
  \href {https://ui.adsabs.harvard.edu/abs/2002AN....323..392H} {323, 392}

\bibitem[\protect\citeauthoryear{{H{\'e}brard}, {Donati}, {Delfosse}, {Morin},
  {Boisse}, {Moutou}  \& {H{\'e}brard}}{{H{\'e}brard}
  et~al.}{2014}]{Hebrard(2014)}
{H{\'e}brard} {\'E}.~M.,  {Donati} J.~F.,  {Delfosse} X.,  {Morin} J.,
  {Boisse} I.,  {Moutou} C.,   {H{\'e}brard} G.,  2014, \mn@doi [mnras]
  {10.1093/mnras/stu1285}, \href
  {https://ui.adsabs.harvard.edu/abs/2014MNRAS.443.2599H} {443, 2599}

\bibitem[\protect\citeauthoryear{{Hempelmann}, {Mittag}, {Gonzalez-Perez},
  {Schmitt}, {Schr{\"o}der}  \& {Rauw}}{{Hempelmann}
  et~al.}{2016}]{Hempelmann(2016)}
{Hempelmann} A.,  {Mittag} M.,  {Gonzalez-Perez} J.~N.,  {Schmitt} J.~H.~M.~M.,
   {Schr{\"o}der} K.~P.,   {Rauw} G.,  2016, \mn@doi [\aap]
  {10.1051/0004-6361/201526972}, \href
  {https://ui.adsabs.harvard.edu/abs/2016A&A...586A..14H} {586, A14}

\bibitem[\protect\citeauthoryear{{Hinkel} et~al.,}{{Hinkel}
  et~al.}{2017}]{Hinkel(2017)}
{Hinkel} N.~R.,  et~al., 2017, \mn@doi [\apj] {10.3847/1538-4357/aa8b0f}, \href
  {https://ui.adsabs.harvard.edu/abs/2017ApJ...848...34H} {848, 34}

\bibitem[\protect\citeauthoryear{{Hyvärinen} \& {Oja}}{{Hyvärinen} \&
  {Oja}}{2000}]{Hyvarinen(2000)}
{Hyvärinen} A.,  {Oja} E.,  2000, {Independent Component Analysis: Algorithms
  and Applications. Neural Networks}

\bibitem[\protect\citeauthoryear{{Isaacson} \& {Fischer}}{{Isaacson} \&
  {Fischer}}{2010}]{Isaacson(2010)}
{Isaacson} H.,  {Fischer} D.,  2010, \mn@doi [apj]
  {10.1088/0004-637X/725/1/875}, \href
  {http://adsabs.harvard.edu/abs/2010ApJ...725..875I} {725, 875}

\bibitem[\protect\citeauthoryear{{Jay}, {Guinan}, {Morgan}, {Messina}  \&
  {Jassour}}{{Jay} et~al.}{1997}]{Jay(1997)}
{Jay} J.~E.,  {Guinan} E.~F.,  {Morgan} N.~D.,  {Messina} S.,   {Jassour} D.,
  1997, in American Astronomical Society Meeting Abstracts \#189. p.~730

\bibitem[\protect\citeauthoryear{{Jurgenson}, {Fischer}, {McCracken}, {Sawyer},
  {Szymkowiak}, {Davis}, {Muller}  \& {Santoro}}{{Jurgenson}
  et~al.}{2016}]{Jurgenson(2016)}
{Jurgenson} C.,  {Fischer} D.,  {McCracken} T.,  {Sawyer} D.,  {Szymkowiak} A.,
   {Davis} A.,  {Muller} G.,   {Santoro} F.,  2016, in {Evans} C.~J.,  {Simard}
  L.,   {Takami} H.,  eds,  Society of Photo-Optical Instrumentation Engineers
  (SPIE) Conference Series Vol. 9908, Ground-based and Airborne Instrumentation
  for Astronomy VI. p. 99086T (\mn@eprint {arXiv} {1606.04413}),
  \mn@doi{10.1117/12.2233002}

\bibitem[\protect\citeauthoryear{{Kaisig} \& {Schroeter}}{{Kaisig} \&
  {Schroeter}}{1983}]{Kaisig(1983)}
{Kaisig} M.,  {Schroeter} E.~H.,  1983, \aap, \href
  {https://ui.adsabs.harvard.edu/abs/1983A&A...117..305K} {117, 305}

\bibitem[\protect\citeauthoryear{{Keil}, {Roudier}, {Cambell}, {Koo}  \&
  {Marmolino}}{{Keil} et~al.}{1989}]{Keil(1989)}
{Keil} S.~L.,  {Roudier} T.,  {Cambell} E.,  {Koo} B.~C.,   {Marmolino} C.,
  1989, in {Rutten} R.~J.,  {Severino} G.,  eds,  NATO Advanced Study Institute
  (ASI) Series C Vol. 263, Solar and Stellar Granulation. p.~273

\bibitem[\protect\citeauthoryear{{Kervella}, {Bigot}, {Gallenne}  \&
  {Th{\'e}venin}}{{Kervella} et~al.}{2017}]{Kervella(2017)}
{Kervella} P.,  {Bigot} L.,  {Gallenne} A.,   {Th{\'e}venin} F.,  2017, \mn@doi
  [aap] {10.1051/0004-6361/201629505}, \href
  {https://ui.adsabs.harvard.edu/abs/2017A&A...597A.137K} {597, A137}

\bibitem[\protect\citeauthoryear{{Kesseli}, {West}, {Veyette}, {Harrison},
  {Feldman}  \& {Bochanski}}{{Kesseli} et~al.}{2017}]{Kesseli(2017)}
{Kesseli} A.~Y.,  {West} A.~A.,  {Veyette} M.,  {Harrison} B.,  {Feldman} D.,
  {Bochanski} J.~J.,  2017, \mn@doi [\apjs] {10.3847/1538-4365/aa656d}, \href
  {https://ui.adsabs.harvard.edu/abs/2017ApJS..230...16K} {230, 16}

\bibitem[\protect\citeauthoryear{{Klein} et~al.,}{{Klein}
  et~al.}{2024}]{Klein(2024)}
{Klein} B.,  et~al., 2024, \mn@doi [\mnras] {10.1093/mnras/stae1313}, \href
  {https://ui.adsabs.harvard.edu/abs/2024MNRAS.531.4238K} {531, 4238}

\bibitem[\protect\citeauthoryear{{Kuridze} et~al.,}{{Kuridze}
  et~al.}{2024}]{Kuridze(2024)}
{Kuridze} D.,  et~al., 2024, \mn@doi [arXiv e-prints]
  {10.48550/arXiv.2402.04545}, \href
  {https://ui.adsabs.harvard.edu/abs/2024arXiv240204545K} {p. arXiv:2402.04545}

\bibitem[\protect\citeauthoryear{{Labonte}}{{Labonte}}{1986}]{Labonte(1986)}
{Labonte} B.~J.,  1986, \mn@doi [\apjs] {10.1086/191137}, \href
  {https://ui.adsabs.harvard.edu/abs/1986ApJS...62..229L} {62, 229}

\bibitem[\protect\citeauthoryear{{Lafarga} et~al.,}{{Lafarga}
  et~al.}{2020}]{Lafarga(2020)}
{Lafarga} M.,  et~al., 2020, \mn@doi [aap] {10.1051/0004-6361/201937222}, \href
  {https://ui.adsabs.harvard.edu/abs/2020A&A...636A..36L} {636, A36}

\bibitem[\protect\citeauthoryear{{Landstreet}, {Kupka}, {Ford}, {Officer},
  {Sigut}, {Silaj}, {Strasser}  \& {Townshend}}{{Landstreet}
  et~al.}{2009}]{Landstreet(2009)}
{Landstreet} J.~D.,  {Kupka} F.,  {Ford} H.~A.,  {Officer} T.,  {Sigut}
  T.~A.~A.,  {Silaj} J.,  {Strasser} S.,   {Townshend} A.,  2009, \mn@doi
  [\aap] {10.1051/0004-6361/200912083}, \href
  {https://ui.adsabs.harvard.edu/abs/2009A&A...503..973L} {503, 973}

\bibitem[\protect\citeauthoryear{{Liebing}, {Jeffers}, {Reiners}  \&
  {Zechmeister}}{{Liebing} et~al.}{2021}]{Liebing(2021)}
{Liebing} F.,  {Jeffers} S.~V.,  {Reiners} A.,   {Zechmeister} M.,  2021,
  \mn@doi [aap] {10.1051/0004-6361/202039607}, \href
  {https://ui.adsabs.harvard.edu/abs/2021A&A...654A.168L} {654, A168}

\bibitem[\protect\citeauthoryear{{L{\"o}hner-B{\"o}ttcher}, {Schmidt},
  {Schlichenmaier}, {Doerr}, {Steinmetz}  \&
  {Holzwarth}}{{L{\"o}hner-B{\"o}ttcher} et~al.}{2018}]{Lohner(2018b)}
{L{\"o}hner-B{\"o}ttcher} J.,  {Schmidt} W.,  {Schlichenmaier} R.,  {Doerr}
  H.~P.,  {Steinmetz} T.,   {Holzwarth} R.,  2018, \mn@doi [\aap]
  {10.1051/0004-6361/201832886}, \href
  {https://ui.adsabs.harvard.edu/abs/2018A&A...617A..19L} {617, A19}

\bibitem[\protect\citeauthoryear{{Lovis} et~al.,}{{Lovis}
  et~al.}{2011}]{Lovis(2011)}
{Lovis} C.,  et~al., 2011, preprint, \href
  {http://adsabs.harvard.edu/abs/2011arXiv1107.5325L} {} (\mn@eprint {arXiv}
  {1107.5325})

\bibitem[\protect\citeauthoryear{{Malavolta}, {Lovis}, {Pepe}, {Sneden}  \&
  {Udry}}{{Malavolta} et~al.}{2017}]{Malavolta(2017)}
{Malavolta} L.,  {Lovis} C.,  {Pepe} F.,  {Sneden} C.,   {Udry} S.,  2017,
  \mn@doi [mnras] {10.1093/mnras/stx1100}, \href
  {https://ui.adsabs.harvard.edu/abs/2017MNRAS.469.3965M} {469, 3965}

\bibitem[\protect\citeauthoryear{{Mart{\'\i}nez Gonz{\'a}lez}, {Asensio Ramos},
  {Carroll}, {Kopf}, {Ram{\'\i}rez V{\'e}lez}  \& {Semel}}{{Mart{\'\i}nez
  Gonz{\'a}lez} et~al.}{2008}]{Martinez2008}
{Mart{\'\i}nez Gonz{\'a}lez} M.~J.,  {Asensio Ramos} A.,  {Carroll} T.~A.,
  {Kopf} M.,  {Ram{\'\i}rez V{\'e}lez} J.~C.,   {Semel} M.,  2008, \mn@doi
  [\aap] {10.1051/0004-6361:200809719}, \href
  {https://ui.adsabs.harvard.edu/abs/2008A&A...486..637M} {486, 637}

\bibitem[\protect\citeauthoryear{{Mayor} et~al.,}{{Mayor}
  et~al.}{2003}]{Mayor(2003)}
{Mayor} M.,  et~al., 2003, The Messenger, \href
  {http://adsabs.harvard.edu/abs/2003Msngr.114...20M} {114, 20}

\bibitem[\protect\citeauthoryear{{Meunier} \& {Delfosse}}{{Meunier} \&
  {Delfosse}}{2009}]{Meunier(2009)}
{Meunier} N.,  {Delfosse} X.,  2009, \mn@doi [aap]
  {10.1051/0004-6361/200911823}, \href
  {https://ui.adsabs.harvard.edu/abs/2009A&A...501.1103M} {501, 1103}

\bibitem[\protect\citeauthoryear{{Meunier} \& {Lagrange}}{{Meunier} \&
  {Lagrange}}{2013}]{Meunier(2013)}
{Meunier} N.,  {Lagrange} A.-M.,  2013, \mn@doi [aap]
  {10.1051/0004-6361/201219917}, \href
  {http://adsabs.harvard.edu/abs/2013A%26A...551A.101M} {551, A101}

\bibitem[\protect\citeauthoryear{{Meunier} \& {Lagrange}}{{Meunier} \&
  {Lagrange}}{2020}]{Meunier(2020)}
{Meunier} N.,  {Lagrange} A.~M.,  2020, \mn@doi [aap]
  {10.1051/0004-6361/201937354}, \href
  {https://ui.adsabs.harvard.edu/abs/2020A&A...638A..54M} {638, A54}

\bibitem[\protect\citeauthoryear{{Meunier}, {Desort}  \& {Lagrange}}{{Meunier}
  et~al.}{2010}]{Meunier(2010)}
{Meunier} N.,  {Desort} M.,   {Lagrange} A.-M.,  2010, \mn@doi [aap]
  {10.1051/0004-6361/200913551}, \href
  {http://adsabs.harvard.edu/abs/2010A%26A...512A..39M} {512, A39}

\bibitem[\protect\citeauthoryear{{Meunier}, {Lagrange}  \& {Cuzacq}}{{Meunier}
  et~al.}{2019}]{Meunier(2019e)}
{Meunier} N.,  {Lagrange} A.~M.,   {Cuzacq} S.,  2019, \mn@doi [\aap]
  {10.1051/0004-6361/201935348}, \href
  {https://ui.adsabs.harvard.edu/abs/2019A&A...632A..81M} {632, A81}

\bibitem[\protect\citeauthoryear{{Meunier}, {Kretzschmar}, {Gravet}, {Mignon}
  \& {Delfosse}}{{Meunier} et~al.}{2022}]{Meunier(2022)}
{Meunier} N.,  {Kretzschmar} M.,  {Gravet} R.,  {Mignon} L.,   {Delfosse} X.,
  2022, \mn@doi [aap] {10.1051/0004-6361/202142120}, \href
  {https://ui.adsabs.harvard.edu/abs/2022A&A...658A..57M} {658, A57}

\bibitem[\protect\citeauthoryear{{Meunier}, {Lagrange}, {Dumusque}  \&
  {Sulis}}{{Meunier} et~al.}{2024}]{Meunier(2024)}
{Meunier} N.,  {Lagrange} A.~M.,  {Dumusque} X.,   {Sulis} S.,  2024, \mn@doi
  [\aap] {10.1051/0004-6361/202449146}, \href
  {https://ui.adsabs.harvard.edu/abs/2024A&A...687A.303M} {687, A303}

\bibitem[\protect\citeauthoryear{{Milbourne} et~al.,}{{Milbourne}
  et~al.}{2021}]{Milbourne(2021)}
{Milbourne} T.~W.,  et~al., 2021, \mn@doi [apj] {10.3847/1538-4357/ac1266},
  \href {https://ui.adsabs.harvard.edu/abs/2021ApJ...920...21M} {920, 21}

\bibitem[\protect\citeauthoryear{{Morris}, {Hebb}, {Davenport}, {Rohn}  \&
  {Hawley}}{{Morris} et~al.}{2017}]{Morris(2017)}
{Morris} B.~M.,  {Hebb} L.,  {Davenport} J. R.~A.,  {Rohn} G.,   {Hawley}
  S.~L.,  2017, \mn@doi [apj] {10.3847/1538-4357/aa8555}, \href
  {https://ui.adsabs.harvard.edu/abs/2017ApJ...846...99M} {846, 99}

\bibitem[\protect\citeauthoryear{{Neckel} \& {Labs}}{{Neckel} \&
  {Labs}}{1984}]{Neckel(1984)}
{Neckel} H.,  {Labs} D.,  1984, \mn@doi [\solphys] {10.1007/BF00173953}, \href
  {https://ui.adsabs.harvard.edu/abs/1984SoPh...90..205N} {90, 205}

\bibitem[\protect\citeauthoryear{{Ning}, {Wise}, {Cisewski-Kehe},
  {Dodson-Robinson}  \& {Fischer}}{{Ning} et~al.}{2019}]{Ning(2019)}
{Ning} B.,  {Wise} A.,  {Cisewski-Kehe} J.,  {Dodson-Robinson} S.,   {Fischer}
  D.,  2019, \mn@doi [aj] {10.3847/1538-3881/ab441c}, \href
  {https://ui.adsabs.harvard.edu/abs/2019AJ....158..210N} {158, 210}

\bibitem[\protect\citeauthoryear{{Oranje}}{{Oranje}}{1983}]{Oranje(1983b)}
{Oranje} B.~J.,  1983, aap, \href
  {https://ui.adsabs.harvard.edu/abs/1983A&A...124...43O} {124, 43}

\bibitem[\protect\citeauthoryear{{Palle} et~al.,}{{Palle}
  et~al.}{2023}]{Palle(2023)}
{Palle} E.,  et~al., 2023, \mn@doi [arXiv e-prints]
  {10.48550/arXiv.2311.17075}, \href
  {https://ui.adsabs.harvard.edu/abs/2023arXiv231117075P} {p. arXiv:2311.17075}

\bibitem[\protect\citeauthoryear{{Palumbo}, {Saar}  \& {Haywood}}{{Palumbo}
  et~al.}{2024}]{Palumbo(2024)}
{Palumbo} Michael~L. I.,  {Saar} S.~H.,   {Haywood} R.~D.,  2024, \mn@doi
  [arXiv e-prints] {10.48550/arXiv.2404.16747}, \href
  {https://ui.adsabs.harvard.edu/abs/2024arXiv240416747P} {p. arXiv:2404.16747}

\bibitem[\protect\citeauthoryear{{Pedregosa} et~al.,}{{Pedregosa}
  et~al.}{2011}]{Pedregosa(2011)}
{Pedregosa} F.,  et~al., 2011, \mn@doi [Journal of Machine Learning Research]
  {10.48550/arXiv.1201.0490}, \href
  {https://ui.adsabs.harvard.edu/abs/2011JMLR...12.2825P} {12, 2825}

\bibitem[\protect\citeauthoryear{{Pepe}, {Ehrenreich}  \& {Meyer}}{{Pepe}
  et~al.}{2014}]{Pepe(2014b)}
{Pepe} F.,  {Ehrenreich} D.,   {Meyer} M.~R.,  2014, \mn@doi [nat]
  {10.1038/nature13784}, \href
  {https://ui.adsabs.harvard.edu/abs/2014Natur.513..358P} {513, 358}

\bibitem[\protect\citeauthoryear{{Pietrow}, {Kiselman}, {Andriienko}, {Petit
  dit de la Roche}, {D{\'\i}az Baso}  \& {Calvo}}{{Pietrow}
  et~al.}{2023}]{Pietrow(2023)}
{Pietrow} A.~G.~M.,  {Kiselman} D.,  {Andriienko} O.,  {Petit dit de la Roche}
  D.~J.~M.,  {D{\'\i}az Baso} C.~J.,   {Calvo} F.,  2023, \mn@doi [\aap]
  {10.1051/0004-6361/202244811}, \href
  {https://ui.adsabs.harvard.edu/abs/2023A&A...671A.130P} {671, A130}

\bibitem[\protect\citeauthoryear{{Pietrow} et~al.,}{{Pietrow}
  et~al.}{2024}]{Pietrow(2024)}
{Pietrow} A.~G.~M.,  et~al., 2024, \mn@doi [\aap]
  {10.1051/0004-6361/202347895}, \href
  {https://ui.adsabs.harvard.edu/abs/2024A&A...682A..46P} {682, A46}

\bibitem[\protect\citeauthoryear{{Pourbaix} \& {Boffin}}{{Pourbaix} \&
  {Boffin}}{2016}]{Pourbaix(2016)}
{Pourbaix} D.,  {Boffin} H. M.~J.,  2016, \mn@doi [\aap]
  {10.1051/0004-6361/201527859}, \href
  {https://ui.adsabs.harvard.edu/abs/2016A&A...586A..90P} {586, A90}

\bibitem[\protect\citeauthoryear{{Przybylski}, {Shelyag}  \&
  {Cally}}{{Przybylski} et~al.}{2015}]{Przybylski(2015)}
{Przybylski} D.,  {Shelyag} S.,   {Cally} P.~S.,  2015, \mn@doi [\apj]
  {10.1088/0004-637X/807/1/20}, \href
  {https://ui.adsabs.harvard.edu/abs/2015ApJ...807...20P} {807, 20}

\bibitem[\protect\citeauthoryear{{Reiners}, {Mrotzek}, {Lemke}, {Hinrichs}  \&
  {Reinsch}}{{Reiners} et~al.}{2016}]{Reiners(2016)}
{Reiners} A.,  {Mrotzek} N.,  {Lemke} U.,  {Hinrichs} J.,   {Reinsch} K.,
  2016, \mn@doi [aap] {10.1051/0004-6361/201527530}, \href
  {http://adsabs.harvard.edu/abs/2016A%26A...587A..65R} {587, A65}

\bibitem[\protect\citeauthoryear{{Rezaei}, {Schlichenmaier}, {Beck}, {Bruls}
  \& {Schmidt}}{{Rezaei} et~al.}{2007}]{Rezaei(2007)}
{Rezaei} R.,  {Schlichenmaier} R.,  {Beck} C.~A.~R.,  {Bruls} J.~H.~M.~J.,
  {Schmidt} W.,  2007, \mn@doi [\aap] {10.1051/0004-6361:20067017}, \href
  {https://ui.adsabs.harvard.edu/abs/2007A&A...466.1131R} {466, 1131}

\bibitem[\protect\citeauthoryear{{Robertson}, {Roy}  \&
  {Mahadevan}}{{Robertson} et~al.}{2015}]{Robertson(2015)}
{Robertson} P.,  {Roy} A.,   {Mahadevan} S.,  2015, \mn@doi [apjl]
  {10.1088/2041-8205/805/2/L22}, \href
  {https://ui.adsabs.harvard.edu/abs/2015ApJ...805L..22R} {805, L22}

\bibitem[\protect\citeauthoryear{{Routray,}, {Ray}  \& {Mishra}}{{Routray,}
  et~al.}{2019}]{Routray2019}
{Routray,} S.,  {Ray} A.,   {Mishra} C.,  2019, \mn@doi [Signal, Image and
  Video Processing] {10.1007/s11760-019-01489-2}, \href
  {https://ui.adsabs.harvard.edu/abs/2019A&A...623A..74D} {13, 1405}

\bibitem[\protect\citeauthoryear{{Saar} \& {Donahue}}{{Saar} \&
  {Donahue}}{1997}]{Saar(1997)}
{Saar} S.~H.,  {Donahue} R.~A.,  1997, \mn@doi [apj] {10.1086/304392}, \href
  {https://ui.adsabs.harvard.edu/abs/1997ApJ...485..319S} {485, 319}

\bibitem[\protect\citeauthoryear{Satopaa, Albrecht, Irwin  \& Raghavan}{Satopaa
  et~al.}{2011}]{Satopaa(2011)}
Satopaa V.,  Albrecht J.,  Irwin D.,   Raghavan B.,  2011, in 2011 31st
  International Conference on Distributed Computing Systems Workshops. pp
  166--171, \mn@doi{10.1109/ICDCSW.2011.20}

\bibitem[\protect\citeauthoryear{{Schwab} et~al.,}{{Schwab}
  et~al.}{2018}]{Schwab(2018)}
{Schwab} C.,  et~al., 2018, in {Evans} C.~J.,  {Simard} L.,   {Takami} H.,
  eds,  Society of Photo-Optical Instrumentation Engineers (SPIE) Conference
  Series Vol. 10702, Ground-based and Airborne Instrumentation for Astronomy
  VII. p. 1070271, \mn@doi{10.1117/12.2314420}

\bibitem[\protect\citeauthoryear{{Shahaf} \& {Zackay}}{{Shahaf} \&
  {Zackay}}{2023}]{Sharaf(2023)}
{Shahaf} S.,  {Zackay} B.,  2023, \mn@doi [\mnras] {10.1093/mnras/stad2742},
  \href {https://ui.adsabs.harvard.edu/abs/2023MNRAS.525.6223S} {525, 6223}

\bibitem[\protect\citeauthoryear{{Shapiro}, {Solanki}, {Krivova}, {Schmutz},
  {Ball}, {Knaack}, {Rozanov}  \& {Unruh}}{{Shapiro}
  et~al.}{2014}]{Shapiro(2014)}
{Shapiro} A.~I.,  {Solanki} S.~K.,  {Krivova} N.~A.,  {Schmutz} W.~K.,  {Ball}
  W.~T.,  {Knaack} R.,  {Rozanov} E.~V.,   {Unruh} Y.~C.,  2014, \mn@doi [aap]
  {10.1051/0004-6361/201323086}, \href
  {http://adsabs.harvard.edu/abs/2014A%26A...569A..38S} {569, A38}

\bibitem[\protect\citeauthoryear{{Shapiro}, {Solanki}, {Krivova}, {Yeo}  \&
  {Schmutz}}{{Shapiro} et~al.}{2016}]{Shapiro(2016)}
{Shapiro} A.~I.,  {Solanki} S.~K.,  {Krivova} N.~A.,  {Yeo} K.~L.,   {Schmutz}
  W.~K.,  2016, \mn@doi [\aap] {10.1051/0004-6361/201527527}, \href
  {https://ui.adsabs.harvard.edu/abs/2016A&A...589A..46S} {589, A46}

\bibitem[\protect\citeauthoryear{{Sharma}, {Kembhavi}, {Kembhavi}, {Sivarani},
  {Abraham}  \& {Vaghmare}}{{Sharma} et~al.}{2020}]{Sharma(2020)}
{Sharma} K.,  {Kembhavi} A.,  {Kembhavi} A.,  {Sivarani} T.,  {Abraham} S.,
  {Vaghmare} K.,  2020, \mn@doi [\mnras] {10.1093/mnras/stz3100}, \href
  {https://ui.adsabs.harvard.edu/abs/2020MNRAS.491.2280S} {491, 2280}

\bibitem[\protect\citeauthoryear{{Simola}, {Dumusque}  \&
  {Cisewski-Kehe}}{{Simola} et~al.}{2019}]{Simola(2019)}
{Simola} U.,  {Dumusque} X.,   {Cisewski-Kehe} J.,  2019, \mn@doi [aap]
  {10.1051/0004-6361/201833895}, \href
  {https://ui.adsabs.harvard.edu/abs/2019A%26A...622A.131S} {622, A131}

\bibitem[\protect\citeauthoryear{{Sindhuja} \& {Singh}}{{Sindhuja} \&
  {Singh}}{2015}]{Sindhuja(2015)}
{Sindhuja} G.,  {Singh} J.,  2015, \mn@doi [Journal of Astrophysics and
  Astronomy] {10.1007/s12036-015-9330-4}, \href
  {https://ui.adsabs.harvard.edu/abs/2015JApA...36...81S} {36, 81}

\bibitem[\protect\citeauthoryear{{Smalley}}{{Smalley}}{2005}]{Smalley(2005)}
{Smalley} B.,  2005, \mn@doi [Memorie della Societa Astronomica Italiana
  Supplementi] {10.48550/arXiv.astro-ph/0509535}, \href
  {https://ui.adsabs.harvard.edu/abs/2005MSAIS...8..130S} {8, 130}

\bibitem[\protect\citeauthoryear{{Solanki}}{{Solanki}}{2003}]{Solanki(2003)}
{Solanki} S.~K.,  2003, \mn@doi [\aapr] {10.1007/s00159-003-0018-4}, \href
  {https://ui.adsabs.harvard.edu/abs/2003A&ARv..11..153S} {11, 153}

\bibitem[\protect\citeauthoryear{{Soubiran}, {Brouillet}  \&
  {Casamiquela}}{{Soubiran} et~al.}{2022}]{Soubiran(2022)}
{Soubiran} C.,  {Brouillet} N.,   {Casamiquela} L.,  2022, \mn@doi [\aap]
  {10.1051/0004-6361/202142409}, \href
  {https://ui.adsabs.harvard.edu/abs/2022A&A...663A...4S} {663, A4}

\bibitem[\protect\citeauthoryear{{Takeda}}{{Takeda}}{2007}]{Takeda(2007)}
{Takeda} Y.,  2007, \mn@doi [\pasj] {10.1093/pasj/59.2.335}, \href
  {https://ui.adsabs.harvard.edu/abs/2007PASJ...59..335T} {59, 335}

\bibitem[\protect\citeauthoryear{{Thompson}, {Watson}, {de Mooij}  \&
  {Jess}}{{Thompson} et~al.}{2017}]{Thompson(2017)}
{Thompson} A.~P.~G.,  {Watson} C.~A.,  {de Mooij} E.~J.~W.,   {Jess} D.~B.,
  2017, \mn@doi [mnras] {10.1093/mnrasl/slx018}, \href
  {http://adsabs.harvard.edu/abs/2017MNRAS.468L..16T} {468, L16}

\bibitem[\protect\citeauthoryear{{Title}, {Tarbell}, {Topka}, {Ferguson},
  {Shine}  \& {SOUP Team}}{{Title} et~al.}{1989}]{Title(1989)}
{Title} A.~M.,  {Tarbell} T.~D.,  {Topka} K.~P.,  {Ferguson} S.~H.,  {Shine}
  R.~A.,   {SOUP Team} 1989, \mn@doi [apj] {10.1086/167026}, \href
  {https://ui.adsabs.harvard.edu/abs/1989ApJ...336..475T} {336, 475}

\bibitem[\protect\citeauthoryear{{Tlatov}}{{Tlatov}}{2024}]{Tlatov(2024)}
{Tlatov} A.~G.,  2024, \mn@doi [Geomagnetism and Aeronomy]
  {10.1134/S0016793223080236}, \href
  {https://ui.adsabs.harvard.edu/abs/2024Ge&Ae..63.1116T} {63, 1116}

\bibitem[\protect\citeauthoryear{{Valenti} \& {Fischer}}{{Valenti} \&
  {Fischer}}{2005}]{Valenti(2005)}
{Valenti} J.~A.,  {Fischer} D.~A.,  2005, \mn@doi [\apjs] {10.1086/430500},
  \href {https://ui.adsabs.harvard.edu/abs/2005ApJS..159..141V} {159, 141}

\bibitem[\protect\citeauthoryear{{Vaughan} \& {Preston}}{{Vaughan} \&
  {Preston}}{1980}]{Vaughan(1980)}
{Vaughan} A.~H.,  {Preston} G.~W.,  1980, \mn@doi [pasp] {10.1086/130683},
  \href {https://ui.adsabs.harvard.edu/abs/1980PASP...92..385V} {92, 385}

\bibitem[\protect\citeauthoryear{{Vernazza}, {Avrett}  \& {Loeser}}{{Vernazza}
  et~al.}{1981}]{Vernazza(1981)}
{Vernazza} J.~E.,  {Avrett} E.~H.,   {Loeser} R.,  1981, \mn@doi [apjs]
  {10.1086/190731}, \href
  {https://ui.adsabs.harvard.edu/abs/1981ApJS...45..635V} {45, 635}

\bibitem[\protect\citeauthoryear{{Veyette}, {Muirhead}, {Mann}, {Brewer},
  {Allard}  \& {Homeier}}{{Veyette} et~al.}{2017}]{Veyette(2017)}
{Veyette} M.~J.,  {Muirhead} P.~S.,  {Mann} A.~W.,  {Brewer} J.~M.,  {Allard}
  F.,   {Homeier} D.,  2017, \mn@doi [\apj] {10.3847/1538-4357/aa96aa}, \href
  {https://ui.adsabs.harvard.edu/abs/2017ApJ...851...26V} {851, 26}

\bibitem[\protect\citeauthoryear{{Wehrhahn}, {Piskunov}  \&
  {Ryabchikova}}{{Wehrhahn} et~al.}{2023}]{Wehrhahn(2023)}
{Wehrhahn} A.,  {Piskunov} N.,   {Ryabchikova} T.,  2023, \mn@doi [\aap]
  {10.1051/0004-6361/202244482}, \href
  {https://ui.adsabs.harvard.edu/abs/2023A&A...671A.171W} {671, A171}

\bibitem[\protect\citeauthoryear{{West} \& {Hawley}}{{West} \&
  {Hawley}}{2008}]{West2008}
{West} A.~A.,  {Hawley} S.~L.,  2008, \mn@doi [\pasp] {10.1086/593024}, \href
  {https://ui.adsabs.harvard.edu/abs/2008PASP..120.1161W} {120, 1161}

\bibitem[\protect\citeauthoryear{{Wilson}}{{Wilson}}{1978}]{Wilson(1978)}
{Wilson} O.~C.,  1978, \mn@doi [apj] {10.1086/156618}, \href
  {https://ui.adsabs.harvard.edu/abs/1978ApJ...226..379W} {226, 379}

\bibitem[\protect\citeauthoryear{{Wise}, {Dodson-Robinson}, {Bevenour}  \&
  {Provini}}{{Wise} et~al.}{2018}]{Wise(2018)}
{Wise} A.~W.,  {Dodson-Robinson} S.~E.,  {Bevenour} K.,   {Provini} A.,  2018,
  \mn@doi [aj] {10.3847/1538-3881/aadd94}, \href
  {http://adsabs.harvard.edu/abs/2018AJ....156..180W} {156, 180}

\bibitem[\protect\citeauthoryear{{Wise}, {Plavchan}, {Dumusque}, {Cegla}  \&
  {Wright}}{{Wise} et~al.}{2022}]{Wise(2022)}
{Wise} A.,  {Plavchan} P.,  {Dumusque} X.,  {Cegla} H.,   {Wright} D.,  2022,
  arXiv e-prints, \href {https://ui.adsabs.harvard.edu/abs/2022arXiv220315161W}
  {p. arXiv:2203.15161}

\bibitem[\protect\citeauthoryear{{Wright}, {Marcy}, {Butler}  \&
  {Vogt}}{{Wright} et~al.}{2004}]{Wright(2004)}
{Wright} J.~T.,  {Marcy} G.~W.,  {Butler} R.~P.,   {Vogt} S.~S.,  2004, \mn@doi
  [apjs] {10.1086/386283}, \href
  {http://adsabs.harvard.edu/abs/2004ApJS..152..261W} {152, 261}

\bibitem[\protect\citeauthoryear{{Zechmeister} \& {K{\"u}rster}}{{Zechmeister}
  \& {K{\"u}rster}}{2009}]{Zechmeister(2009)}
{Zechmeister} M.,  {K{\"u}rster} M.,  2009, \mn@doi [aap]
  {10.1051/0004-6361:200811296}, \href
  {https://ui.adsabs.harvard.edu/abs/2009A&A...496..577Z} {496, 577}

\bibitem[\protect\citeauthoryear{{Zhao}, {Hogg}, {Bedell}  \& {Fischer}}{{Zhao}
  et~al.}{2021}]{Zhao(2021)}
{Zhao} L.~L.,  {Hogg} D.~W.,  {Bedell} M.,   {Fischer} D.~A.,  2021, \mn@doi
  [aj] {10.3847/1538-3881/abd105}, \href
  {https://ui.adsabs.harvard.edu/abs/2021AJ....161...80Z} {161, 80}

\makeatother
\end{thebibliography}

%\begin{thebibliography}{99}
%\end{thebibliography}

%%%%%%%%%%%%%%%%%%%%%%%%%%%%%%%%%%%%%%%%%%%%%%%%%%

%%%%%%%%%%%%%%%%% APPENDICES %%%%%%%%%%%%%%%%%%%%%

\appendix

\section{Simulation of the validity of the FF' framework.}
\label{app:nonlinear}

In Sect.~\ref{sec:method}, we proposed an extension for the ICB term in the FF' model given a more general description of the stellar surface than a single long-lived rotating active region. 
Our model (Eq.\ref{eq:mono}) led us to the conclusion that, for a purely radial convective flow model, the ICB RV signal  for a general distribution of active regions could be approximated at first order by: 

$$\Delta\text{RV}_{\text{ICB}}(t) \simeq   \Delta v_c \cdot \bar{\mu}(t) \cdot f_{\text{tot}}(t)$$

We further argued that, in this equation, the temporal evolution of the average location of active regions $\bar{\mu}$ was converging rapidly to a fix value as soon as several active regions were present simultaneously as observed for the Sun during its cycle, meaning that the RV signal was mostly linearly dependent on the total filling factor $f_{\text{tot}}$ and not quadrically. 

\begin{figure}
	
	\centering
	\includegraphics[width=9cm]{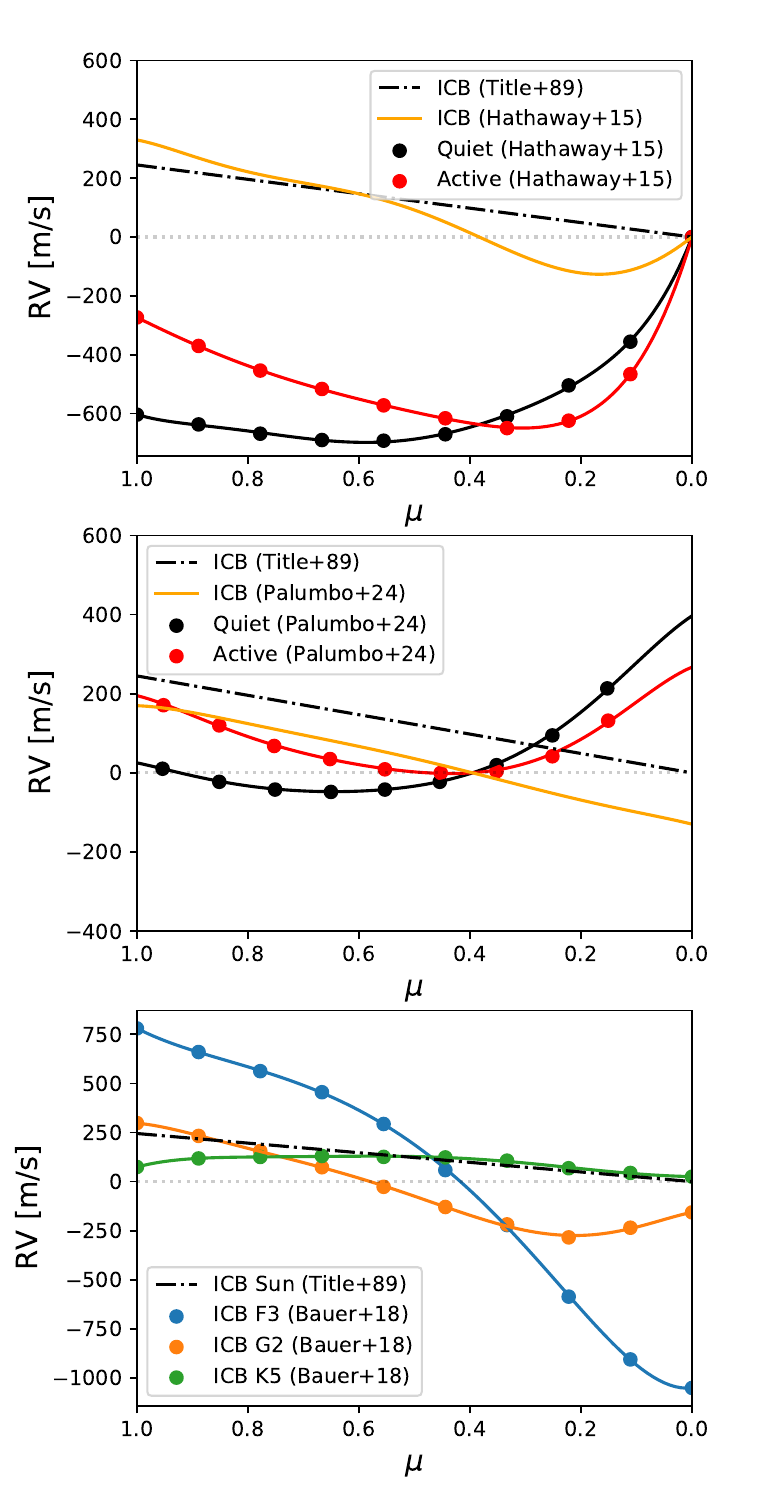}
	\caption{ICB laws $\Delta v_r(\mu)$ as provided by different sources in litterature. \textbf{Top:} ICB law from SDO measurements as provided by  \citet{Hathaway(2015b)}. Star center is at the left and limb on the right. The ICB law (orange curve) is obtained as the difference of the convective flows velocities measured in the quiet Sun (black surve) and inside AR (red curve). The purely radial flow model described in \citet{Title(1989)} is shown as a comparison (black dashed line). Note that some convective redshift inhibition (or convective blueshift enhancement) is visible close to the limb ($\mu<0.4$). \textbf{Middle:} Same as top for the recent analysis of \citet{Palumbo(2024)}. \textbf{Bottom:} ICB laws $\Delta v_r(\mu)$ for three different spectral types (F3, G2 and K5) as provided by MHD simulations of \citet{Bauer(2018)}. Again the purely radial flow model is plotted as a comparison (black dashed line). Convective flows amplitudes are strongly reduced along the spectral type sequence.}
	\label{FigFlow1}
\end{figure}

We propose in this section to prove this aspect and compare our toy model approximation with numerical disk-integration of different ICB laws $\Delta v_r(\mu)$ provided in the literature. If a stellar disk with a total surface $S_{\text{tot}}$ is covered by ARs and an ICB law is known, it is indeed also possible to use the numerical integration to directly obtain the RV counterpart:

\begin{equation}
\label{eq:exact}
\Delta \text{RV}_{\text{ICB}}(t) = \sum_{i=\{AR\}} \frac{\Delta v_r(\mu_i)}{S_{\text{tot}}}
\end{equation}

where such equation is theoretically valid for a single wavelength element of a spectrum a priori, but integrating over the wavelength would lead to a similar relation since no chromatic ICB convective laws can be found in the literature. 
We implemented the ICB laws of \citet{Hathaway(2015b)} and \citet{Palumbo(2024)} for the Sun, obtained from a single iron line in the red part of the spectrum as measured by SDO, and ICB laws given by \citet{Bauer(2018)} for three different spectral types and extacted from MHD simulations. The ICB laws are represented in Fig.\ref{FigFlow1}. We note that despite the fact that the authors use similar data as \citet{Hathaway(2015b)}, \citet{Palumbo(2024)} did not discuss or compare their result with them and so the difference in the observed offset is not clear. However, only the curvature and not the offset is relevant for the exercice.

The deviation from a linear law for a solar type-star is mainly arising for $\mu<0.30$ while all the models (including \citet{Title(1989)}) agree for a ICB at star center of $\sim250$ m/s. In contrary to the purely blueshifted radial flow model, both works have a non-negligible convective redshift flows component towards the limb ($\mu<0.4$) likely due to the horizontal flows contribution. 

%\begin{figure}
	
%	\centering
%	\includegraphics[width=9cm]{Images/CB_Flow_Hathaway.pdf}
%	\caption{ICB law $\Delta v_r(\mu)$ as provided in \citet{Hathaway(2015)b} from SDO measurements. Star center is at the left and limb on the right. The ICB law (orange curve) is obtained as the difference of the convective flows velocities measured in the quiet Sun (black surve) and inside AR (red curve). The purely radial flow model described in \citet{Title(1989)} is shown as a comparison (black dashed line). Note that some convective redshift inhibition (or convective blueshift enhancement) is visible close to the limb ($\mu<0.4$).}
%	\label{FigFlow1}
%\end{figure}

%\begin{figure}
	
%	\centering
%	\includegraphics[width=9cm]{Images/CB_Flow_Bauer.pdf}
%	\caption{ICB laws $\Delta v_r(\mu)$ for three different spectral types (F3, G2 and K5) as provided by MHD simulations of \citet{Bauer(2018)}. Again the purely radial flow model is plotted as a comparison (black dashed line). Convective flows amplitudes are strongly reduced along the spectral type sequence.}
%	\label{FigFlow2}
%\end{figure}

We now describe the simulation of the stellar surface ARs location and evolution. For this exercise, we do not need to precisely reconstruct the solar surface statistical behaviour since our argument is based on a purely geometrical consideration independent on the precise statistical properties of the active regions. As a consequence, we only aimed to reproduce the solar cycle qualitatively.  

\begin{figure*}
	
	\centering
	\includegraphics[width=18cm]{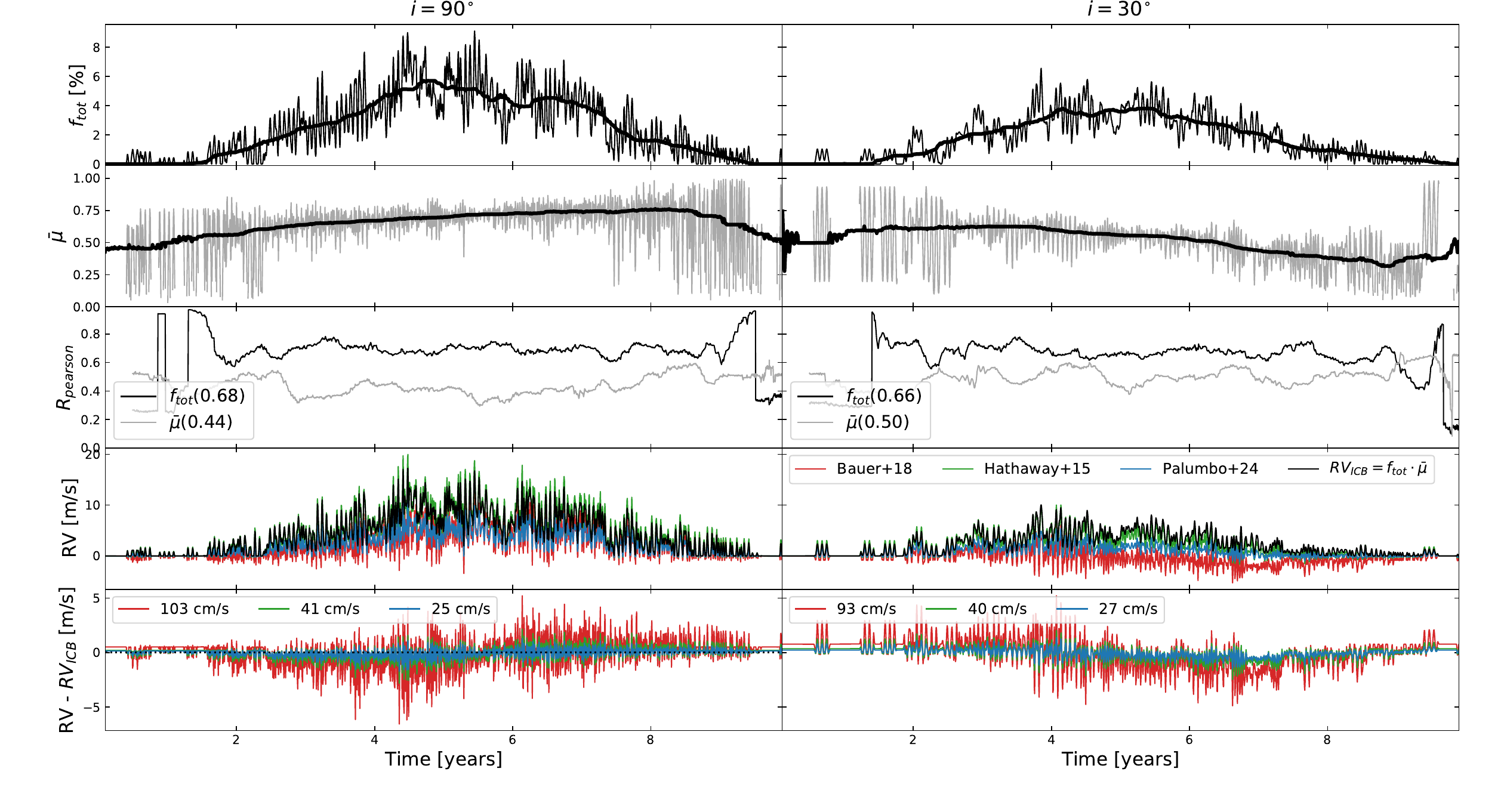}
	\caption{Simulations demonstrating the approximation validity of the toy model used for the ICB in this work. We simulated a stellar surface over a magnetic cycle (see plain text for details on the simulation). Both equator-on inclination (left column) and $i=30^{\circ}$ (right column) were studied for the same configuration of active regions. \textbf{First row:} Total filling factor time-serie $f_{\text{tot}}(t)$. The moving average over a year is over-plotted with a thick black curve. \textbf{Second row:} Average $\mu$ location $\bar{\mu}(t)$ of the active regions at the stellar surface. \textbf{Third row:} Pearson correlation coefficient $R$ between the RV signal extracted with Eq.\ref{eq:mono} and the individual time-series $f_{\text{tot}}(t)$ and $\bar{\mu}(t)$ computed into a moving window of a year. The RV signal correlates with a coefficient correlation of $R=0.68$ with the total filling factor and slightly less with the average location of the active regions $\bar{\mu}(t)$. \textbf{Fourth row:} RV  obtained from cell grid integration (Eq.\ref{eq:exact}) using different ICB laws are compared with our toy model single-point approximation describing the ICB term in the generalized FF' framework. \textbf{Fifth row:} RV residual between the approximation and the exact cell integration. RV dispersion as measured by the MAD is displayed in the label. Naturally, when using a purely radial flows model (linear ICB law) the toy model approximation exactly match the disk-integrated value. For any non-linear ICB law, the FF' framework will deviate from the cell disk-integrated value. As an example, the approximation hold down to 41 cm/s compared to the ICB law provided by \citet{Hathaway(2015)b}.}
	\label{FigSimu}
 
\end{figure*}

We simulated a solar surface over a baseline of 10 years with a time sampling of 1 day by randomly creating evolving active regions. A rectangular grid of 512 $\times$ 512 pixels was used, for a solar radii of 256 pixels, creating a total surface of $S_{\text{tot}} = \pi \cdot (256)^2 \simeq 205887$ pixels. A solid-body rotation was used with a rotational period of 25 days. The maximum latitude for the active regions were limited between $\pm 40$ and we included the progressive migration of the active latitudes along the cycle to recreate the solar Butterfly diagram, but not the active longitudes that are not relevant for the geometrical consideration. The lifetime of the active regions was randomly draw between two values of 20 and 70 days to represent short versus long-lived active regions respectively where the functional law was an exponential decaying law. The emergence was fixed for all the active regions with a 5-day linear law. The size of the active regions was randomly draw within three filling factor values of 0.5, 1 and $3\%$. A sinusoid probability density function was used for the occurrence rate such that a magnetic cycle behaviour was reproduced. A total number of 500 active regions were generated according to the previous description. This number was selected such that, for our lifetime parameters and filling factor values, the total filling factor $f_{\text{tot}}$ along the magnetic cycle was roughly around $5\%$ in order to represent a moderate solar-like cycle. 

We show in the Fig.\ref{FigSimu} one random realisation given by our simulation setup. Note that the choice of the realisation was irrelevant for the qualitative consideration raised below and we checked that other realisations were leading to the same conclusions. We also used the same active regions configurations to recreate the stellar surface for a different inclination angle ($i=30^{\circ}$).

Once the stellar surface time-serie created, we extracted the time-series of interest, namely the total filling factor $f_{\text{tot}}(t)$, average $\bar{\mu}(t)$ location and also the exact $\Delta RV$ given by Eq.\ref{eq:exact}. 

\begin{figure}
	
	\centering
	\includegraphics[width=8.5cm]{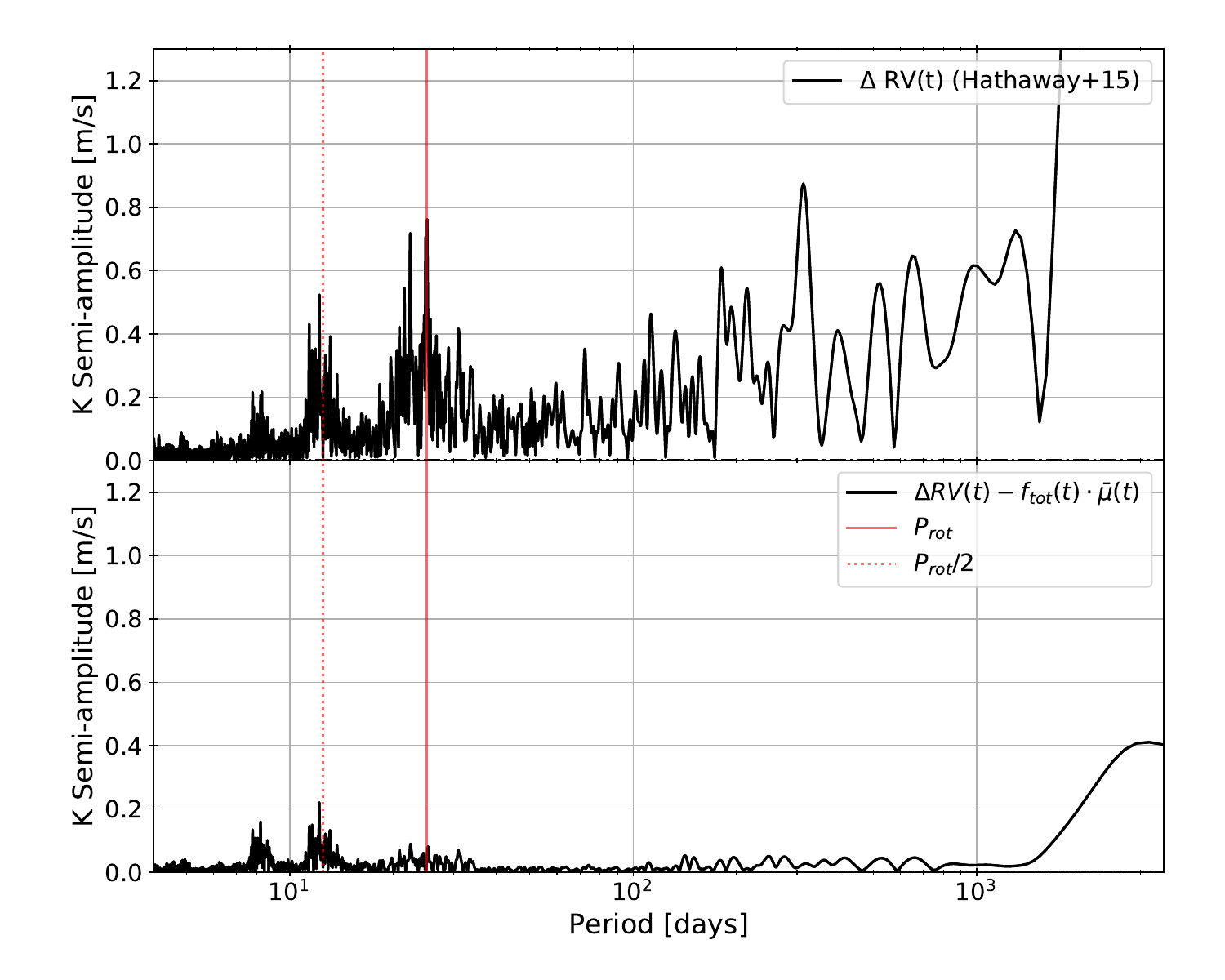}
	\caption{GLS analysis of the RV time-series for the equator-on simulation. \textbf{Top:} RV from the cell integration using the \citet{Hathaway(2015)b} ICB law (green time series forth panel in Fig.\ref{FigSimu}). \textbf{Bottom:} RV residuals (green time series in the bottom panel Fig.\ref{FigSimu}) with $RV_{\text{ICB}} = f_{\text{tot}}(t) \cdot \bar{\mu}(t)$. While the rotational period (vertical red line) is well mitigated, some power at its first harmonic (vertical dotted line) persists.}
	\label{FigSimu2}
 
\end{figure}

In our simulation, the magnetic cycle is very clear as much as the rotational modulation on top of it. The saturation of the average $\bar{\mu}(t)$ location is notable as soon as the total filling factor reach 2\% (between year=2 and year=8) that represents the mid-activity level of our simulated star. A small positive trend is visible for $\bar{\mu}(t)$ along the cycle due to the progressive decrease in latitude (from $40^\circ$ to $0^\circ$) of the active regions with the Butterfly diagram. This trend however depends on the stellar inclination\footnote{as well as on the maximum latitude parameter as shown in \citet{Meunier(2019e)}} as demonstrated here with the $i=30^\circ$ inclination simulation, where the slope is now negative since the active regions are pushed further and further toward the limb along the cycle. A star without Butterfly would trivially present no linear trend in the average location of the active regions.

Because in our approximated model, the RV is just the product of the two time-series, the time modulation of the signal is mainly driven by $f_{\text{tot}}(t)$. This aspect is highlighted by the Pearson coefficient $R$ parameter that measure the degree of colinearity between two variables. The correlation is stronger with the $0^{\rm th}$ statistical moment $f_{\text{tot}}(t)$ than with the $1^{\rm st}$ moment ($R\simeq0.68$ and $R\simeq0.44$ respectively) since the time-variation amplitude is stronger on low statistical moments. 

However, we observed that the value is also changing with time since this latter depends on extra parameters such as the repartition of the active regions and their evolution. Correlations as large as $R=0.80$ and as low as $R=0.60$ are typically observed along the cycle and no dependency were observed between the time evolution of both Pearson coefficients. Interestingly, even if the $\bar{\mu}(t)$ time-series seems to converge when several ARs are present simultaneously, the correlation with the $f_{\text{tot}}(t)$ is not enhanced. 

\begin{figure*}
	
	\centering
	\includegraphics[width=18cm]{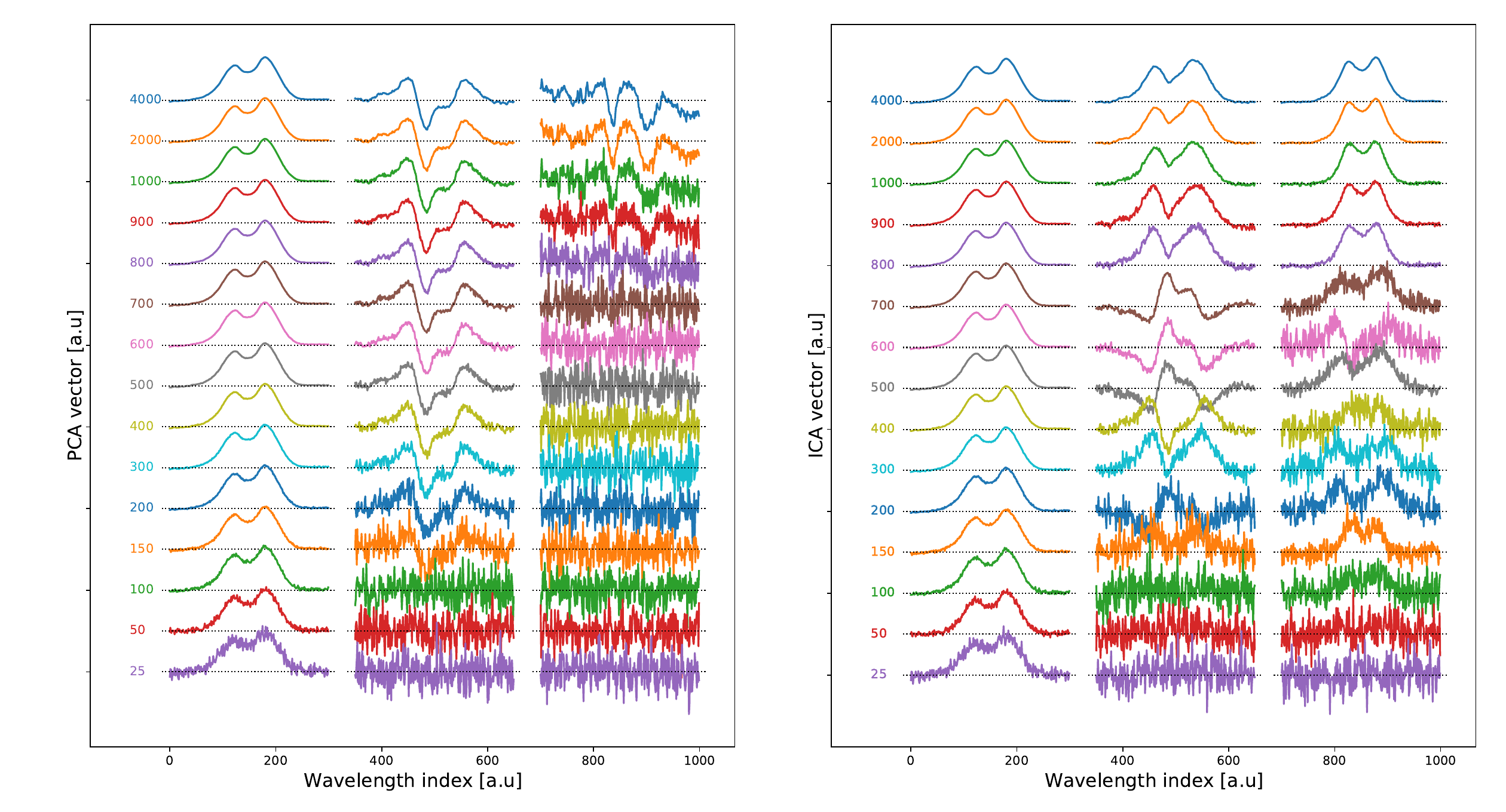}
	\caption{Illustration of the components degradation as a function of the $S/N_{\text{cont}}$ (from $S/N_{\text{cont}}=4000$ to $S/N_{\text{cont}}=25$). An arbitrary vertical offset was added to each S/N realisation for graphical consideration while the $S/N_{\text{cont}}$ value at 4000 $\ang{}$ is indicated on the left. First, second and third components are respectively drawn on the left, middle and right. \textbf{Left:} PCA $E_j(\lambda)$ weighting profiles. While the first component $E_1(\lambda)$ is recovered at all the S/N, the second component $E_2(\lambda)$ is significantly altered for $S/N_{\text{cont}}<300$, and $E_3(\lambda)$ below $S/N_{\text{cont}}<1000$. \textbf{Right:} Same as left for the ICA. Results are qualitatively similar to the PCA results. We observe that below $S/N_{\text{cont}}<800$ a similar mixing of the components, as the one observed for the PCA for all the S/N values, begin to occur.}
	\label{FigSNRcomp}
 
\end{figure*}

We then compared our approximated RV time-series with disk-integrated values using different ICB laws. All the time-series provide similar amplitude signals with a peak-to-peak of 15 m/s roughly and a dispersion around 4.5 m/s. We showed in last panel the residuals between our RV approximation and the others disk-integrated laws and computed the median absolute difference (MAD) of the residuals multiplied by 1.48 to convert it in usual $1\sigma$ root-mean square dispersion. 

When using a purely radial blueshifted  model ($\Delta v_r(\mu) = \Delta v_0 \cdot \mu$) such as \citet{Title(1989)}, the approximation leads exactly to the disk-integrated surface at the numerical precision of our simulation. However, when using non-linear ICB laws, we see that the FF' model approximation becomes less accurate, leading to a dispersion of 41 cm/s with solar observations of \citet{Hathaway(2015b)} for instance. Results are even worse for \citet{Bauer(2018)} since the quadratic dependency of the ICB law is even higher in MHD simulation. We aim to think that the truth is closer to the real solar observations than the MHD simulations at the moment, even if some caveat exists for the extrapolation of SDO results. Since the law provided by \citet{Palumbo(2024)} is even closer to a linear relationship, this one provided the better approximation with residuals of 25 cm/s. Even if this result is the most encouraging, Fig.4 top left panel in \citet{Palumbo(2024)} clearly show that this relation is only an average behaviour and some region-to-region variation is expected.

Analysing the residuals of the RVs with the ICB law from \citet{Hathaway(2015b)}, a generalized Lomb-Scargle periodogram (see Fig.\ref{FigSimu2}) revealed that the 41 cm/s dispersion is made of signals around the first harmonic of $P_{\text{rot}}$ ($\simeq 13$ days) with amplitudes of 20 cm/s. It indicates that while $P_{\text{rot}}$ may be correctly mitigated by the model, the approximation is not sufficient to correct its first harmonics. Furthermore a residual of the magnetic cycle is visible and mitigated from 5 m/s (not visible in the top panel) down to 40 cm/s.

We then discuss the results obtained when strongly decreasing the stellar inclination. The global conclusions are very similar to the equator-on case. In agreement with the result of \citet{Borgniet(2015)}, the smaller inclination displace the ARs towards the limb (due to the $\pm 40^{\circ}$ latitudes range) which decreases their contribution by decreasing the total filling factor and the rotational modulation signal. As expected, the saturation value of $\bar{\mu}$ is lower since the "AR latitude band" is displayed toward the limb. %On the opposite to the equator-on case, the convergence of $\bar{\mu}$ toward its saturation regime is less strong. Our approximation is still in agreement with ICB laws (40 cm/s of dispersion for \citet{Hathaway(2015b)}), that is coherent with the fact that the non-linear regime of the ICB laws mainly arise close to the limb (see Fig.\ref{FigFlow1}) where most ARs are now present.  

\section{S/N dependency of the PCA and ICA decomposition}
\label{app:SN}

In Sect.~\ref{sec:SNcomp}, we investigated how the photonic S/N was impacting the recovery of the components by the PCA and ICA. We illustrate in the Fig.\ref{FigSimu} how the components $E_j(\lambda)$ were affected both for the PCA (left panel) and the ICA (right panel) as a function of the continuum S/N at 4000 $\ang{}$. The noise-injection were spanning $S/N_{\text{cont}}=4000$ to $S/N_{\text{cont}}=25$.

The noise model used is a simple white noise without time-dependency and those results are therefore optimistic considering that instrumental heterogeneous noise exists at low S/N (in particular in the extreme blue).

The $E_1(\lambda)$ weighting profile is easily detected even down to $S/N_{\text{cont}}=50$ since the data are dominated by the plage component and the number of observations remains large. Very rapidly any information is lost from the third component as soon as $S/N_{\text{cont}}<1000$ and similarly for the second component as soon as $S/N_{\text{cont}}<200$. Interestingly, while the PCA component tend to preserve their "shape" even at lower S/N, more versatility is observed with the ICA, highlighting the instability of the decomposition that sometimes even predict profile close to the PCA decomposition (see e.g the $S/N_{\text{cont}}=400$ decomposition). While the ICA remains an useful analysis to perform that provide different information in comparison to the PCA, its instability is difficult to conciliate with the reproducibility necessity of science and we advise to always present both PCA and ICA results jointly. 

\section{Time-series table of the $\alpha$ Cen B $\log R^{\prime}_{\text{HK}}$}
\label{app:Table}

We provided here below the table of the $\log R^{\prime}_{\text{HK}}$ time-series as processed and extracted by ourself. As an extra comment, this extraction is now a classical output of the YARARA post-processing reduction and from the few preliminary tests performed, no instrumental offset nor scaling factors is required between the instruments.

\begin{table}
 \caption{Time-series of $\log R^{\prime}_{\text{HK}}$ observations derived for $\alpha$ Cen B over 30 years. The values provided are the seasonal mean for a given instrument. The X-ray luminosity of \citet{Dewarf(2010)} was scaled in amplitude using overlapping time region with HARPS observation.}
 \label{tab:example1}
 \begin{tabular}{llllll}
  \hline
 & Instrument & Date [year] & $\log R^{\prime}_{\text{HK}}$ & $ \sigma_{\text{HK}}$ & Ref.\\
  \hline

0  &   CORALIE14 &  2017.55 &  -4.946 &       0.030 &  This work \\
1  &   CORALIE14 &  2019.42 &  -4.814 &       0.030 &  This work \\
2  &   CORALIE14 &  2021.36 &  -4.891 &       0.030 &  This work \\
3  &   CORALIE14 &  2022.15 &  -4.932 &       0.030 &  This work \\
4  &   CORALIE98 &  2000.42 &  -4.956 &       0.015 &  This work \\
5  &   CORALIE98 &  2001.73 &  -4.909 &       0.018 &  This work \\
6  &   CORALIE98 &  2005.19 &  -4.857 &       0.026 &  This work \\
7  &  ESPRESSO19 &  2023.50 &  -4.942 &       0.010 &  This work \\
8  &  ESPRESSO19 &  2024.24 &  -4.910 &       0.014 &  This work \\
9  &     HARPS03 &  2003.32 &  -4.868 &       0.020 &  This work \\
10 &     HARPS03 &  2004.23 &  -4.887 &       0.039 &  This work \\
11 &     HARPS03 &  2005.63 &  -4.922 &       0.039 &  This work \\
12 &     HARPS03 &  2007.07 &  -4.943 &       0.039 &  This work \\
13 &     HARPS03 &  2008.32 &  -4.968 &       0.020 &  This work \\
14 &     HARPS03 &  2009.37 &  -4.935 &       0.021 &  This work \\
15 &     HARPS03 &  2010.33 &  -4.897 &       0.035 &  This work \\
16 &     HARPS03 &  2011.29 &  -4.902 &       0.022 &  This work \\
17 &     HARPS03 &  2012.32 &  -4.856 &       0.025 &  This work \\
18 &     HARPS03 &  2013.39 &  -4.879 &       0.030 &  This work \\
19 &     HARPS03 &  2015.39 &  -4.958 &       0.039 &  This work \\
20 &     HARPS15 &  2015.41 &  -4.942 &       0.019 &  This work \\
21 &     HARPS15 &  2016.52 &  -4.940 &       0.011 &  This work \\
22 &     HARPS15 &  2018.34 &  -4.839 &       0.019 &  This work \\
23 &     HARPS15 &  2023.36 &  -4.946 &       0.011 &  This work \\
24 &     HARPS15 &  2024.16 &  -4.918 &       0.014 &  This work \\
25 &        XLum &  1996.02 &  -4.903 &       0.022 &  DeWarf+10 \\
26 &        XLum &  1996.65 &  -4.918 &       0.023 &  DeWarf+10 \\
27 &        XLum &  1998.12 &  -4.948 &       0.038 &  DeWarf+10 \\
28 &        XLum &  2003.12 &  -4.876 &       0.038 &  DeWarf+10 \\
29 &        XLum &  2003.87 &  -4.867 &       0.038 &  DeWarf+10 \\
30 &        XLum &  2004.82 &  -4.903 &       0.038 &  DeWarf+10 \\
31 &        XLum &  2005.84 &  -4.927 &       0.038 &  DeWarf+10 \\
32 &        XLum &  2006.65 &  -4.955 &       0.038 &  DeWarf+10 \\
33 &        XLum &  2007.70 &  -4.957 &       0.038 &  DeWarf+10 \\
34 &        XLum &  2008.67 &  -4.954 &       0.038 &  DeWarf+10 \\
  
  \hline
 \end{tabular}
\end{table}

\section{BSS decomposition of $\alpha$ Cen B}
\label{app:bss}

\begin{figure*}
	
	\centering
	\includegraphics[width=18cm]{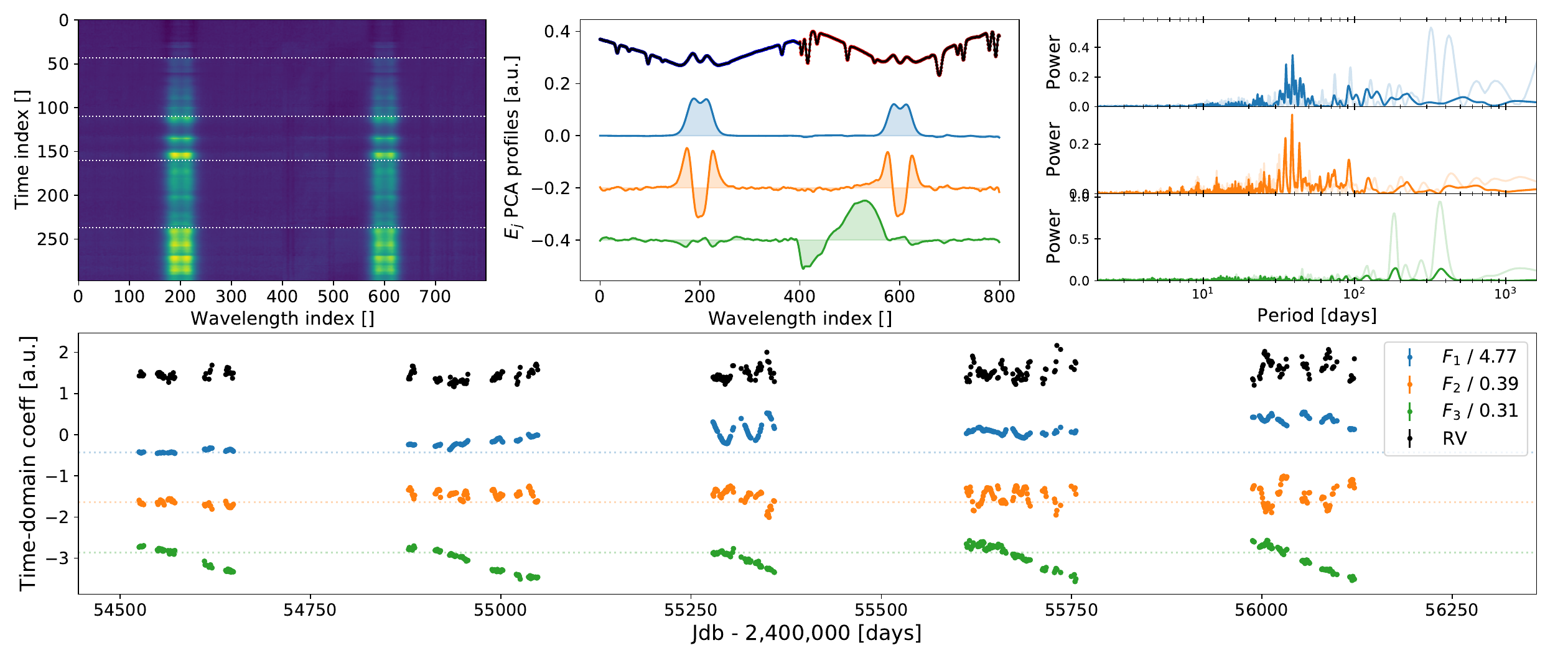}
	\caption{Same as Fig.\ref{FigPCA2} applied on the spectra time-series with the YARARA ghost correction injected back. The third component clearly exhibits the 1-year signal on the left-side of the CaII H line.}
	\label{FigPCA}
 
\end{figure*}

In Sect.~\ref{sec:cenb}, we applied the PCA decomposition on the YARARA corrected spectra and discovered that the third component was likely a ghost residual signature. We confirmed it here by analysing the same spectra with the ghost correction of YARARA injected back (Fig.\ref{FigPCA}). The third weighting profile  $E_3(\lambda)$ is exhibiting the same signature on the left side of the core of the H line, while the $F_3(t)$ time-series shows the characteristic 1-year signature related to the relative movement on the detector of the ghosts and main spectral order along the year induces by the Barycentric Earth RV. 

\begin{figure*}
	
	\centering
	\includegraphics[width=18cm]{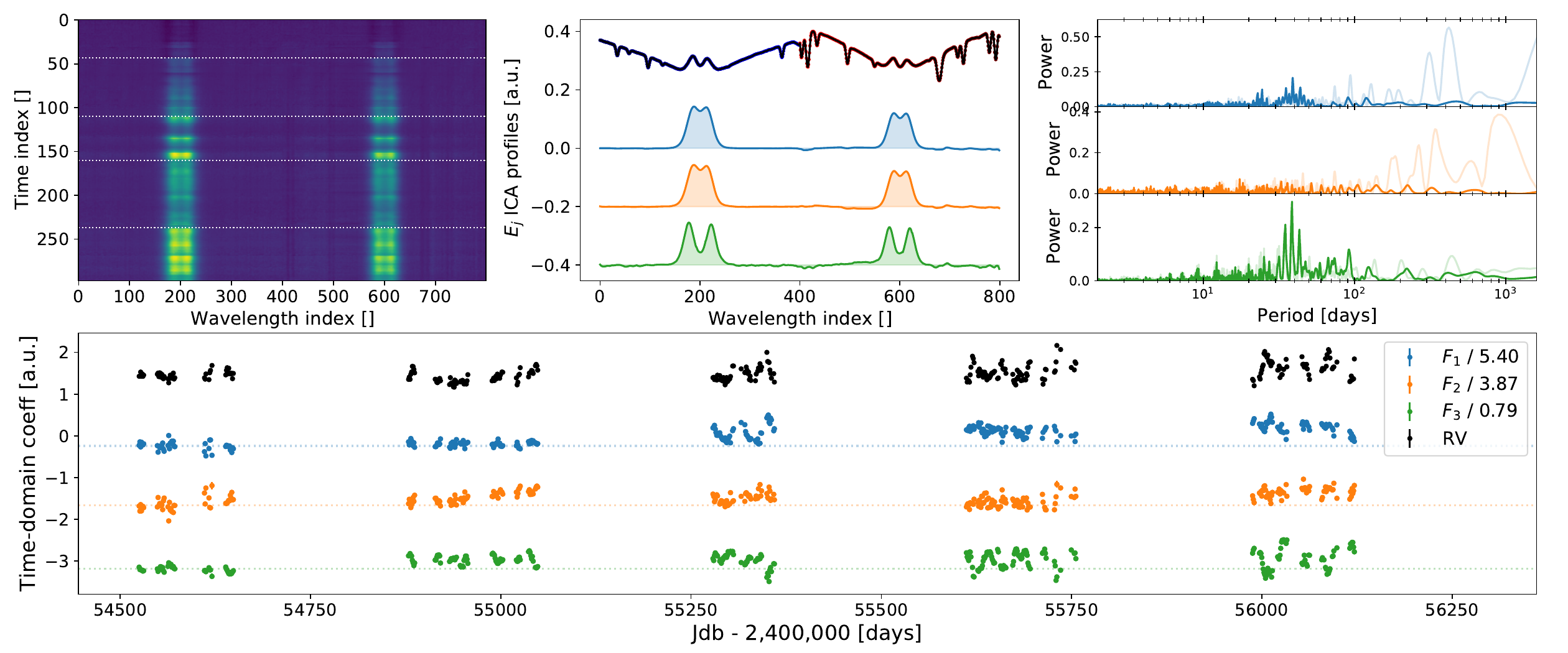}
	\caption{Same as Fig.\ref{FigPCA2} using a 3-components ICA. The ghost contamination is not correctly detected and the main component that contained usually the magnetic cycle is split into two noisier sub-components. Only two components are in fact needed to fit the data, but doing so would preserve the mixing with the residual ghost signature. }
	\label{FigICA2}
 
\end{figure*}

We also displayed in Fig.\ref{FigICA2} the analysis of the spectra time-series (ghost corrected) with a 3-component ICA similarly to the PCA decomposition in Fig.\ref{FigPCA2}. Opposite to the PCA, the ICA is not able to recover the residual ghost contamination and extra tests revealed that only 2 components are required to fit the data. Adding a third component has for effect to split the previous first component into two "sub-components" more noisy. Since neither the 2 nor the 3-component model is correcting for the ghost residual (that is know to be present), we sticked in the paper to the 2-component ICA decomposition from the 2-component PCA subspace described in Sect.~\ref{sec:cenb}. Note that the time-series provided by both algorithm are very similar and we mainly used the ICA to have an alternative view of the decomposition and in particular of the weighting profiles $E_j(\lambda)$.

\section{Implementation of a recipe in YARARA for the recovery of stellar atmospheric parameters}
\label{app:atmos}
For the present paper, a crucial element to obtain is the stellar effective temperature of the star $T_{\text{eff}}$, since we expect that the stellar spectrum follows a smooth evolution along the temperature variable and therefore solar results can be generalized for other spectral types easily. A rapid overview of the different methods existing today is presented in \citet{Smalley(2005)}. Generally, there exists three main classes of methods that provide estimate for the stellar effective temperature based on spectroscopy measurements: 1) the measurement of the continuum, 2) the modelling of stellar lines, and 3) equivalent-width (EW) calibrations. Each method has advantages and disadvantages. 

The method 1) can not be used from high-resolution ground based facilities since as already discussed in paper I, most high-resolution spectrograph are not diffracted limited, and a significant amount of the light goes out of the fiber optics. Furthermore, this loss of light is chromatic given that the PSF is chromatic as well as the refraction index of the atmosphere. As a consequence, different airmass observations will change the "color" of the spectrum and such corrections are difficult to establish and correct a posteriori since they can also depends on the nightly onsite seeing conditions.  

Today the method the most widely used in high-resolution spectroscopy is the method 2), that relies on the modelling of photospheric stellar lines after removing the continuum. When doing so, all the stellar parameters such as the temperature ($T_{\text{eff}}$), the surface gravity (log(g)), the metallicity (Fe/H), the macroturbulence ($v_{M}$), the microturbulence ($v_{m}$), and the projected equatorial velocity ($v \sin i$), are fit using hundreds of stellar lines simultaneously. However the method is often model dependent and assume that atomic parameters of the lines or instrumental parameter, such as oscillator strength or the spectrograph PSF, are known accurately. Furthermore, we did not find any public available Python code that could be easily implemented as a YARARA module for an homogeneous processing of the data. As an example, we found PyHammer \citep{Kesseli(2017)} but this latter is working for low-resolution spectra ($R\sim2000$). Similarly, PySME was recently published \citep{Wehrhahn(2023)}, but this latter also contains C++ and FORTRAN compilation codes, while the ultimate goal of YARARA is to be converted toward a full Python code in the near future.

The last method to derive temperature consists to use EW of specific lines \citep{Gray(2005)} and use their calibration with atmospheric quantities obtained by methods 1) or 2). Such a method was widely used in the past since the line depth ratio of two stellar lines could be used to provide for instance a good effective temperature estimate. Unfortunately, such calibration were mainly done for the derivation of a single atmospheric parameter ($T_{\text{eff}}$) and solar-like metallicity and have not been tabulated for the full parameter space.  

We propose hereafter to rederive those calibrations based on EW methods where the atmospheric parameters values were obtained from other previous published catalogues. This method was used recently by several authors. For instance, \citet{Veyette(2017)} used the EW of TiI and FeI lines to derive some atmospheric parameters. Similarly, \citet{Malavolta(2017)} used the area of CCFs of FeI lines with different excitation potential to establish empirical calibrations with the stellar atmospheric parameters. Also such empirical mapping between spectra and atmospheric parameters was also done via machine learning algorithms \citep{Sharma(2020),Flores(2021)}. 

We first gathered an extensive list of stellar atmospheric parameters catalogues obtained with different methods \citep{Valenti(2005),Takeda(2007),Brewer(2016),Hinkel(2017),Gomes(2021),Soubiran(2022)}. We then estimated the accuracy of the catalogues by performing pair-to-pair catalogue comparison for all stars in common between each pair. In such a way, we would be able to report any potential bias in one of them. No systematic bias was detected neither in $T_{\text{eff}}$ nor FeH. However, we notice larger dispersion for the surface gravity log(g) between the catalogues. We also noticed a bias in particular with the \citet{Gomes(2021)} catalogue that contains a small offset compared to the other most significant catalogues \citep{Valenti(2005),Brewer(2016),Soubiran(2022)}. There is a priori no reason to know which catalogues are right and wrong, in particular we point out that \citet{Gomes(2021)} is the catalogue obtained from the highest resolution spectra. Nevertheless, we note that this catalogue is an updated of previous catalogues \citep{Adibekyan(2012),Delgado(2017)} that were already pointed out by \citet{Gomes(2021)} themselves to contain some bias in surface gravity. Since we are not yet interested in an accurate empirical relation but mostly into a precise one, we added an offset value of $\log(g)=+0.09$ on the reported values of \citet{Gomes(2021)}, that is the mean offset value with the other catalogues. 

We also corrected manually the atmospheric parameters of HD114330\footnote{Reported in their catalogue as "tet Vir"} from \citet{Gomes(2021)} that was clearly an outlier in our plots. Indeed, the A0 star was reported with a low temperature ($T_{\text{eff}}=4369$K), while this star is likely the hottest star in our sample with $T_{\text{eff}}=9250$K \citep{Landstreet(2009)}. 

All the catalogues provides a typical dispersion agreement of $\Delta T=\pm50$ K, $\Delta$FeH $=\pm$0.04 dex and $\Delta\log(g) = \pm0.10$. As a consequence, since we will calibrate our relations using those catalogues, we expect to achieve a similar precision than the one reported here. For all the stars processed, if a star was catalogued by multi-instruments, we extracted the average of the stellar parameters. 

\begin{figure}
	
	\centering
	\includegraphics[width=9cm]{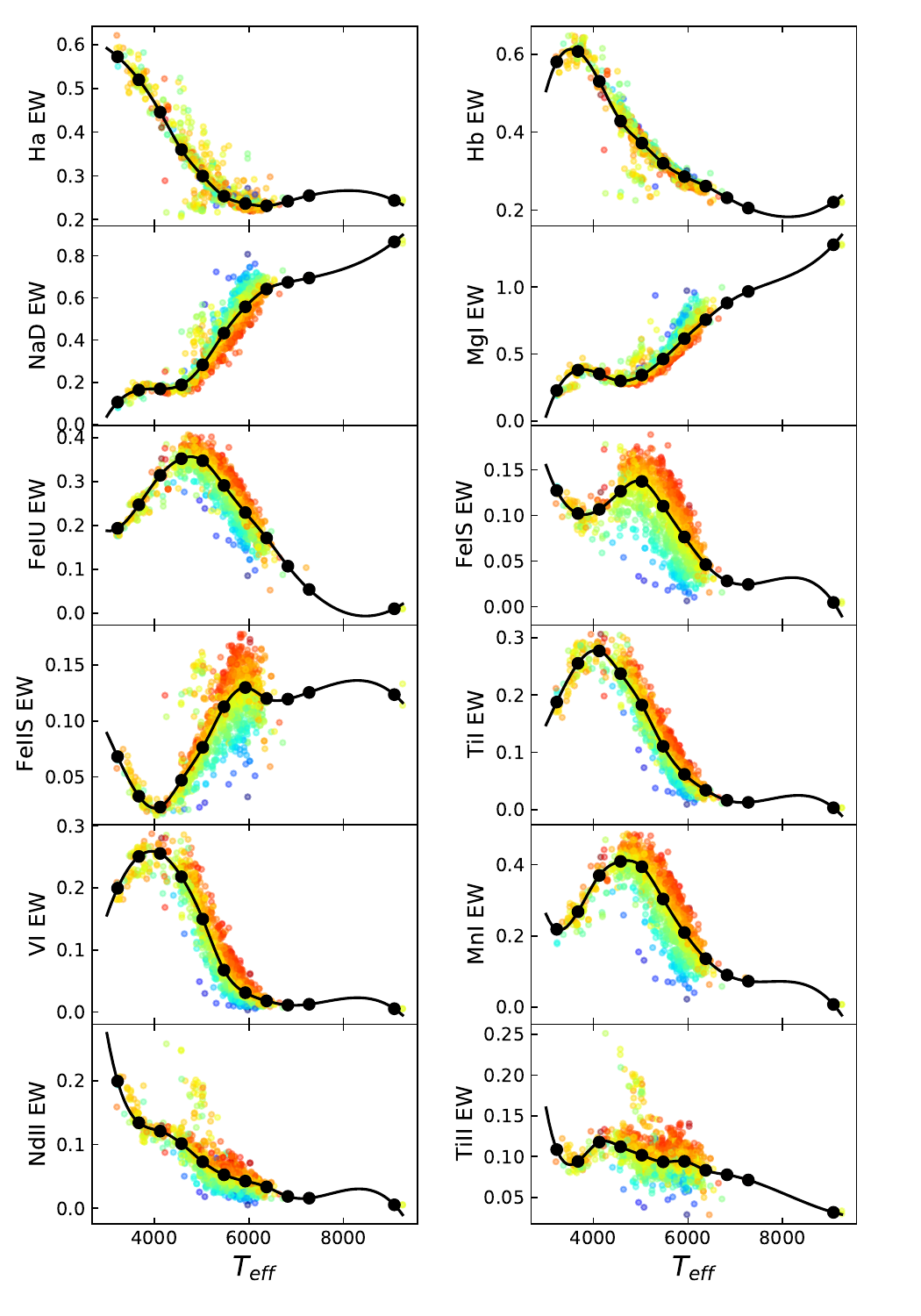}
	\caption{Empirical relations between the EW of 11 species and the effective temperature $T_{\text{eff}}$. FeH is represented by the color code. We also displayed the spline interpolation (black curve) between different temperature binned (black dots). We point out that such 1:1 calibration will not be used in this paper and their representation is purely for qualitative illustrative purpose of the behaviour of the line with the effective temperature and metallicity.}
	\label{FigTeff}
\end{figure}

Next step consists in finding a suitable set of lines to probe the atmospheric parameters. We mainly followed the recommendations raised in \citet{Gray(2005)} and extracted the EW of 12 different species or transitions (see Fig.\ref{FigTeff}). The reason to use EW rather than line depth was to remove the v $\sin$ $i$ dependency, since the rotational broadening preserves the EW of the lines. YARARA is measuring the EW in the core of some of these lines by default to monitor the stellar activity in different chromospheric lines. To simplify the notation, we will refer hereafter to those EW directly by the name of the lines. We also extracted the EW for 7 other species formed in the photosphere (FeI, FeII, TiI, TiII, VI, MnI, NdII). The detailed list of atomic transitions used are listed in Table.\ref{tab:example}. We also produced two FeI line lists by selecting lines strongly sensitive to metallicity effect (FeIS) and lines unsensitive (FeIU), where this selection was done by comparing the spectra of two stars with similar effective temperature, but different metallicities.

The EWs were computed on the cross correlation functions (CCFs) using those line lists, where all the weights of the CCF mask was set to unity to derive an average line profile. The velocity range used for the EW computation was set between $\pm 3 \sigma$, with $\sigma$ the $\sigma$-width of CCFs obtained from the width parameter of a Gaussian fit.

\begin{table}
 \caption{List of the atomic transitions used for the computation of the CCF EW. The half windows (HW) used for the EW computation are provided either in $\ang{}$ for the chromospheric lines or in km/s (sigma width).}
 \label{tab:example}
 \begin{tabular}{lll}
  \hline
  Elem. & HW & Wavelength Air $\lambda$ [$\ang{}$]\\
  \hline
  $H_{\alpha}$ & $0.5$ & 6562.79 \\
  $H_{\beta}$ & $0.5$ & 4861.35 \\
  NaD & $0.5$ & 5889.95, 5895.92 \\
  MgI & $0.5$ & 5167.32, 5172.68, 5183.60 \\

  MnI & $3 \sigma$ & 4709.71, 4754.04, 4762.38, 4766.43, 5377.61, 5394.68, \\ & & 5420.35, 5432.54, 5516.77 \\
  NdII & $3 \sigma$ & 4446.39, 4811.35, 5092.80, 5293.16 \\  
  TiI & $3 \sigma$ & 5866.46, 6064.63, 6126.22, 6258.11, 6261.11, 6336.10, \\ & & 6358.69, 6554.23, 6556.07, 6599.11, 6743.13 \\
  TiII & $3 \sigma$ & 4719.51, 4996.37, 5381.03, 5418.77, 5490.7 \\ 
  VI & $3 \sigma$ & 5604.94, 5627.63, 5646.11, 5703.58, 6039.73, 6090.21,\\ & & 6150.15, 6199.19, 6285.17, 6531.43 \\

  FeIU & $3 \sigma$ & 6200.32, 6252.56, 6297.80, 6421.36, 6430.85, 6546.25,\\ & &  6592.92, 6593.88, 6677.99 \\

  FeIS & $3 \sigma$ & 6187.99, 6574.23, 6597.56, 6627.55, 6703.57, 6705.11,\\ & &  6710.32, 6726.67, 6752.71    \\
  FeII & $3 \sigma$ & 5197.57, 5234.63, 5256.94, 5264.81, 5284.11, 5414.07, \\ & &  5425.25, 5534.84, 5991.37, 6084.11, 6149.25, 6238.39,\\  & &    6247.56, 6369.46, 6432.68, 6456.39, 6516.08 \\
  
  \hline
 \end{tabular}
\end{table}

We explain here below how the combination of all those EWs allow to recover atmospheric parameters. For instance, Hydrogen Balmer lines are very good diagnostic for temperature cooler than 7000K with a very poor sensitivity in metallicity and surface gravity \citep{Smalley(2005)}. However, the peak of their sensitivity\footnote{Given by the largest $|\frac{\partial \text{EW}}{\partial T_{\text{eff}}}|$ value in Fig.\ref{FigTeff}} is around 4500K and the sensitivity decrease rapidly for hotter stars. As a consequence, the use of others lines is required to probe this region in temperature and provide complementary information. Such lines could be iron lines FeI or strong lines such as the Sodium Doublet (NaD) or Magnesium Triplet (MgI) for which the largest sensitivity is around 5800K, even if those latters also contains metallicity dependency as expected. On the opposite, some lines are barely unsensitive to temperature and can therefore be used to track change in metallicity or surface gravity. This is for instance the case of TiII, where the sensitivity to surface gravity of ionized lines is due to the combination of Botzmann and Saha equation and the pressure dependency of the population levels.

Even if Fig.\ref{FigTeff} shows some 1:1 calibration, it is very clear that the optimal way to solve the problem is to use all the EW information simultaneously to fit the three atmospheric parameters. 

\begin{equation}
 F(EW_1,EW_2,...) = (T_{\text{eff}},\text{FeH},\log(g))
\end{equation}

\begin{figure}
	
	\centering
	\includegraphics[width=9cm]{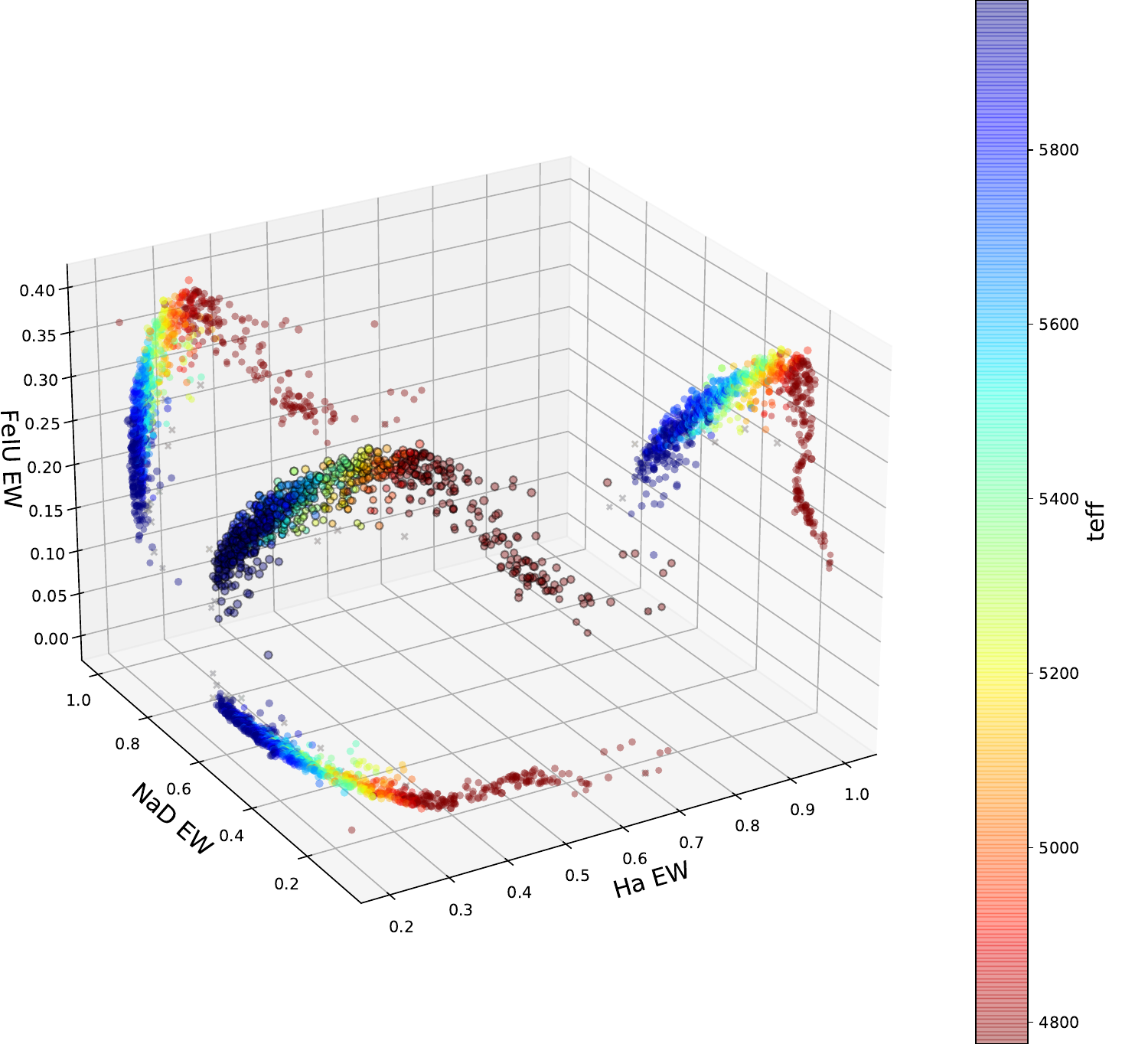}
	\caption{Representation of the 5D cloud (NaD, $H_{\alpha}$, FeIU, $T_{\text{eff}}$, FeH). Metallicity is represented by the marker size and the effective temperature by the color. This space is by designed build to enhanced temperature effect and have a low sensitivity to [Fe/H] by using $H_{\alpha}$ and unsensitive iron lines (FeIU). Projection of the 5D cloud into the 4D plans are also depicted with symbol without edgecolor.}
	\label{Fig5Da}
\end{figure}

\begin{figure}
	
	\centering
	\includegraphics[width=9cm]{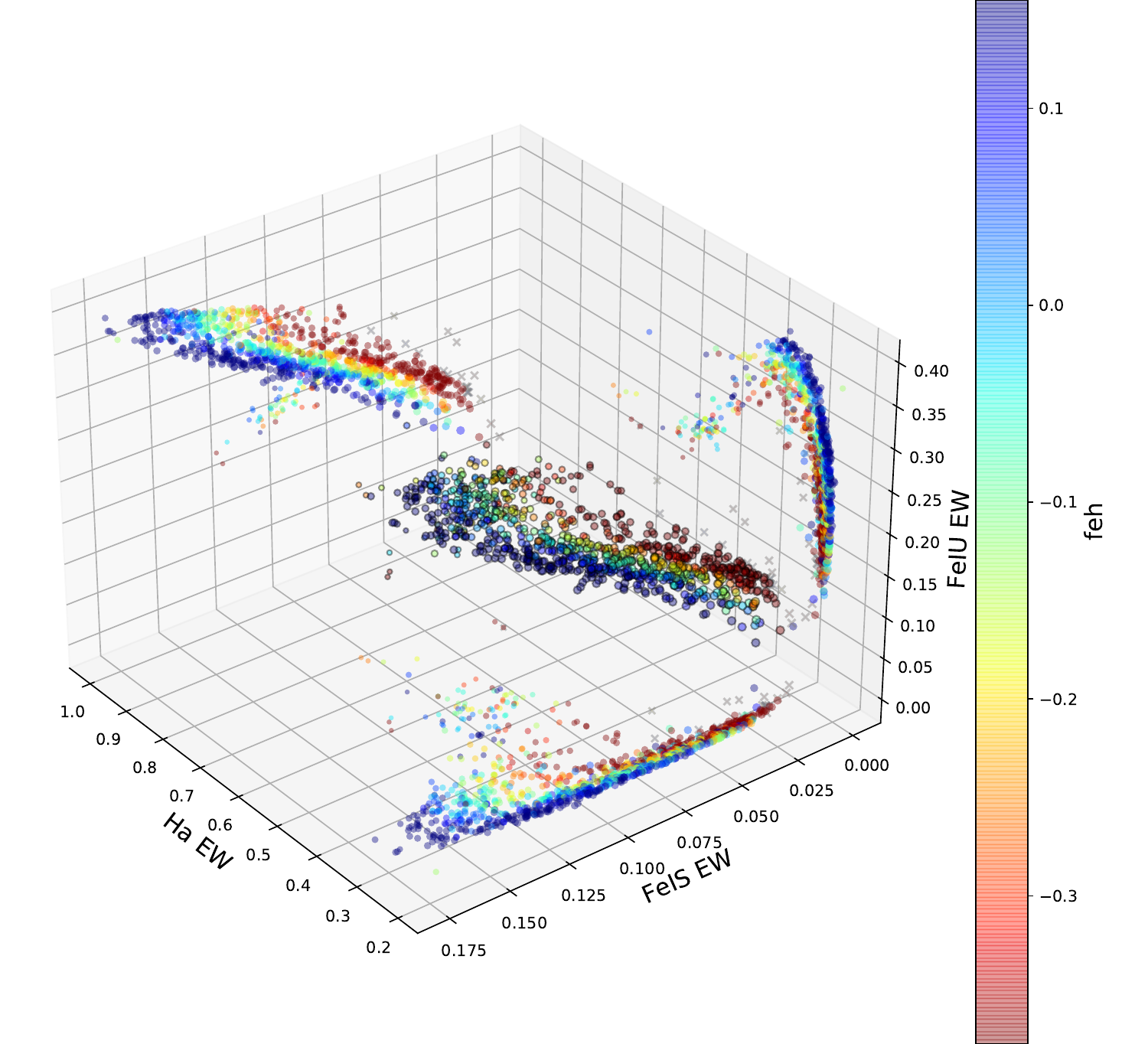}
	\caption{Representation of the 5D cloud ($H_{\alpha}$, FeIS, FeIU, FeH, $T_{\text{eff}}$). Temperature is represented by the marker size and the metallicity by the color. The use of iron lines sensitive to abundance effect (FeIS) allow to fit for the metallicity dependency. }
	\label{Fig5Db}
\end{figure}

\begin{figure}
	
	\centering
	\includegraphics[width=9cm]{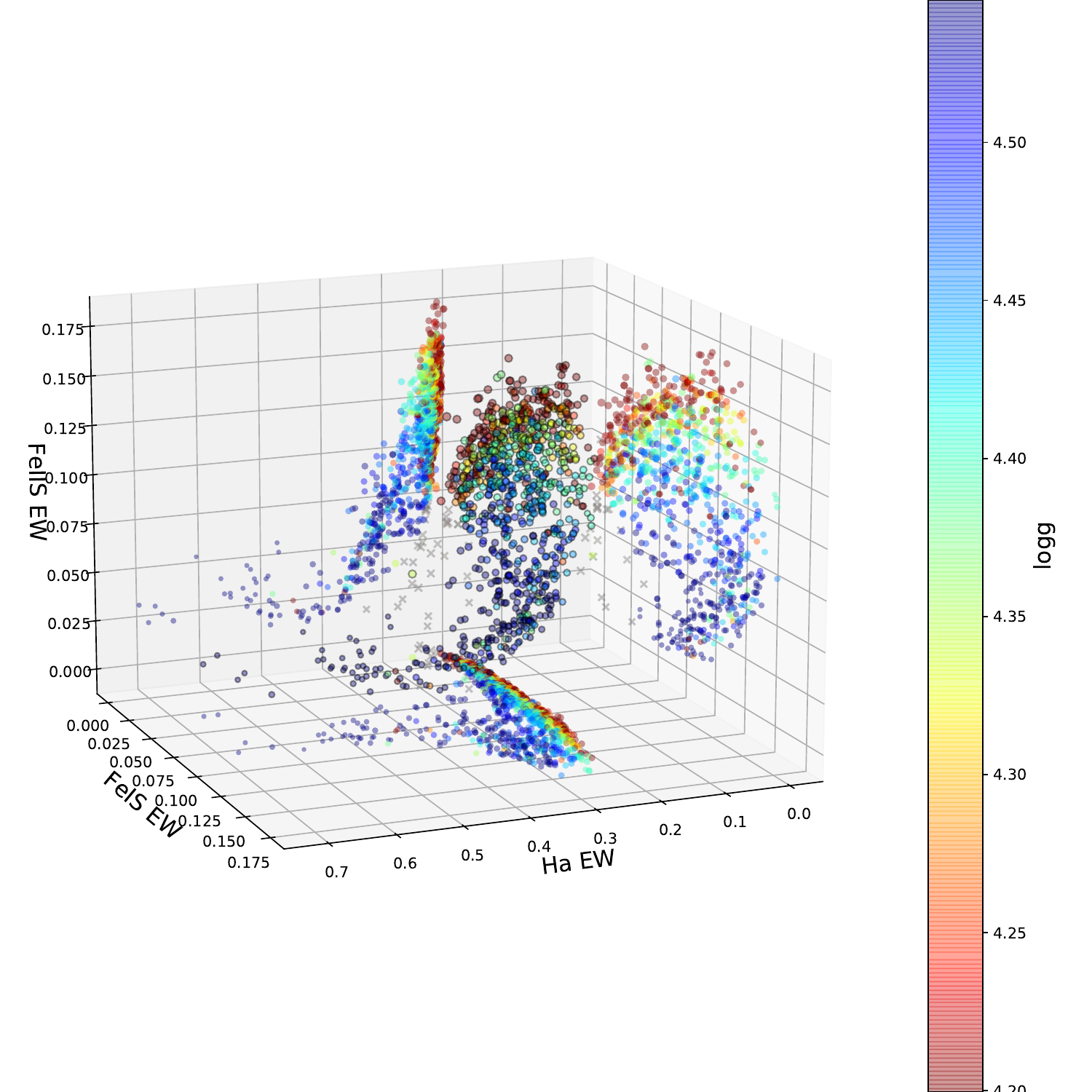}
	\caption{Representation of the 5D cloud ($H_{\alpha}$, FeIS, FeIIS, $\log(g)$, $T_{\text{eff}}$). Temperature is represented by the marker size and the surface gravity by the color. The use of ionised iron lines sensitive to pressure effect (FeIIS) allow to fit for the surface gravity dependency.}
	\label{Fig5Dc}
\end{figure}

We displayed for instance in Fig.\ref{Fig5Da}, Fig.\ref{Fig5Db} and Fig.\ref{Fig5Dc} different subspaces that highlight how, by increasing the dimension of the parameter space, some dependencies become easier to identify. The main question to solve is how to fit the mapping function $F$ on the data. 

It is not straightforward to perform the mapping between the 12D space toward the 3D atmospheric parameter space. We naturally expect that the model has to be "smooth", but except this condition there is a priori no good analytical model to fit the data.  Furthermore, we can see that the data are not randomly distributed, but rather follow a high-dimensional curvy structure as visible for the temperature in Fig.\ref{Fig5Da}. This implies that the parameter space is strongly inhomogeneous which prevent the use of polynomial functions for instance. 

Based on this consideration, we rather used a non-analytical model such as those given by machine learning algorithms. We used a random forest from the XGBoost library \citep{Chan(2016)} with the MultipleOutputRegressor object. We splited our data into a training and testing set using 75\% and 25\% of the data respectively. We set the hyperparameters of the random forest by searching for the parameters that were providing the smallest residuals of the training set, but also by looking for the closest dispersion in the residuals between the train and test sample in order to avoid the overfitting issue. We only optimized the number of tree $N$ ($N=60$), the depth of the tree $D$ ($D=3$) and the learning rate $\eta$ ($\eta=0.1$)

Given that some of our EWs are strongly collinear, we could think to first reduce the dimensionality of the space by applying a PCA before the random forest,
however we did not notice any improvement by doing so, which could be explained by the low number of features ($\sim$10) compared to the size of the dataset ($\sim$1000) and the low noise in the data.

Our precision, as measured from the test set, is $\Delta T=\pm70$ K, $\Delta$FeH $=\pm$0.08 dex and $\Delta\log(g) = \pm0.07$. Applying the model on the solar HARPS-N spectrum \citep{Dumusque(2021)}, the recovered solar atmospheric parameters ($T_{\text{eff},\odot}$, FeH$_{\odot}$, $\log(g)_{\odot}$) are: (5724, $0.01$, 4.39) which is not far from the expected solar values (5777, 0.00, 4.44) provided in \citet{Smalley(2005)} and compatible at $1\sigma$ of our uncertainties. 

Given that some of these lines are also used for stellar activity monitoring, we could wonder if the stellar activity level could not affect the recovery of the stellar parameters. We prove here below that this is not the case given that the line deformation induced by activity and effective parameters are completely different order of magnitude. As a proof of it, we computed the atmospheric parameters using one spectrum of $\alpha$ Cen B at the lowest and highest activity level. The effect of activity on the stellar lines is so small, that the recovered parameters are exactly the same, except for the metallicity that change from $\text{FeH}=0.10$ to $\text{FeH}=0.11$.

%%%%%%%%%%%%%%%%%%%%%%%%%%%%%%%%%%%%%%%%%%%%%%%%%%

% Don't change these lines
\bsp	% typesetting comment
\label{lastpage}
\end{document}